\documentclass[runningheads,a4paper]{llncs}

\usepackage{amsfonts}
\usepackage{stmaryrd}
\usepackage{amsmath,amsfonts,amssymb}
\usepackage{graphicx}

\usepackage{tabularx}

\usepackage{hyperref}
\usepackage{url}
\usepackage{fca}

\usepackage{algorithmic}
\usepackage{algorithm}

\usepackage{tikz}
\usepackage{tikz-qtree}

\usepackage{colortbl}
\usepackage{xcolor}

\definecolor{tomato}{RGB}{255,99,71}

\definecolor{lightseagreen}{RGB}{32,178,170}

\definecolor{darkkhaki}{RGB}{189,183,107}

\definecolor{Gray}{gray}{0.9}

\definecolor{obj-red}{rgb}{1,0,0}
\definecolor{obj-blue}{rgb}{0,0,1}
\definecolor{obj-green}{rgb}{0,0.5,0}
\definecolor{grey}{gray}{.5}

\newcommand{\bi}{\begin{itemize}}
\newcommand{\ei}{\end{itemize}}
\newcommand{\be}{\begin{enumerate}}
\newcommand{\ee}{\end{enumerate}}

\newcommand{\bd}{\begin{definition}}
\newcommand{\ed}{\end{definition}}

\newcommand{\K}{\context}

\newcommand{\fcaX}{$\times$}

\begin{document}

\title{Introduction to Formal Concept Analysis and Its Applications in Information Retrieval and Related Fields}
\titlerunning{Introduction to Formal Concept Analysis and Its Applications in IR}
\author{Dmitry I. Ignatov}

\institute{National Research University Higher School of Economics, Moscow\\
dignatov@hse.ru\\
}

\maketitle

\begin{abstract}
This paper is a tutorial on Formal Concept Analysis (FCA) and its applications. FCA is an applied branch of Lattice Theory, a mathematical discipline which enables formalisation of concepts as basic units of human thinking and analysing data in the object-attribute form. Originated in early 80s, during the last three decades, it became a popular human-centred tool for knowledge representation and data analysis with numerous applications. Since the tutorial was specially prepared for RuSSIR 2014, the covered FCA topics include Information Retrieval with a focus on visualisation aspects, Machine Learning, Data Mining and Knowledge Discovery, Text Mining and several others.

\end{abstract}

\section{Introduction}

According to \cite{Manning:2008}, ``information retrieval (IR) is finding material
(usually documents) of an unstructured nature (usually text) that satisfies an information
need from within large collections (usually stored on computers).'' In the past,
only specialized professions such as librarians had to retrieve information on a regular
basis. These days, massive amounts of information are available on the Internet and
hundreds of millions of people make use of information retrieval systems such as web
or email search engines on a daily basis. Formal Concept Analysis (FCA) was introduced
in the early 1980s by Rudolf Wille as a mathematical theory \cite{wille:1982,Ganter:1999} and
became a popular technique within the IR field. FCA is concerned with the formalisation of
concepts and conceptual thinking and has been applied in many disciplines such as
software engineering, machine learning, knowledge discovery and ontology construction during the last
20-25 years. Informally, FCA studies how objects can be hierarchically grouped together with their common attributes.

\begin{figure}
	\centering
		\includegraphics[width=1\textwidth]{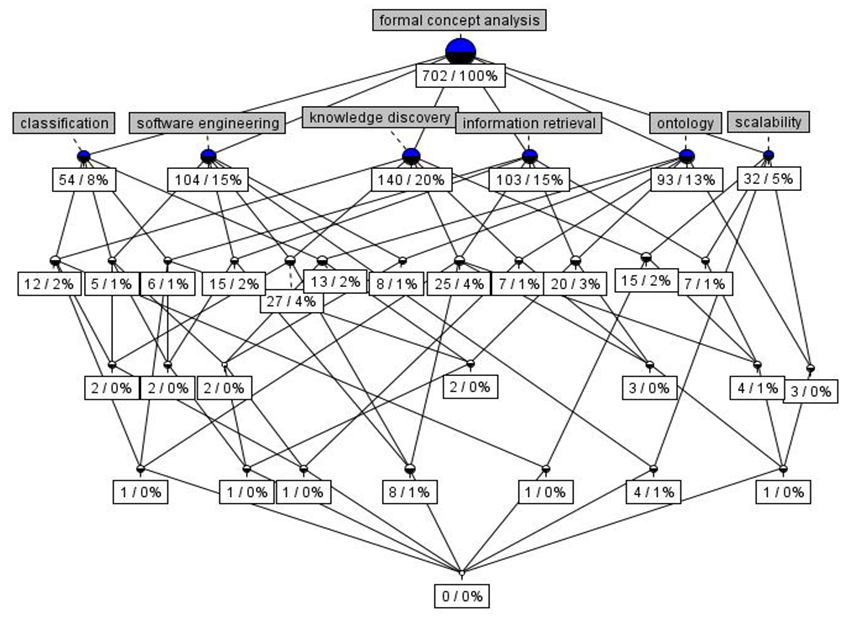}
	\caption{The lattice diagram representing collection of 702 papers on FCA including 103 papers on FCA and IR (2003-2009).}
	\label{fig:IRLatt}
\end{figure}

The core contributions of this tutorial from IR perspective are based on our surveys~\cite{Poelmans:2012,Poelmans:2013a,Poelmans:2013b} and experiences in both fields, FCA and IR. In our surveys we visually represented the literature on FCA and IR using concept lattices as well as several related fields, in which the objects are the scientific
papers and the attributes are the relevant terms available in the title, keywords
and abstract of the papers. You can see an example of such a visualisation in Figure \ref{fig:IRLatt} for papers published between 2003 and 2009. We developed a toolset with a central FCA component that we used to index the papers with a thesaurus containing terms related to FCA research
and to generate the lattices. The tutorial also contains a partial overview of the papers on using FCA in Information Retrieval with a focus on visualisation.

In 2013 European Conference on Information Retrieval \cite{DBLP:conf/ecir/2013} hosted a thematic workshop \href{http://fcair.hse.ru/}{FCA meets IR} which was devoted to two main issues:
\begin{itemize}
	\item How can FCA support IR activities including but not limited to query analysis, document representation, text classification and clustering, social network mining, access to semantic web data, and ontology engineering?
	\item How can FCA be extended to address a wider range of IR activities, possibly including new retrieval tasks?
\end{itemize}

Claudio Carpineto delivered an invited lecture at the workshop -- ``FCA and IR: The Story So Far''.
The relevant papers and results presented there are also discussed in the tutorial. 

Since the tutorial preparations were guided by the idea to present the content at a solid and comprehensible level accessible even for newcomers, it is a balanced combination of theoretical foundations, practice and relevant applications. Thus we provide intro to FCA, practice with main tools for FCA, discuss FCA in Machine Learning and Data Mining, FCA in Information Retrieval and Text Mining, FCA in Ontology Modeling and other selected applications. Many of the used examples are real-life studies conducted by the course author.

The target audience is Computer Science, Mathematics and Linguistics students, young scientists, university teachers and researchers who want to use FCA models and tools in their IR and data analysis tasks.

The course features five parts. Each part is placed in a separate section and contains a short highlight list to ease the navigation within the material. An archive with supplementary files for exercises and examples is available at~\footnote{\url{http://bit.ly/RuSSIR2014FCAtut}}. Section 2 contains introduction to FCA and related notions of Lattice and Order Theory.
In Section 3, we describe selected FCA tools and provide exercises.  	
	Section 4 provides an overview of FCA-based methods and applications in Data Mining and Machine Learning, and describes an FCA-based tool for supervised learning, QuDA (Qualitative Data Analysis). Section 5 presents the most relevant part of the course, FCA in Information Retrieval and Text Mining. Penultimate Section 6 discusses FCA in Ontology Modeling and gives an example of FCA-based Attribute Exploration  technique on building the taxonomy of transportation means. Section 7 concludes the paper and briefly outlines prospects and limitations of FCA-based models and techniques.  
	
\section{Introduction to FCA}\label{sec:intro}

Even though that many disciplines can be dated back to Aristotle’s time, more closer prolegomena of FCA can be found, for example, in the Logic of Port Royal (1662)\cite{Arnauld:1996}, an old philosophical concept logic, where a concept  was treated as a pair  of its extent and its intent (yet without formal mathematical apparatus). 

Being a part of lattice theory, concept lattices are deeply rooted in works of Dedekind, Birkgoff~\cite{Birkhoff:1967} (Galois connections and ``polarities''), and Ore~\cite{Ore:1944} (Galois connections), and, later, on Barbut\&Monjardet~\cite{Barbut:1970} (treillis de Galois, i.e. Galois lattices). 

In fact, the underlying structure, Galois connection, has a strong impact in Data Analysis\cite{Duquenne:1999,Wolski:2004,Kuznetsov:2005a,Carpineto:2005}.

In this section, we mainly reproduce basic definitions from Ganter\&Wille's book on Formal Concept Analysis \cite{Ganter:1999}. However, one can find a good introductory material, more focused on partial orders and lattices, in the book of Davey and Priestly \cite{D&P:2002}.
An IR-oriented reader may also find the following books interesting and helpful~\cite{Carpineto:2005,Dominich:2008}.

There were several good tutorials with notes in the past, for example, a basic one~\cite{Wolff:1993} and more theoretical with algorithmic aspects~\cite{Belohlavek:2008}.

We also  refer the readers to some online materials that might be suitable for self-study purposes \footnote{\url{http://www.kbs.uni-hannover.de/~jaeschke/teaching/2012w/fca/}},\footnote{\url{http://www.upriss.org.uk/fca/fcaintro.html}},\footnote{\url{http://ddll.inf.tu-dresden.de/web/Introduction_to_Formal_Concept_Analysis_(WS2014)/en}}.

A short section summary:	
	
\begin{itemize}

\item Binary Relations, Partial Orders, Lattices, Line (Hasse) Diagram.
\item Galois Connection, Formal Context, Formal Concept, Concept Lattice.
\item Concept Lattice drawing. Algorithms for concept lattices generation (na\"{i}ve, Ganter's algorithm, Close-by-One).
\item Attribute Dependencies: implications, functional dependencies. Armstrong Rules. Implication bases (Stem Base, Generator base).
\item Many-valued contexts. Concept scaling.
		
\end{itemize}

\subsection{Binary Relations, Partial Orders, Lattices, Hasse Diagram}

The notion of a \textbf{set} is fundamental in mathematics. In what follows, we consider only finite sets of objects. 

\begin{definition} A \textbf{binary relation} $R$ between two sets $A$ and $B$ is a set of all pairs $(a,b)$ with $a \in A$ and $b \in B$., i.e., a subset of their Cartesian product $A \times B$, the set of all such pairs.
\end{definition}

Sometimes it is convenient to write $aRb$ instead $(a,b)\in R$ for brevity. 	If $A=B$ then $R \subseteq A \times A$ is called a binary relation on the set $A$.

\begin{definition} A binary relation $R$ on a set $A$  is called a \textbf{partial order relation} (or shortly a partial order), if it satisfies the following conditions for all elements $a, b, c \in A$:

\begin{enumerate}
\item  $aRa$ (reflexivity)
\item  $aRb$ and $a\neq b \implies$ not $aRb$ (antisymmetry)
\item $aRb$ and $bRc \implies aRc$ (transitivity)
\end{enumerate}

\end{definition}  

We use symbol $\leq$ for partial order, and in case $a \leq b$ and $a \neq b$ we write $a \le b$. We read $a \leq b$ as ``$a$ is  less of equal to $b$''. A partially ordered set (or poset) is a pair $(P,\leq)$, where $P$ is a set and $\leq$ is an partial order on $P$.

\begin{definition} Given a poset $(P,\leq)$, an element $a$ is called a \textbf{lower neighbour} of $b$, if $a \le b$ and there is no such $c$ fulfilling $a \le c \le b$. In this case, $b$ is also an \textbf{upper neighbour} of $a$, and we write $a \prec b$.
\end{definition} 

Every finite ordered poset  $(P, \leq)$ can be depicted as a \textbf{line diagram} (many authors call it Hasse diagram). Elements of $P$ are represented by small circles in the plane. If $a \leq b$, the circle corresponding to $a$ is depicted higher than the circle corresponding to $b$, and the two circles are connected by a line segment. One can check whether some $a \leq b$ if there is an ascending path from $b$ to $a$ in the diagram.

\begin{example} The poset $P$ is given by its incidence cross-table, where $\times$ in a cell means that the corresponding pair of row  and column elements $x$ and $y$ are related as follows $x\leq y$.

\begin{minipage}{0.30\linewidth}
	\begin{tabular}{|c|ccccc|}
  \hline
  $\leq$ &  a & b & c & d & e\\
  \hline
	a & $\times$ &  &  & &\\
  b & $\times$ & $\times$ & $\times$ & &\\
  c &  &  & $\times$ &  &\\
  d & $\times$ & $\times$ & $\times$ & $\times$ & $\times$\\
	e &  &  &  & & $\times$\\
  \hline
  \end{tabular}
		\end{minipage}
	\hfill	
	\begin{minipage}{0.30\linewidth}
		 \includegraphics[width=1.0\textwidth]{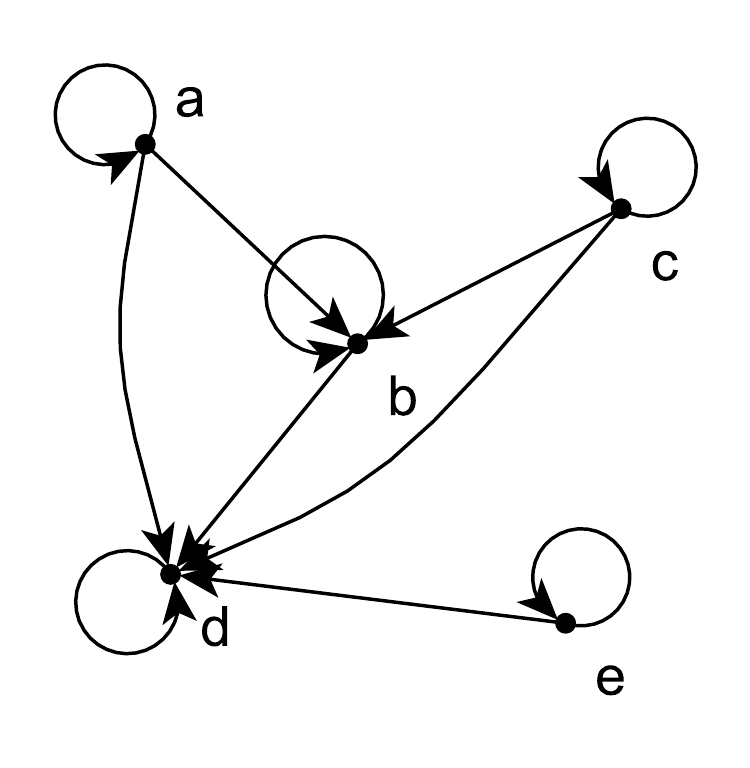}\\
		The graph of $P$.
			\end{minipage}
	\hfill
		\begin{minipage}{0.30\linewidth}
			\includegraphics[width=1.0\textwidth]{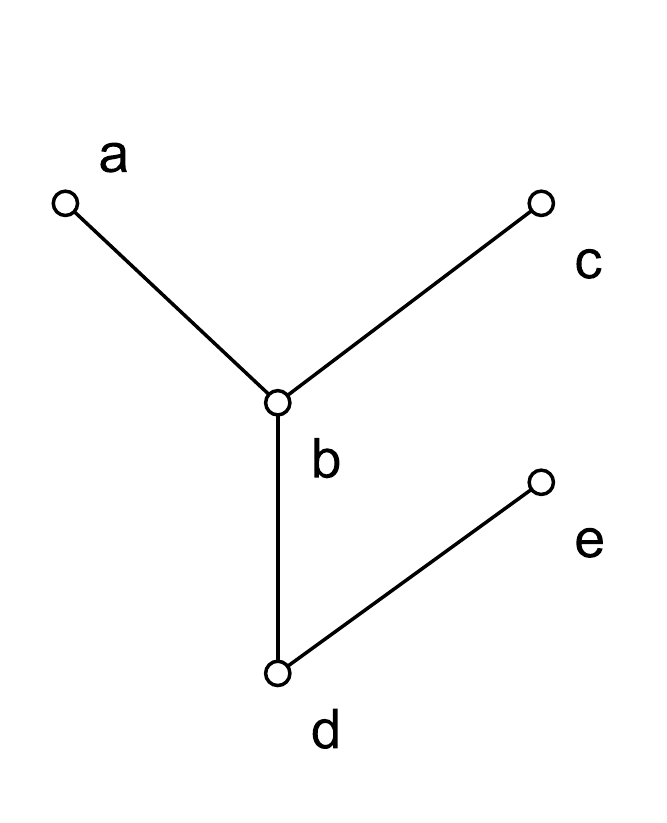}\\
			The line diagram  of $P$.
	\end{minipage}
	$\square$
\end{example}

\begin{definition} Let  $(P,\leq)$ be a poset and $A$ a subset of $P$. A \textbf{lower bound} of $A$ is an element $l$ of $P$ with $l \leq A$ for all $a \in A$. An \textbf{upper bound} of $A$ is defined dually. If there is a largest element in the set of all lower bounds of $A$, it is called the \textbf{infimum} of $A$ and is denoted by $inf A$ or $\bigwedge A$. Dually, if there is a smallest element in the set of all upper bounds, it is called \textbf{supremum} and denoted by $sup A$ or $\bigvee A$. 
\end{definition} 
For $A=\{a,b\}$ we write $x\wedge y$ for $inf A$ and  $x\vee y$ for $sup A$. Infimum and supremum are also called \textbf{meet} and \textbf{join}.

\begin{definition} A poset   $\mathbf{L}=(L,\leq)$  is a \textbf{lattice}, if for any two elements $a$  and $b$ in $L$ the supremum $a\vee b$  and the infimum $a\wedge b$  always exist. $\mathbf{L}$ is called a \textbf{complete lattice}, if the supremum $\bigvee X$ and the infimum $\bigwedge X$ exist for any subset $A$ of $L$. For every complete lattice $\mathbf{L}$ there exist its largest element, $\bigvee L$, called the \textbf{unit element }of the lattice, denoted by $\mathbf{1}_L$. Dually, the smallest element $\mathbf{0}_L$ is called the \textbf{zero element}.
\end{definition}

\begin{example} In Fig.~\ref{fig:posetandlatt} there are the line diagrams of the poset $P$, which is not a lattice, and the lattice $L$.
It is interesting that $P$ has its largest and smallest elements, $\mathbf{1}_P$ and $\mathbf{0}_P$; the pair of its elements, $s$ and $t$, has its infumum, $s \vee t=\mathbf{0}_P$, but there is no a supremum for it. In fact, ${p,t}$ does not have a smallest element in the set of all its upper bounds.

\begin{figure}[h]
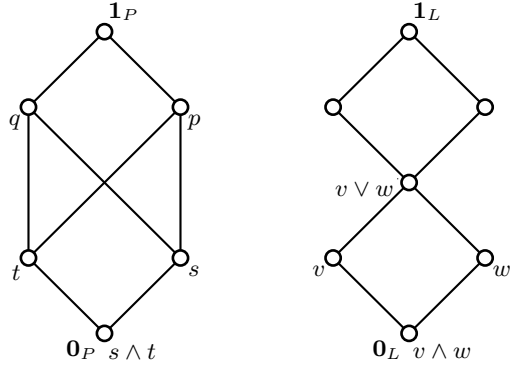

\centering
{\unitlength .5mm
\begin{diagram}{50}{90}
\Node{1}{20}{5}  
\Node{2}{40}{25}  
\Node{3}{0}{25}  
\Node{4}{40}{65}  
\Node{5}{0}{65}  
\Node{6}{20}{85}
\Edge{1}{2}
\Edge{1}{3}
\Edge{2}{5}
\Edge{3}{4}
\Edge{2}{4}
\Edge{3}{5}
\Edge{4}{6}
\Edge{5}{6}
\rightObjbox{2}{2}{2}{$s$}  
\leftObjbox{3}{2}{2}{$t$}  
\rightObjbox{4}{2}{2}{$p$}  
\leftObjbox{5}{2}{2}{$q$}
\leftObjbox{1}{2}{2}{$\mathbf{0}_P$}  
\rightAttbox{6}{1}{3}{$\mathbf{1}_P$}
\rightObjbox{1}{1}{3}{$s \wedge t$}   
\end{diagram}}\hspace*{14mm}
{\unitlength .5mm
\begin{diagram}{50}{90}
\Node{1}{20}{5}  
\Node{2}{40}{25}  
\Node{3}{0}{25}  
\Node{4}{20}{45}
\Node{5}{40}{65}  
\Node{6}{0}{65}  
\Node{7}{20}{85}
\Edge{1}{2}
\Edge{1}{3}
\Edge{2}{4}
\Edge{3}{4}
\Edge{4}{5}
\Edge{4}{6}
\Edge{5}{7}
\Edge{6}{7}
\leftObjbox{1}{2}{2}{$\mathbf{0}_L$}  
\rightObjbox{2}{2}{2}{$w$}  
\leftObjbox{3}{2}{2}{$v$} 
\leftObjbox{4}{4}{0}{$v \vee w$}   
\rightAttbox{7}{1}{3}{$\mathbf{1}_L$}  
\rightObjbox{1}{1}{3}{$v \wedge w$}  
\end{diagram}}
  \caption{The line diagrams of the order, which is not a lattice (left), and the order, which is a lattice (right)}
  \label{fig:posetandlatt}
\end{figure}

$\square$

\end{example}

\section{Galois Connection, Formal Context, Formal Concept, Concept Lattice}\label{sec:basics}

\begin{definition} Let $\varphi: P \to Q $  and $\psi: Q \to P$ be maps between two posets $(P,\leq)$ and $(Q,\leq )$. Such a pair of maps is called a \textbf{Galois connection} between the ordered sets if:

\begin{enumerate}
\item $p_1 \leq p_2  \Rightarrow \varphi p_1 \geq \varphi p_2$ 
\item $q_1 \leq q_2  \Rightarrow \psi q_1 \geq \psi q_2$ 
\item  $p \leq \psi\varphi p  \Rightarrow q \leq \varphi\psi q$. 
\end{enumerate}

\end{definition}

\begin{exercise}
Prove that a pair $(\varphi, \psi)$ of maps is a Galois connection if  and only if  $p \leq \psi q 	\Leftrightarrow q \leq \psi p$. $\square$
\end{exercise}

\begin{exercise}
Prove that for every Galois connection $(\varphi, \psi)$ 
$$\psi= \psi\varphi\psi \mbox{ and } \varphi= \varphi\psi\varphi.$$ $\square$
\end{exercise}

\begin{definition} A \textbf{formal context} $\context=(G,M,I)$ consists of two sets $G$ and $M$ and a relation  $I$ between $G$ and $M$. The elements of $G$ are called the \textbf{objects} and the elements of  $M$ are called the \textbf{attributes }of the context. The notation $gIm$ or $(g,m) \in I$ means that the object $g$ has attribute $m$. 
\end{definition} 

\begin{definition} 

For $A \subseteq G$, let

$$A^\prime := \{m \in M | (g,m) \in I \mbox{ for all } g \in A\}$$

and, for $B \subseteq M$, let  

$$B^\prime := \{ g \in G | (g,m) \in I \mbox{ for all } m \in B\}.$$

These operators are called \textbf{derivation operators} or \textbf{ concept-forming operators} for $\context=(G,M,I)$. 

\end{definition}

\begin{proposition} Let $(G,M,I)$ be a formal context, for subsets $A, A_1, A_2 \subseteq G$ and $B\subseteq M$ we have
\label{primeprop}
\begin{enumerate}
\item $A_1 \subseteq A_2$ iff $A_2^\prime \subseteq A_1^\prime$,
\item $A \subseteq A^{\prime\prime}$,
\item $A = A^{\prime\prime\prime}$ (hence, $A'''' = A''$),
\item $(A_1 \cup A_2)' = A'_1 \cap A'_2$,
\item $A\subseteq  B'\Leftrightarrow B\subseteq A' \Leftrightarrow A\times B \subseteq I$.
\end{enumerate}

Similar properties hold for subsets of attributes.

\end{proposition}

\begin{exercise}
Prove properties of operator $(\cdot)^\prime$ from proposition~\ref{primeprop}. $\square$
\end{exercise}

\begin{definition}

A \textbf{closure operator} on set $G$ is a mapping $\varphi \colon 2^G \to 2^G$ with the following properties:

\begin{enumerate}
\item[1.] ${\varphi{\varphi X}} = {\varphi X}$ (\textbf{idempotency})

\item[2.] $X\subseteq {\varphi X}$ (\textbf{extensity})

\item[3.] $X\subseteq Y\Rightarrow {\varphi X}\subseteq {\varphi Y}$ (\textbf{monotonicity})
\end{enumerate}

For a closure operator $\varphi$ the set  $\varphi X$ is called \textbf{closure} of $X$.

A subset $X\subseteq G$ is called \textbf{closed} if ${\varphi X} = X$.

\end{definition}

\begin{exercise} Let $(G,M,I)$ be a context, prove that operators
$$(\cdot)''\colon 2^G\to 2^G,\ (\cdot)''\colon 2^M\to 2^M$$
are closure operators. $\square$
\end{exercise}

\begin{definition} 
A \textbf{formal concept} of a formal context $\context=(G,M,I)$ is a pair $(A,B)$ with $A \subseteq G$,
$B \subseteq M$, $A^\prime = B$ and $B^\prime = A$. The sets $A$ and $B$ are called the extent and the intent
of the formal concept $(A,B)$, respectively. The \textbf{subconcept-superconcept relation} is
given by $(A_1,B_1) \leq (A_2,B_2)$ iff $A_1 \subseteq A_2$ ($B_1 \subseteq B_2$).
\end{definition}

This definition says that every formal concept has two parts, namely, its extent and
intent. This follows an old tradition in the \textit{Logic of Port Royal (1662)}, and is in line with the International Standard ISO 704 that formulates the following definition: ``A concept is considered to be a unit of thought constituted of two parts: its extent and its intent.''

\begin{definition} The set of all formal concepts of a context $\context$ together with the order relation $I$ forms a complete lattice, called the \textbf{concept lattice } of $\context$ and denoted by $\BV(\context)$.
\end{definition}

\begin{example}\label{ex:geomcxt} The context with four geometric figures and four attributes is below.

\begin{minipage}[h]{0.30\linewidth}
\begin{tabular}{|c|c||cccc|}
  \hline
  \multicolumn{2}{|c|}{G $\setminus$ M} & a & b & c & d\\
  \hline
	\hline
  1 & \begin{tikzpicture}
      \shadedraw [shading=axis] (0,0) -- (0.4,0) -- (0.2,0.36) -- (0,0);
      \end{tikzpicture}
      &\fcaX &  & & \fcaX\\
  2 & \begin{tikzpicture}
      \shadedraw [shading=axis] (0,0) -- (0.4,0) -- (0,0.4) -- (0,0);
      \end{tikzpicture} &\fcaX &  & \fcaX &\\
  3 & \begin{tikzpicture}
      \shadedraw [shading=axis] (-0.4,0) -- (0.4,0) -- (0.4,0.4) -- (-0.4,0.4) -- (-0.4,0);
      \end{tikzpicture} &  &\fcaX & \fcaX &\\
  4 & \begin{tikzpicture}
      \shadedraw [shading=axis] (-0.2,0) -- (0.2,0) -- (0.2,0.4) -- (-0.2,0.4) -- (-0.2,0);
      \end{tikzpicture} & & \fcaX &\fcaX & \fcaX\\
  \hline
  \end{tabular}
\end{minipage}
\hfill  
\begin{minipage}[h]{0.30\linewidth}  
    \begin{tabular}{p{5cm}}
  \bi
  \item[] Objects:
  \item[\textbf{1}] -- equilateral triangle,
  \item[\textbf{2}] -- rectangle triangle,
  \item[\textbf{3}] -- rectangle,
  \item[\textbf{4}] -- square.

  \ei
  \end{tabular}
 \end{minipage} 
\hfill
\begin{minipage}[h]{0.30\linewidth}
    \begin{tabular}{p{5cm}}
  \bi
  \item[] {Attributes:}
  \item[\textbf{a}] -- has 3 vertices,
  \item[\textbf{b}] -- has 4 vertices,
  \item[\textbf{c}] -- has a direct angle,
  \item[\textbf{d}] -- equilateral.
  \ei
  \end{tabular}
  \end{minipage} 
$\square$
\end{example}

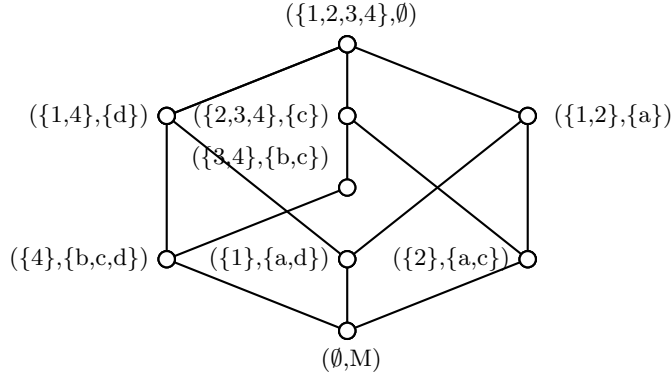
\begin{figure}
	\centering
		
\begin{tikzpicture}
\begin{scope}[scale=0.95]
\draw (4,5) node[label=above:{{ (\{1,2,3,4\},$\emptyset)$ }}]{};
\draw (1.5,4) node[label=left:{{ (\{1,4\},\{d\})}}] {};
\draw (4,4) node[label=left:{{ (\{2,3,4\},\{c\}) }}] {};
\draw (6.5,4) node[label=right:{{ (\{1,2\},\{a\}) }}] {};
\draw (4,3) node[label=above left:{{ (\{3,4\},\{b,c\}) }}]{};
\draw (4,2) node[label=left:{{(\{1\},\{a,d\}) }}] {};
\draw (4,1) node[label=below:{{ ($\emptyset$,M) }}] {};
\draw (1.5,2) node[label=left:{{ (\{4\},\{b,c,d\}) }}] {};
\draw (6.5,2) node[label=left:{{ (\{2\},\{a,c\}) }}] {};
\tikzstyle{every node}=
  [fill=white,draw=black,circle,scale=0.70]
\tikzstyle{every path}=[thick]
\draw (4,5) node (A) {} -- (1.5,4) node (B) {};
\draw (4,5) node (A) {} -- (4,4) node (C) {};
\draw (4,5) node (A) {} -- (6.5,4) node (D) {};
\draw (4,4) node (C) {} -- (4,3) node (E) {};
\draw (1.5,4) node (B) {} -- (1.5,2) node (F) {};
\draw (6.5,4) node (D) {} -- (6.5,2) node (H) {};
\draw (4,2) node (G) {} -- (4,1) node (I) {};
\draw (1.5,2) node (F) {} -- (4,1) node (I) {};
\draw (6.5,2) node (H) {} -- (4,1) node (I) {};
\draw (1.5,4) node (B) {} -- (4,2) node (G) {};
\draw (6.5,4) node (D) {} -- (4,2) node (G) {};
\draw (4,3) node (E) {} -- (1.5,2) node (F) {};
\draw (4,4) node (C) {} -- (6.5,2) node (H) {};
\draw (4,5) node (A) {} -- (1.5,4) node (B) {};
\end{scope}
\end{tikzpicture}
			
	\caption{The line diagram of the concept lattice for the context of geometric figures}
	\label{fig:geomlatt}
\end{figure}

\bd
For every two formal concepts $(A_1,B_1)$ and $(A_2,B_2)$ of a certain formal context their \textbf{greatest common subconcept} is defined as follows:

$$(A_1,B_1) \wedge (A_2,B_2) = (A_1 \cap A_2, (B_1 \cup B_2)'').$$

The \textbf{least common superconcept} of $(A_1,B_1)$ and $(A_2,B_2)$ is given as

$$(A_1,B_1) \vee (A_2,B_2) = ((A_1 \cup A_2)'', B_1 \cap B_2).$$
\ed

We say supremum instead of ``least common superconcept'', and instead of  ``greatest common subconcept'' we use the term infimum.

\begin{figure}[h]
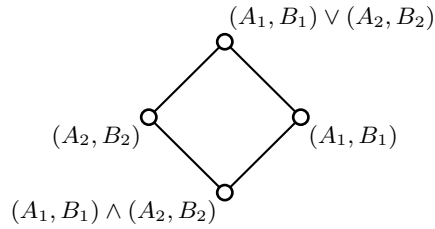

\centering
{\unitlength .5mm
\begin{diagram}{50}{50}
\Node{1}{20}{5}  
\Node{2}{40}{25}  
\Node{3}{0}{25}  
\Node{4}{20}{45}
\Edge{1}{2}
\Edge{1}{3}
\Edge{2}{4}
\Edge{3}{4}
\leftObjbox{1}{2}{2}{$(A_1,B_1) \wedge (A_2,B_2)$}  
\rightObjbox{2}{2}{2}{$(A_1,B_1)$}  
\leftObjbox{3}{2}{2}{$(A_2,B_2)$}  
\rightAttbox{4}{1}{3}{$(A_1,B_1)\vee (A_2,B_2)$}  
\end{diagram}}
  \caption{Supremum and infimum of two concepts}
  \label{fig:father}
\end{figure}

It is possible to define supremum and infumum operations for an arbitrary set of concepts of a certain context. 
This is done in the first part of Theorem~\ref{thrm:fca}.

\bd
A subset $X\subseteq L$ of lattice $(L,\leq)$ is called \textbf{supremum-dense} if any lattice element $v\in L$ can be represented as
$$v = \bigvee \{x\in X \mid x\leq v\}.$$
Dually for  \textbf{infimum-dense} subsets.
\ed

The Basic Theorem of Formal Concept Analysis below defines not only supremum and infimum of arbitrary sets of concepts; it also answer the question whether concept lattices have any special properties. In fact, the answer is ``no'' since every concept lattice is (isomorphic to some) complete lattice. That is one can compose a formal context with objects $G$, attributes $M$ and binary relation $I \subset G \times M$ such that the original complete lattice is isomorphic $\BVGMI$. Even though the theorem does not reply how such a context can be built, but rather describes all possibilities to do this.

\begin{theorem}{\emph{\textbf{Basic Theorem of Formal Concept Analysis} ([Wille 1982],[Ganter, Wille 1996])}}\label{thrm:fca}

Concept lattice $\BV (G,M,I)$ is a complete lattice. For arbitrary sets of formal concepts
$$\{(A_j, B_j)\mid j\in J\}\subseteq \BV(G,M,I)$$
their infimum and supremum are given in the following way:
$$\bigwedge_{j\in J}(A_j,B_j)=(\bigcap_{j\in J}A_j,
(\bigcup_{j\in J}B_j)''),$$
$$\bigvee_{j\in J}(A_j,B_j)=
((\bigcup_{j\in J}A_j)'', \bigcap_{j\in J}B_j).$$
\noindent A complete lattice $L$ is isomorphic to a lattice  $\BV (G,M,I)$ iff there are mappings
$\gamma\colon G\to V$ and $\mu\colon M\to V$ such that
$\gamma(G)$ is supremum-dense in
$\mathbf{L}$, $\mu(M)$ is infimum-dense in  $\mathbf{L}$, and $gIm\Leftrightarrow$  $\gamma g\leq \mu m$ for all
$g\in G$ and all $m\in M$.
In particular, $\mathbf{L}$ is isomorphic to $\BV(L,L,\leq)$.
\end{theorem}

 An interested reader may refer to Ganter\&Wille's book on FCA~\cite{Ganter:1999} for further detailed and examples.

\subsection{Concept Lattice drawing and algorithms for concept lattices generation}

One can obtain the whole set of concepts of a particular context $\context$ simply by definition, i.e. it is enough to enumerate all subsets of objects $A\subseteq G$ (or attributes $B \subseteq M$) and apply derivation operators to them. For example, for the context from example~\ref{ex:geomcxt} and empty set of objects, $A=\emptyset$, one may obtain $A^\prime=\emptyset^\prime=\{a,b,c,d\}=B$, and then by applying $(\cdot)^\prime$ second time $B^\prime=\emptyset$. Thus, the resulting concept is $(A,B)=(\emptyset, M)$.

\begin{proposition}\label{prop:congen}
Every formal concept of a context $(G,M,I)$ has the form $(X^{\prime\prime},X^\prime)$ for some
subset $X \subseteq G$ and the form $(Y^{\prime}, Y^{\prime\prime})$ for some subset $Y \subseteq M$. Vice versa all such
pairs of sets are formal concepts.
\end{proposition}

One may follow the na\"{i}ve algorithm below:

\be
\item $\CGMI := \emptyset$

\item For every subset $X$ of $G$, add $(X^{\prime\prime},X^\prime)$ to $\CGMI$.
\ee

\begin{exercise}
1. Prove proposition~\ref{prop:congen}.
2. For the context of geometric figures from example~\ref{ex:geomcxt} find all formal concepts. $\square$
\end{exercise} 

Since the total number of formal concept is equal to $2^{\min(|G|,|M|)}$ in the worst case, this na\"{i}ve approach is quite inefficient even for small contexts. However, let us assume that now we know how find concepts and we are going to build the line diagram of a concept lattice. 

\be
\item Draw a rather small circle for each formal concept such that a circle for a concept is always depicted higher
than the all circles for its subconcepts.
\item Connect each circle with the circles of its lower neighbors.
\ee

To label concepts by attribute and object names in a concise form, we need the notions of object and attributes concepts.

\bd\label{def:OAconc}
Let $(G,M,I)$ be a formal context,  then for each object $g \in G$ there is the \textbf{object concept} $(\{g\}^{\prime\prime}, \{g\}^\prime)$  and for each attribute $m \in M$ the \textbf{attribute concept} is given by $(\{m\}^\prime, \{m\}^{\prime\prime})$.
\ed
 
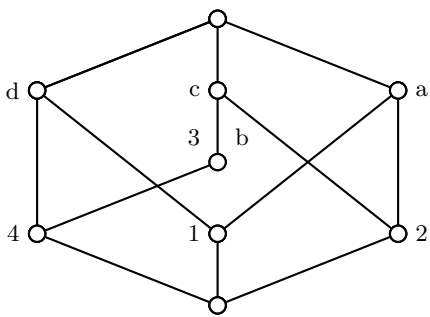
\begin{figure}
	\centering
		
\begin{tikzpicture}
\begin{scope}[scale=0.95]
\draw (4,5) node[label=above:{{}}]{};
\draw (1.5,4) node[label=left:{{d}}] {};
\draw (4,4) node[label=left:{{c}}] {};
\draw (6.5,4) node[label=right:{{a}}] {};
\draw (4,3) node[label=above left:{{3}}, label=above right:{{b}}]{};
\draw (4,2) node[label=left:{{1}}] {};
\draw (4,1) node[label=below:{{ }}] {};
\draw (1.5,2) node[label=left:{{4}}] {};
\draw (6.5,2) node[label=right:{{2}}] {};
\tikzstyle{every node}=
 [fill=white,draw=black,circle,scale=0.70]
\tikzstyle{every path}=[thick]
\draw (4,5) node (A) {} -- (1.5,4) node (B) {};
\draw (4,5) node (A) {} -- (4,4) node (C) {};
\draw (4,5) node (A) {} -- (6.5,4) node (D) {};
\draw (4,4) node (C) {} -- (4,3) node (E) {};
\draw (1.5,4) node (B) {} -- (1.5,2) node (F) {};
\draw (6.5,4) node (D) {} -- (6.5,2) node (H) {};
\draw (4,2) node (G) {} -- (4,1) node (I) {};
\draw (1.5,2) node (F) {} -- (4,1) node (I) {};
\draw (6.5,2) node (H) {} -- (4,1) node (I) {};
\draw (1.5,4) node (B) {} -- (4,2) node (G) {};
\draw (6.5,4) node (D) {} -- (4,2) node (G) {};
\draw (4,3) node (E) {} -- (1.5,2) node (F) {};
\draw (4,4) node (C) {} -- (6.5,2) node (H) {};
\draw (4,5) node (A) {} -- (1.5,4) node (B) {};
\end{scope}
\end{tikzpicture}
			
	\caption{An example of reduced labeling for the lattice of geometric figures}
	\label{fig:geomlattRL}
\end{figure}

So, if one has finished a line diagram drawing for some concept lattice, it is possible to label the diagram with attribute names:
one needs to attach the attribute $m$ to the circle representing the concept $(\{m\}^\prime, \{m\}^{\prime\prime})$.
Similarly for labeling by object names: one needs to attach each object $g$ to the circle representing
the concept $(\{g\}^{\prime\prime}, \{g\}^\prime)$. An example of such reduced labeling is given in Fig.~\ref{fig:geomlattRL}.

The na\"{i}ve concept generation algorithm is not efficient since it enumerates all subsets of $G$ (or $M$). For homogeneity, in what follows we reproduce the pseudocodes of the algorithms from~\cite{Kuznetsov:2002}. There are different algorithms that compute closures for only some subsets of $G$ and use an efficient test to check whether the current concept is generated first time (canonicity test). Thus, Ganter's Next Closure algorithm does not refer the list of generated concepts and uses little storage space.

Since the extent of a concept defines its intent in a unique way, to obtain the set of all formal concepts, it is enough to find closures either of subsets of objects or subsets of attributes.

We assume that there is a linear order ($<$) on $G$. The algorithm starts by examining the set consisting of the
object maximal with respect to $<$ ($max(G)$), and finishes when the canonically
generated closure is equal to $G$. Let $A$ be a currently examined subset of $G$. The
generation of $A^{\prime\prime}$ is considered canonical if $A^{\prime\prime}\setminus A$ does not contain $g < max(A)$. If the
generation of $A^{\prime\prime}$ is canonical (and $A^{\prime\prime}$ is not equal to $G$), the next set to be examined
is obtained from $A^{\prime\prime}$ as follows:

$$A^{\prime\prime} \cup \{g\} \setminus \{h | h \in  A^{\prime\prime} \wedge g < h \}, \mbox{ where } g=max(\{h | h \in G \setminus A^{\prime\prime} \}).$$

Otherwise, the set examined at the next step is obtained from $A$ in a similar way, but
the added object must be less (w.r.t. $<$) than the maximal object in $A$:

$$A^{\prime\prime} \cup \{g\} \setminus \{h | h \in  A \wedge g < h \}, \mbox{ where } g=max(\{h | h \in G \setminus A \wedge h < max(A) \}).$$

The pseudocode code is given in Algorithm~\ref{nextcl-alg} and the generation protocol of \textsc{NextClosure} for the context of geometric figures is given in Table~\ref{tbl:nextcloprot}.

\begin{algorithm}
\caption{NextClosure}\label{nextcl-alg}
    \begin{algorithmic}[1]
      	\REQUIRE $\mathbb{K}=(G,M,I)$ is a context\\
    	 	\ENSURE $L$ is the concept set
    	\STATE $L:=\emptyset, A:=\emptyset, g:=max(G)$
    	\WHILE{$A \neq G$}
			\STATE{$A:=A^{\prime\prime} \cup \{g\} \setminus \{h | h \in  A \wedge g < h \}$}
			\IF{$\{h | h \in  A \wedge g \leq h \}=\emptyset$}
			\STATE{$L:=L\cup \{(A^{\prime\prime},A^\prime)\}$}
			\STATE{$g:=g=max(\{h | h \in G \setminus A^{\prime\prime} \})$}
			\STATE{$A:=A^{\prime\prime}$}
			\ELSE
			\STATE{$g=max(\{h | h \in G \setminus A \wedge h < g \})$}
			\ENDIF
			\ENDWHILE
			\RETURN $L$
    \end{algorithmic}
\end{algorithm}

The NextClosure algorithm produces the set of all concepts in time $O(|G|^2|M||L|)$
and has polynomial delay $O(|G|^2|M|)$.

\begin{table}
	\centering
	\caption{Generation protocol of NextClosure for the context of geometric figures }\label{tbl:nextcloprot}
		\begin{tabular}{|c|c|c|c|}
\hline			
$g$ & $A$ & $A^{\prime\prime}$ & formal concept $(A,B)$\\
\hline
4 & $\{4\}$ & $\{4\}$ & ($\{4\}$, $\{2, 4\}$)\\
3 & $\{3\}$ & $\{3\}$ & ($\{3\}$, $\{2, 3\}$)\\
4 & $\{3, 4\}$ & $\{3, 4\}$ & ($\{3, 4\}$, $\{2\}$)\\
2 & $\{2\}$ & $\{1, 2\}$ & non-canonic generation\\
1 & $\{1\}$ & $\{1\}$ & ($\{1\}$, $\{1, 3, 4\}$)\\
4 & $\{1, 4\}$ & $\{1, 4\}$ & ($\{1, 4\}$, $\{4\}$)\\
3 & $\{1, 3\}$ & $\{1, 2, 3\}$ & non-canonic generation\\
2 & $\{1, 2\}$ & $\{1, 2\}$ & ($\{1, 2\}$, $\{1, 3\}$)\\
4 & $\{1, 2, 4\}$ & $\{1, 2, 3, 4\}$ & non-canonic generation\\
3 & $\{1, 2, 3\}$ & $\{1, 2, 3\}$ & ($\{1, 2, 3\}$, $\{3\}$)\\
4 & $\{1, 2, 3, 4\}$ &  $\{1, 2, 3, 4\}$ & ($\{1, 2, 3, 4\}$, $\{\}$)\\
\hline
\end{tabular}
\end{table}

We provide a simple recursive version of \textsc{CbO}. The algorithm generates concepts according to the lectic (lexicographic)
order on the subsets of $G$ (concepts whose extents are lectically
less are generated first). By definition  $A$ is lectically less than $B$ if $A \subseteq B$, or $B \not \subset A$
and $min((A \cup B) \setminus (B \cap A)) \in A$. Note that the \textsc{NextClosure} algorithm computes concepts
in a different lectic order: $A$ is lectically less than $B$ if
$min((A \cup B) \setminus (B \cap A)) \in B$. The order in which concepts are generated by \textsc{CbO} is beneficial when the line diagram is constructed: the first generation of the concept is always canonical, which makes it possible to find a concept in the tree and to draw appropriate diagram edges. \textsc{NextClosure}-like lectic order allows binary search, which is helpful
when the diagram graph has to be generated after the generation of all concepts.

\begin{algorithm}
\caption{Close by One}\label{cbo}
    \begin{algorithmic}[1]
      	\REQUIRE $\mathbb{K}=(G,M,I)$ is a context\\
    	 	\ENSURE $L$ is the concept set
    	\STATE{ $L:=\emptyset$}
    	\FORALL{$g \in G$}
			\STATE{Process($\{g\},g,(\{g\}^{\prime\prime},g)$)}
			\ENDFOR
			\RETURN $L$
    \end{algorithmic}
\end{algorithm}

\begin{algorithm}
\caption{Process$(A, g, (C, D))$ with $C = A^{\prime\prime}$ and $D = A^\prime$ and < the lexical order on object names}\label{process}
    \begin{algorithmic}[1]
      	\REQUIRE $\mathbb{K}=(G,M,I)$ is a context\\
    	 	\ENSURE $L$ is the concept set
			 $C=A^{\prime\prime}, D=A^\prime$
						\IF{$\{h | h \in  C\setminus A  \wedge g < h \}=\emptyset$}
						\STATE{ $L:=L \cup \{(C,D)\}$}
						\ENDIF
    	
    	\FORALL{$f \in \{h | h \in  G\setminus A  \wedge g < h \}$}
			\STATE{$Z:=C \cup \{f\}$}
			\STATE{$Z:=D \cap \{f\}^{\prime}$}
			\STATE{$X:=Y^{\prime}$}
			\STATE{Process$(Z,f,(X,Y))$}
			\ENDFOR
			
    \end{algorithmic}
\end{algorithm}

The time complexity of \textsc{Close by One} (\textsc{CbO}) is $O(|G|^2|M||L|)$, and its polynomial delay is
$O(|G|^3|M|)$.

\begin{figure}
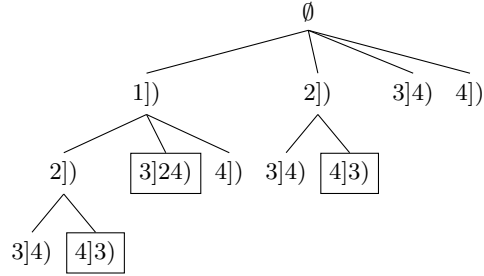

	\centering
		\Tree [.$\emptyset$ [.1$\rbrack$) [.2$\rbrack$) [.3$\rbrack$4) ] [.\node[draw]{4$\rbrack$3)}; ] ] [.\node[draw]{3$\rbrack$24)}; ] [.4$\rbrack$) ]] [.2$\rbrack$) [.3$\rbrack$4) ] [.\node[draw]{4$\rbrack$3)}; ]] [.3$\rbrack$4) ] [.4$\rbrack$) ]]
	\caption{The tree of CbO protocol for the context of geometric figures. Non-canonic generations are drawn in boxes.}
	\label{fig:cboprotoc}
\end{figure}

The generation protocol of \textsc{CbO} in a tree-like form is given in Fig.~\ref{fig:cboprotoc}. Each closed set of objects (extent) can be read from the tree by following the path from the root to the corresponding node. Square bracket ] means that first prime operator has been applied after addition of the lectically next object $g$ to the set $A$ of the parent node and bracket ) shows which object have been added  after application of second prime operator, i.e. between ] and ) one can find $(A\cup\{g\})^{\prime\prime}\setminus (A\cup\{g\})$.
A non-canonic generation can be identified by simply checking whether there is an object between ] and ) that less than $g$ w.r.t. $<$. 
One can note that the traverse of the generation tree is done in a depth-first search manner.

After the inception of the first batch algorithms, the broadened FCA inventory includes efficient incremental algorithms~\cite{Kourie:2009} and the distributed versions of NextClosure and CbO for MapReduce~\cite{Krajca:2009,Xu:2012}.    

\subsection{Many-valued contexts and concept scaling}

\bd
A \textbf{many-valued context} $(G,M,W,I)$ consists of sets $G$, $M$
and $W$ and a ternary relation $I$ between those three sets, i.e.
$I \subseteq G \times M \times W$, for which it holds that
$(g,m,w) \in I$ and $(g,m,v) \in I$ always imply $w = v$
The fact $(g,m,w) \in I$ means ``the attribute $m$ takes value $w$
for object $g$'', simply written as $m(g) = w$.
\ed

\begin{table}
	\centering
	\caption{Many-valued context of university subjects}\label{tbl:subj}
		\begin{tabular}{|c||c|c|c|c|}
		\hline
		G / M & Gender & Age & Subject & Mark\\
		\hline
		\hline
		1 & M &  19 & Math & 8\\
		2 & F & 20 & CS & 9\\
		3 & F & 19 & Math & 7\\
		4 & M & 20 & CS & 10\\
		5 & F & 21 & Data Mining & 9\\
		\hline
		\end{tabular}
\end{table}

\bd
A (\textbf{conceptual) scale} for the attribute $m$ of a many-valued
context is a (one-valued) context $S_m = (G_m, M_m, I_m)$ with
$m(G) = \{ m(g) | \forall g \in G \} \subseteq G_m$. The objects of a scale are
called scale values, the attributes are called scale attributes.
\ed

\textbf{Nominal scale} is defined by the context $(W_m, W_m, =)$. 

This type of scaling is suitable for binary representation of nominal (categorical) attributes like color. For the context of university subjects, the subjects can be scaled by nominal scaling as below. 

\begin{center}

\begin{cxt}%
\cxtName{=}%
\atr{Math}%
\atr{CS}%
\atr{DM}%
\obj{x..}{Math} \obj{.x.}{CS} \obj{..x}{DM}
\end{cxt}

\end{center}

A particular case of nominal scaling is the so called \textbf{dichotomic scaling}, which is suitable for attributes with two mutually exclusive values like ``yes'' and ``no''. In our example, the attribute Gender can be scaled in this way.

\begin{center}
\begin{cxt}%
\cxtName{}%
\att{$M$}%
\att{$F$}%
\obj{x.}{$M$} \obj{.x}{$F$}
\end{cxt}  

\end{center}

\textbf{Ordinal scale} is given by the context $(W_m,W_m,\leq)$ where
 denotes classical real number order. For our example, the attributes age and mark can be scaled by this type of scale.

\begin{table}
	\centering
		\begin{tabular}{ccc}
			\begin{cxt}%
\cxtName{}%
\att{$\leq$ 21}%
\att{$\leq$ 20}%
\att{$\leq$ 19}%
\obj{xxx}{19} \obj{.xx}{20} \obj{..x}{21}%
\end{cxt} & \qquad \qquad \qquad &
\begin{cxt}%
\cxtName{}%
\att{$\leq$ 7}%
\att{$\leq$ 8}%
\att{$\leq$ 9}%
\att{$\leq$ 10}%
\obj{xxxx}{7} \obj{.xxx}{8} \obj{..xx}{9}
\obj{...x}{10}
\end{cxt}

		\end{tabular}
\end{table}

\textbf{Interordinal scale} is given by $(W_m,W_m,\leq)|(W_m,W_m,\geq)$
where $|$ denotes the apposition of two contexts. 

This type of scale can be used as an alternative for ordinal scaling like in example below. 

\begin{center}

\begin{cxt}%
\cxtName{}%
\att{$\leq$ 7}%
\att{$\leq$ 8}%
\att{$\leq$ 9}%
\att{$\leq$ 10}%
\att{$\geq$ 7}%
\att{$\geq$ 8}%
\att{$\geq$ 9}%
\att{$\geq$ 10}%
\obj{xxxxx...}{7} \obj{.xxxxx..}{8} \obj{..xxxxx.}{9}
\obj{...xxxxx}{10}
\end{cxt}

\end{center}

In some domains, e.g., in psychology or sociology there is similar biordinal (bipolar) scaling, which is a good representation of  attributes with so called polarvalues ``agree'', ``rather agree'', ``disagree'', and ``rather disagree''.

There is a special type of scale, \textbf{contranominal scale}, which is rare case in real data, but has important theoretical meaning.
Its context is given by inequality relation, i.e.  $(\{1,\ldots,n\},\{1,\ldots,n\},\neq)$, and the example for $n=4$ is given below.

\begin{center}
\begin{cxt}%
\cxtName{$\neq$}%
\att{1}%
\att{2}%
\att{3}%
\att{4}%
\obj{.xxx}{1} \obj{x.xx}{2} \obj{xx.x}{3} \obj{xxx.}{4}
\end{cxt}
\end{center}

In fact, this type of contexts gives rise to $2^n$ formal concepts and can be used for testing purposes.

The resulting scaled (or plain) context for our university subjects example is below. Note that the Mark attribute is scaled by interordinal scale.
 
\begin{center}

\begin{cxt}%
\cxtName{}%
\att{M}%
\att{F}%
\att{$\leq$ 19}%
\att{$\leq$ 20}%
\att{$\leq$ 21}%
\att{Math}%
\att{CS}%
\att{DM}%
\att{$\leq$ 7}%
\att{$\leq$ 8}%
\att{$\leq$ 9}%
\att{$\leq$ 10}%
\att{$\geq$ 7}%
\att{$\geq$ 8}%
\att{$\geq$ 9}%
\att{$\geq$ 10}%

\obj{x.xxxx...xxxxx..}{1}
\obj{.x.xx.x...xxxxx.}{2}
\obj{.xxxxx..xxxxx...}{3}
\obj{x..xx.x....xxxxx}{4}
\obj{.x..x..x..xxxxx.}{5}
\end{cxt}

\end{center}

\subsection{Attribute Dependencies}

\bd
\textbf{Implication} $A\to B$, where $A, B\subseteq M$ holds in context $(G,M,I)$ if
$A'\subseteq B'$, i.e., each object having all attributes from
$A$ also has all attributes from  $B$.
\ed

\begin{example} For the context of geometric figures one may check that the following implication holds:
 $abc \to d,$  $b \to c,$  $cd \to b.$
Note that for brevity we have omitted curly brackets around and commas between elements of a set attributes.$\square$

\end{example}

\begin{exercise} Find three more implications for the context of geometric figures. $\square$
\end{exercise}

Implications satisfy  \textbf{Armstrong rules} or inference axioms \cite{Armstrong:1974,Maier:83}:

$${{}\over{X\to X}} \mbox{ (reflexivity)}, \quad {{X\to Y}\over{X\cup Z\to Y}} \mbox{ (augmentation)},$$ 
        
				$${{X\to Y, Y\cup Z\to W}\over{X\cup Z\to W}} \mbox{ (pseudotransitivity)}.$$

An inference axiom is a rule that states if certain implications are valid in the context, then certain other implications are valid.

\begin{example}
Let us check that the first and second Armstrong axioms fulfill for implication over attributes.

Since $X^\prime \subseteq X^\prime$ it is always true that $X\to X$.

For the second rule we have $X^\prime \subseteq Y^\prime$.
Applying property 4 from Proposition~\ref{primeprop} we have: $(X \cup Z)^\prime=X^\prime \cap Z^\prime$.
Since $X^\prime \cap Z^\prime \subseteq X^\prime$, we prove that $X^\prime \cap Z^\prime \subseteq Y^\prime$.
This implies $X\cup Z\to Y$.$\square$
\end{example}

\begin{exercise} 1. Prove by applying Armstrong rules that $A_1 \to B_1$ and $A_2 \to B_2$ imply  $A_1 \cup A_2 \to B_1 \cup B_2$.
2. Check the third axiom by using implication definition. $\square$
\end{exercise}

\bd
An \textbf{implication cover} is a subset of
implications from which all other implications can be derived by means of Armstrong rules.

An \textbf{implication base} is a minimal (by inclusion) implication cover.
\ed

\bd

A subset of attributes $D\subseteq M$ is a \textbf{generator} of a closed subset of attributes
$B\subseteq M$, $B'' = B$ if $D\subseteq B$, $D'' = B = B''$.

A subset $D\subseteq M$ is a \textbf{minimal generator} if for any $E\subset D$ one has $E'' \neq D'' = B''$.

Generator $D\subseteq M$ is called \textbf{nontrivial} if $D\neq D'' = B''$.

Denote the set of all nontrivial minimal generators of $B$ by nmingen$(B)$.

\textbf{Generator implication cover} looks as follows:

$\{ F \to (F''\setminus F) \mid F\subseteq M, F \in  \mbox{ nmingen }(F'')\}.$

\ed

\begin{example}
For the context of geometric figures one may check that $b$ is a minimal nontrivial generator for $bc$,   
The set $ab$ is a minimal nontrivial generator for $abcd$, but $abc$, $abd$, and $acd$ are its nontrivial generators. $\square$
\end{example}		

\begin{exercise} For the context of geometric figures find all minimal generators and obtain its generator implication cover. $\square$
\end{exercise}

\bd

The \textbf{Duquenne-Guigues base} is an implication base where each implication is a pseudo-intent \cite{Guigues:1986}.

A subset of attributes  $P\subseteq M$ is called a \textbf{pseudo-intent} if
$P\not=P''$ and for any pseudo-intent  $Q$ such that $Q\subset P$ one has
$Q''\subset P$.

The Duquenne-Guigues base looks as follows:

$\{ P \to (P''\setminus P) \mid P$ is a pseudo-intent $\}.$

\ed

The Duquenne-Guigues base is a minimum (cardinality minimal) implication base.

\begin{example} Let us find all pseudo-intents for the context of geometric figures. We build a table (Table~\ref{tbl:psint}) with $B$ and ${B^\prime\prime}$; it is clear that all closed sets are not pseudo-intents by the definition. Since we have to check the containment of a pseudo-intent in the generated pseudo-intents recursively, we should start with the smallest possible set,  i.e. $\emptyset$.

\begin{table}
	\centering
	\caption{Finding pseudo-itents for the context of geometric figures}\label{tbl:psint}
		\begin{tabular}{|c|c|c|c|}
\hline			
$B$ & $B'$ & $B^{\prime\prime}$ & $B$ is pseudo-intent?\\
\hline
$\emptyset$ & $1234$  & $\emptyset$  & No, it's not. \\
$a$ & 12 & $a$ & No, it's not.\\
$b$ &  34 & $bc$ & Yes, it is.\\
$c$ & 234 & $c$ & No, it's not. \\
$d$ & 14 & $d$ & No, it's not. \\
$ab$ & $\emptyset$ & $abcd$ & No, it's not. \\
$ac$ & 2 & $ac$ & No, it's not. \\
$ad$ & 1 & $ad$ & No, it's not. \\
$bc$ & 34 & $bc$ & No, it's not.\\
$bd$ &  4 & $bcd$ & No, it's not.\\
$cd$ & 4 & $bcd$ & Yes, it is.\\
$abc$ & $\emptyset$ & $abcd$ & Yes, it is. \\
$abd$ & $\emptyset$ & $abcd$ & No, it's not. \\
$acd$ & $\emptyset$ & $abcd$ & No, it's not. \\
$bcd$ & 4 & $bcd$ & No, it's not. \\
$abcd$ & $\emptyset$ & $abcd$ & No, it's not.\\

\hline
\end{tabular}
\end{table}

Thus, $\{b\}$ is the first non-closed set in our table and the second part of pseudo-intent definition fulfills trivially -- there is no another pseudo-intent contained in $\{b\}$.
So, the whole set of pseudo-intents is $\{b, cd, abc\}$. $\square$

\end{example}

\begin{exercise}
Write down the Duquenne-Guigues base for the context of geometric figures. Using Armstrong rules and the obtained Duquenne-Guigues base, deduce the rest implications of the original context. $\square$
\end{exercise}

For recent efficient algorithm of finding the Duquenne-Guigues base see~\cite{Bazhanov:2014}.

\paragraph{Implications and functional dependencies}

Data dependencies are one way to reach two primary purposes of databases: to attenuate data redundancy and
enhance data reliability \cite{Maier:83}. These dependencies are mainly used for data normalisation, i.e. their proper decomposition into interrelated tables (relations). 
The definition of functional dependency~\cite{Maier:83} in terms of FCA is as follows:

\bd
$X\to Y$ is a \textbf{functional dependency} in a
complete many-valued context $(G,M,W,I)$ if the
following holds for every pair
of objects $g, h\in G$:
$$(\forall m\in X\  m(g) = m(h))\Rightarrow (\forall n\in Y\  n(g) = n(h)).$$
\ed

\begin{example}\label{ex:fd}
For the example given in Table~\ref{tbl:subj} the following functional dependencies hold:
$Age \to Subject$, $Subject \to Age$, $Mark \to Gender.$$\square$
\end{example}

The first two functional dependencies may have sense since students of the same year may study the same subjects. However, the last one says Gender is functionally dependent by Mark and looks as a pure coincidence because of the small dataset.

The reduction of functional dependencies to implications:

\begin{proposition}
For a many-valued context $(G,M,W,I)$, one defines the context
$\context_N: = ({\cal P}_2(G), M, I_N)$, where ${\cal P}_2(G)$ is the set of all
pairs of different objects from $G$ and $I_N$ is defined by
$$\{g,h\}I_Nm: \Leftrightarrow m(g) = m(h).$$
Then a set $Y\subseteq M$ is functionally dependent
on the set $X\subseteq M$ if and only if the implication $X\to Y$ holds in the context
$\context_N$.
\end{proposition}

\begin{example} Let us construct the context $\context_N$ for the many-valued context of geometric figures.

\begin{center}
\begin{cxt}%
\cxtName{}%
\atr{{G}ender}%
\atr{{A}ge}%
\atr{{S}ubject}%
\atr{{M}ark}%
\obj{....}{\{1,2\}}
\obj{.xx.}{\{1,3\}}
\obj{x...}{\{1,4\}}
\obj{....}{\{1,5\}}
\obj{x...}{\{2,3\}}
\obj{.xx.}{\{2,4\}}
\obj{x..x}{\{2,5\}}
\obj{....}{\{3,4\}}
\obj{x...}{\{3,5\}}
\obj{....}{\{4,5\}}
\end{cxt}
\end{center}

One may check that the following implications hold: $Age \to Subject$, $Subject \to Age$, $Mark \to Gender$, which are the functional dependencies that we so in example~\ref{ex:fd}. $\square$

\end{example}

An inverse reduction is possible as well.

\begin{proposition} For a context $\context=(G,M,I)$ one can construct
a many-valued context $\context_W$ such that an implication $X\to Y$ holds
if and only if  $Y$ is functionally dependent on $X$ in $\context_W$.
\end{proposition}

\begin{example} To fulfill the reduction one may build the corresponding many-valued context in the following way:

1. Replace all ``$\times$'' by 0s. 2. In each row, replace empty cells by the row number starting from 1.
3. Add a new row filled by 0s.

\begin{center}
\begin{tabular}{|c||cccc|}
  \hline
  &  a & b & c & d\\
  \hline
	\hline
  1 & 0 & 1 & 1 & 0\\
  2 & 0 & 2 & 0 & 2\\
  3 & 3 & 0 & 0 & 3\\
  4 & 4 & 0 & 0 & 0\\
	5 & 0 & 0 & 0 & 0\\
  \hline
  \end{tabular}
\end{center}

$\square$
\end{example}

\begin{exercise}
Check the functional dependencies from the previous example coincide with the implications of the context of geometric figures. $\square$
\end{exercise}

More detailed tutorial on FCA and fuctional dependencies is given in~\cite{Baixeries:2014}.

\section{FCA tools and practice} 
	In this section, we provide a short summary of ready-to-use software that supports basic functionality of \FCA.
	
\begin{itemize}
\item Software for FCA: Concept Explorer, Lattice Miner, ToscanaJ, Galicia, FCART etc.
	\item Exercises.
\end{itemize}

\paragraph{Concept Explorer.} ConExp \footnote{\url{http://conexp.sourceforge.net/}} is probably one of the most user-friendly FCA-based tools with basic functionality; it was developed in Java by S.~Yevtushenko under Prof.~T.~Taran supervision in the beginning of 2000s \cite{Yevtushenko:2000}. Later on it has been improved several times, especially from lattice drawing viewpoint \cite{Yevtushenko:2004}. 

Now the features the following functionality:
\bi
\item Context editing (tab separated and csv formats of input files are supported as well);
\item Line diagrams drawing (allowing their import as image snapshots and even text files with nodes position, edges and attributes names, but vector-based formats are not supported);
\item Finding the Duquenne-Guigues base of implications;
\item Finding the base of association rules that are valid in a formal context;
\item Performing attribute exploration.
\ei
 
\begin{figure}
	\centering
		\includegraphics[scale=1.0]{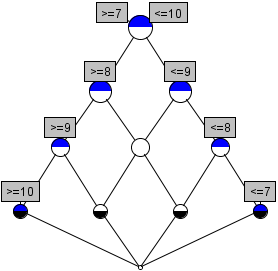}
	\caption{The line diagram of the concept lattice for the interordinal scale of student marks drawn by ConExp.}
	\label{fig:IntOrdClCE}
\end{figure}

It is important to note that the resulting diagram is not static and one may perform exploratory analysis in an interactive manner selecting interesting nodes, moving them etc. In Fig.~\ref{fig:IntOrdClCE}, the line diagram of the concept lattice of interordinal scale  for attribute Mark drawn by ConExp is shown. See more details in Fig.~\cite{Yevtushenko:2006}.

There is an attempt to reincarnate ConExp~\footnote{\url{https://github.com/fcatools/conexp-ng/wiki}} by modern open software tools.

\paragraph{ToscanaJ.}

The ToscanaJ\footnote{\url{http://toscanaj.sourceforge.net/}} project is a result of collaboration between two groups from the Technical University of Darmstadt and the University of Queensland, which aim was declared as ``to give the FCA community a platform to work with'' \cite{Becker:2004} and ``the creation
of a professional tool, coming out of a research environment and still supporting research'' \cite{Becker:2005}.

This open project has a long history with several prototypes~\cite{Vogt:94} and now it is a part of an umbrella framework for conceptual knowledge processing, Tockit\footnote{\url{http://www.tockit.org/}}. As a result, it is developed in Java, supports different types of database connection via JDBC-ODBC bridge and contains an embedded database engine~\cite{Becker:2005}. Apart from ConExp, it features work with multi-valued contexts, conceptual scaling, and nested line diagrams. 
	
\begin{figure}
	\centering
		\includegraphics[width=0.60\textwidth]{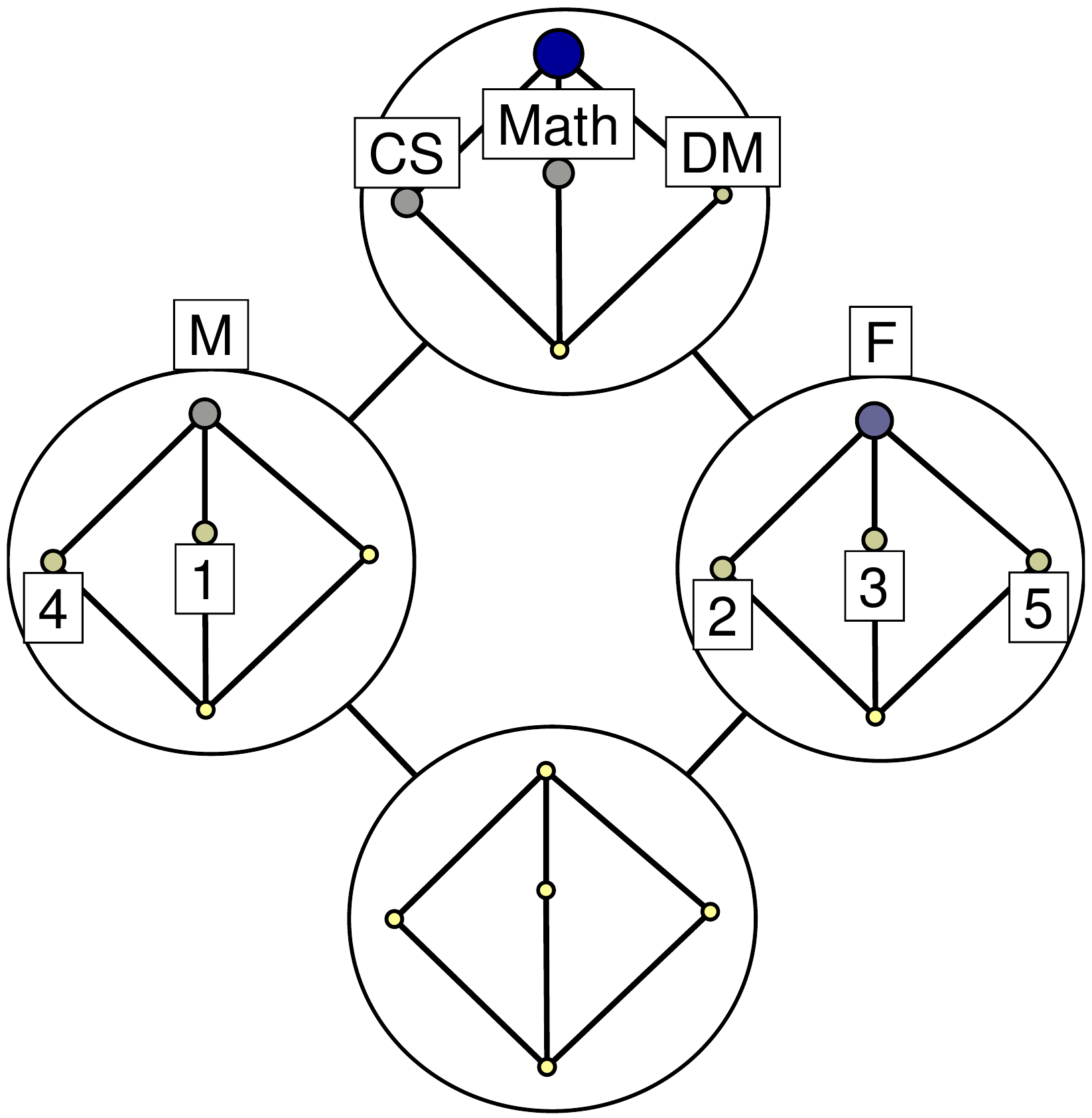}
	\caption{The nested line diagram for the two one-attribute subcontexts of the context of university subjects. The outer diagram is for Gender attribute, and the inner one is for Subject.}
	\label{fig:NestedSexSubjTJ}
\end{figure}

In Fig.~\ref{fig:NestedSexSubjTJ} one can see the nested line diagram for two scales from the university subjects multi-valued context, namely for two attributes, Gender and Subject. Via PDF printing facilities it is possible to print out line diagrams in a vector graphic form.

\paragraph{Galicia.}

Galicia\footnote{\url{http://www.iro.umontreal.ca/~galicia/}} was ``intended as an integrated software platform including components for the key operations on lattices that might be required in practical applications or in more theoretically-oriented studies''. Thus in addition to basic functionality of ConExp, it features work with multi-valued contexts and conceptual scaling, iceberg lattices (well-known in Data Mining community), Galois hierarchies and relational context families, which are popular in  software engineering~\cite{valtchev:2003}. The software is open and its implementation in Java is cross-platform aimed at ``adaptability, extensibility and reusability''.

\begin{figure}
	\centering
		\includegraphics[width=1.0\textwidth]{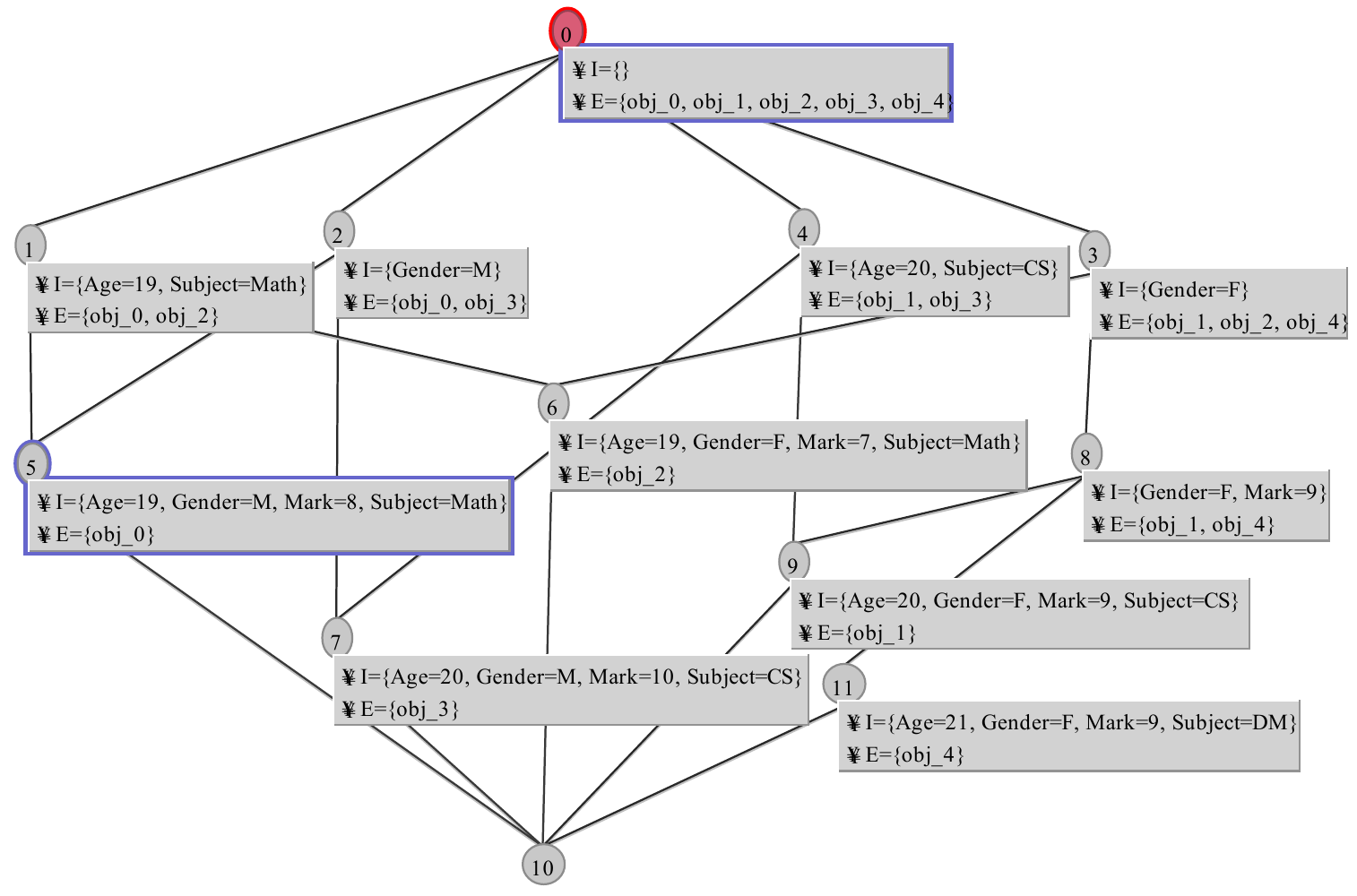}
	\caption{The line diagram of concept lattice for the context of university subjects drawn by Galicia.}
	\label{fig:GalSubjLatt}
\end{figure}

It is possible to navigate through lattice diagrams in an interactive manner; the resulting diagrams contain numbered nodes and this is different from the traditional way of line diagrams drawing. Another Galicia's unique feature is 3D lattice drawing. The diagram of the university subjects context after nominal scaling of all its attributes obtained in Galicia is depicted in Fig.~\ref{fig:GalSubjLatt}. Galicia supports vector-based graphic formats, SVG and PDF. The authors of the program paid substantial attention to algorithmic aspects and incorporated batch and incremental algorithms into it. Various bases of implications and association rules can be generated by the tool. Nested line diagrams are in the to do list.

\paragraph{Lattice Miner.} This is another attempt to establish basic FCA functionality and several specific features to the FCA community \footnote{\url{http://sourceforge.net/projects/lattice-miner/}}~\cite{Lahcen:2010}.

The initial objective of the tool was ``to focus on visualization mechanisms for the representation of concept lattices, including nested line diagrams'' \footnote{\url{https://en.wikipedia.org/wiki/Lattice_Miner}}. Thus, its interesting feature is multi-level nested line diagrams, which can help to explore comparatively large lattices.

After more than a decade of development, FCA-based software having different features produced a lot of different formats thus requiring  interoperability. To this end, in analogy to Rosetta Stone, FcaStone \footnote{\url{http://fcastone.sourceforge.net/}} was proposed. It supports convertation  between commonly used FCA file formats (cxt, cex, csc, slf, bin.xml, and csx) and comma separated value (csv) files as well as 
convertation concept lattices into graph formats (dot, gxl, gml, etc. for use by graph editors such as yEd, jgraph, etc.) or into vector graphics formats (fig, svg, etc. for use by vector graphics editors such as Xfig, Dia, Inkscape, etc.). It can also be incorporated into a webpage script for generating lattices and line diagrams online. Another example of a web-based ported system with basic functionality including attribute exploration is OpenFCA\footnote{\url{https://code.google.com/p/openfca/}}.

\paragraph{FCART.}

Many different tools have been created and some of the projects are not developing anymore but the software is still available; an interested reader can refer Uta Priss's webpage to find dozens of tools\footnote{\url{http://www.fcahome.org.uk/fcasoftware.html}}. However, new challenges such as handling large heterogeneous datasets (large text collections, social networks and media etc.) are coming and the community, which put a lot of efforts in the development of truly cross-platform and open software, needs a new wave of tools that adopts modern technologies and formats.

Inspired by the successful application of FCA-based technologies in text mining for criminology domain~\cite{Poelmans:2012a}, in the Laboratory for Intelligent Systems and Structural Analysis, a tool named Formal Concept Analysis Research Toolbox (FCART) is developing.

FCART follows a methodology from~\cite{Poelmans:2011} to formalise iterative ontology-driven data analysis process and to implement several basic principles:

\be
\item	Iterative process of data analysis using ontology-driven queries and interactive artifacts such as concept lattice, clusters, etc. 
\item	Separation of processes of data querying (from various data sources), data preprocessing (via local immutable snapshots), data analysis (in interactive visualizers of immutable analytic artifacts), and results presentation (in a report editor).
\item	Three-level extendability: settings customisation for data access components, query builders, solvers and visualizers; writing scripts or macros; developing components (add-ins).
\item	Explicit definition of analytic artifacts and their types, which enables integrity of session data and links artifacts for the end-users.
\item	Availability of integrated performance estimation tools.
\item	Integrated documentation for software tools and methods of data analysis.

\ee

\begin{figure}
	\centering
		\includegraphics[width=1.0\textwidth]{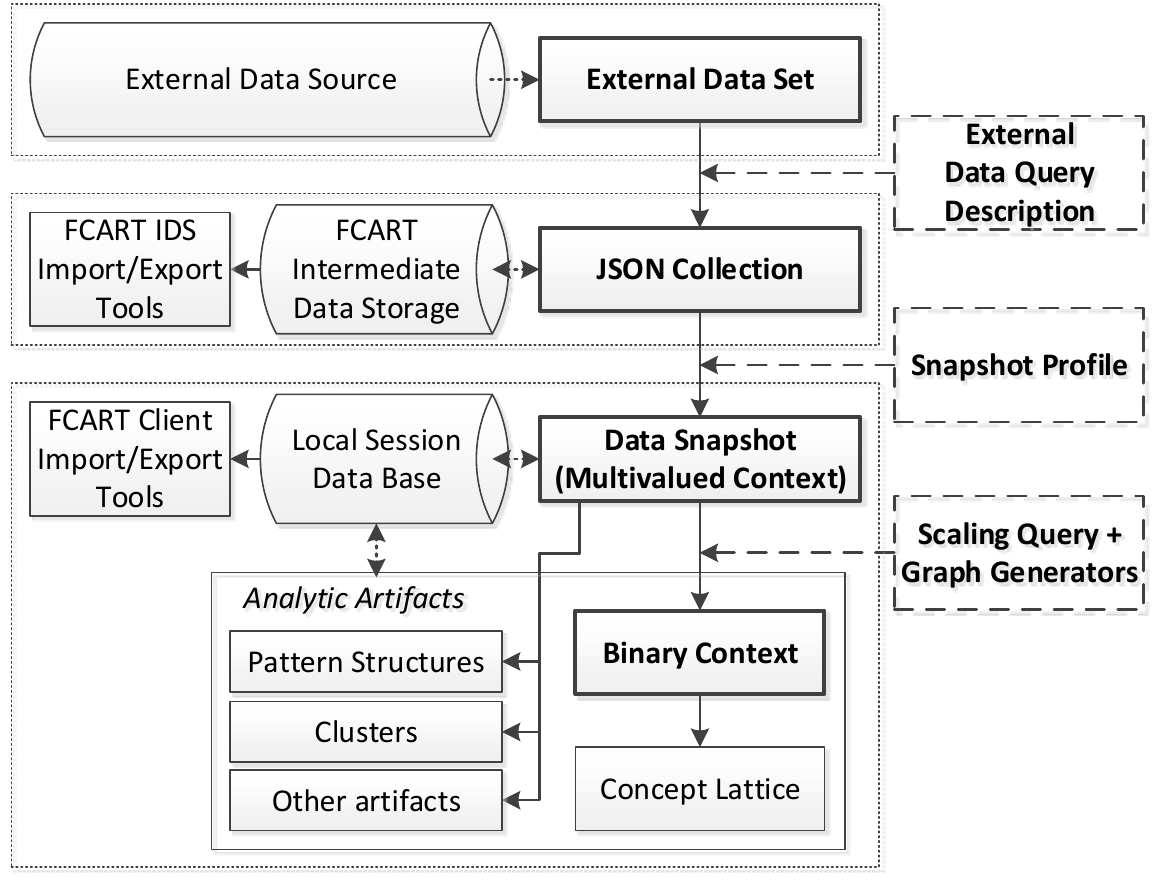}
	\caption{FCART workflow}
	\label{fig:FCARTWorkflow}
\end{figure}

Originally, it was yet another FCA-based ``integrated environment for knowledge and data engineers with a set of research tools based on Formal Concept Analysis''~\cite{Neznanov:2013,Neznanov:2014} featuring in addition work with unstructured data (including texts with various metadata) and Pattern Structures~\cite{Buzmakov:2013}. In its current distributed version, FCART consists of the following parts:

\be
\item	AuthServer for authentication and authorisation.
\item	Intermediate Data Storage (IDS) for storage and preprocessing of big datasets. 
\item	Thick Client for interactive data processing and visualisation in integrated graphical multi-document user interface.
\item	Web-based solvers for implementing independent resource-intensive computations.

\ee

The workflow is shown in Fig.~\ref{fig:FCARTWorkflow}.

The main questions are the following: Whether the product has only technological advantages or it really has fruitful methodology? Can it become open in addition to its extendability? Can it finally handle big volumes of heterogeneous data in a suitable way for an FCART analyst?  
The answers to these posed questions seem to be forthcoming challenging steps.

\paragraph{CryptoLatt.} This tool\footnote{\url{http://www.cs.unic.ac.cy/florent/software.htm}} was developed to help students and researchers from neighbouring domains (e.g., Data Mining) to recognise cryptomorphisms in lattice-based problems, i.e. to realise that a particular problem in one domain is ``isomorphic'' to some other in terms of lattice theory~\cite{Domenach:2013}. Thus, one of the well-known cryptomorphisms in the FCA community is established between a lattice and a binary relation, also known as the basic theorem of FCA. Note that even a particular formal context, its concept lattice and set of implications represent the same information about the underlying dataset but in a different way.

\begin{exercise}\label{ex:ConExp} Practice with Concept Explorer:

\noindent 1. Input the context of geometric figures, build its concept lattice diagram and find the Duquenne-Guigues base. Check whether the obtained base coincide with the base found before. Play with different layouts and other drawing options like labeling or node size. 

\noindent 2. Find real datasets where objects are described by nominal attributes and select about 10 objects and 10 attributes from it. Prepare the corresponding context, build the lattice diagram and find its implication base. Try to interpret found concepts and dependencies. $\square$
\end{exercise}	

\begin{exercise} Practice with ToscanaJ:

\noindent 1. Use Elba tool from the latest version of ToscanaJ for creating two scaled contexts for any two attributes of the context of university subjects. Save the contexts. Then upload them into ToscanaJ and draw their nested line diagram. The result should be similar to Fig.~\ref{fig:NestedSexSubjTJ}. $\square$
\end{exercise}

\begin{exercise} Practice with Galicia:

\noindent 1. Perform tasks from exercise~\ref{ex:ConExp}.
\noindent 2. Compose the context of university subjects. Scale it via \emph{Algorithms}$\to$\emph{Multi-FCA}$\to$\emph{Interactive Multi-FCA} and build the lattice diargam. 
The result should be identical to Fig.~\ref{fig:GalSubjLatt}. $\square$

\end{exercise}

\section{FCA in Data Mining and Machine Learning}
	
	\begin{itemize}
	\item Frequent Itemset Mining and Association Rules: FCA did it even earlier~\cite{Agrawal:94,Luxenburger:91}
			
	\item Multimodal clustering (biclustering and triclustering)~\cite{Jaschke:2006,Ignatov:2011,Ignatov:2013b}
	
	\item FCA in Classification: JSM-method, version spaces$\ast$\footnote{not covered here}, and decision trees$\ast$ \cite{Kuznetsov:2004a}
	
	\item Pattern Structures for data with complex descriptions~\cite{Ganter:2004,Kuznetsov:2013p}
	
	\item FCA-based Boolean Matrix Factorisation~\cite{Belohlavek:2010}
	
	\item Educational Data Mining case study~\cite{Romashkin:2011}
	
	\item Exercises with JSM-method in QuDA (Qualitative Data Analysis): solving classification task~\cite{Grigoriev:2004}
		
	\end{itemize}

	\subsection{Frequent Itemset Mining and Association Rules}
	
Knowledge discovery in databases (KDD) is introduced as the non-trivial extraction
of valid, implicit, potentially useful and ultimately understandable information
in large databases~\cite{Han:2000}.
Data mining is a main step in KDD, and in its turn association rules and frequent itemset mining are among the key techniques in Data Mining.  The original problem for association rules mining is market basket analysis. In early 90s, since the current level of technologies made it possible to store large amount of transactions of purchased items, companies started their attempts to use these data to facilitate their typical business decisions concerning ``what to put on sale, how to design coupons, how to place merchandise on shelves in order to maximize the profit''\cite{agrawal:1993}. So, firstly this market basket analysis problem was formalised in~\cite{agrawal:1993} as a task of finding frequently bought items together in a form of rules ``if a customer buys items $A$, (s)he also buys items $B$''. One of the first and rather efficient algorithms of that period was proposed in \cite{Agrawal:94}, namely Apriori. From the very beginning these rules are tolerant to some number of exceptions, they were not strict as implications in FCA. However,  several years before, in \cite{Luxenburger:91}, Michael Luxenburger introduced partial implications motivated by more general problem statement, ``a generalisation of the theory of implications between attributes to partial implications'' since ``in data analysis the user is not only interested in
(global) implications, but also in ``implications with a few exceptions''''. The author proposed theoretical treatment of the problem in terms of Formal Concept Analysis and was guided by the idea of characterisation of ``sets of partial implications which arise from real data''  and ``a possibility of an ``exploration'' of partial implications by a computer''. In addition, he proposed a minimal base of partial implications known as Luxenburger's base of association rules as well.

\bd
Let  $\mathbb{K}:=(G, M, I)$ be a context, where $G$ is a set of objects, $M$ is a set of attributes (items), $I\subseteq G\times M$
\textbf{An association rule} of the context $\mathbb{K}$ is an expression $A\rightarrow B$, where $A,B \subseteq M$ and (usually) $A \cap B=\emptyset$.
\ed

\bd
\emph{(Relative) support} of an association rule  $A\rightarrow B$ defined as

 $$supp(A\rightarrow B)=\frac{|(A\cup B)'|}{|G|}.$$

\ed
		
The value of $supp(A\rightarrow B)$ shows which part of $G$ contains $A\cup B$. Often support can be given in $\%$.
    
\bd						
\emph{(Relative) confidence} of an association rule $A\rightarrow B$ defined as

 $$conf(A\rightarrow B)=\frac{|(A\cup B)'|}{|A'|}.$$
\ed

This value  $conf(A\rightarrow B)$ shows which part of objects that possess  $A$ also contains $A\cup B$.  Often confidence can be given in $\%$.

\begin{example} An object-attribute table of transactions.

\begin{center}
\begin{cxt}%
\cxtName{}%
\atr{Beer}%
\atr{Cakes}%
\atr{Milk}%
\atr{M\"usli}%
\atr{Chips}%
\obj{x...x}{$c_1$}
\obj{.xxx.}{$c_2$}
\obj{x.xxx}{$c_3$}
\obj{xxx.x}{$c_4$}
\obj{.xxxx}{$c_5$}
\end{cxt}
\end{center}


\bi
\item  $supp(\{$Beer, Chips$\})=3/5$
\item  $supp(\{$Cakes, M\"usli $\} \rightarrow \{$ Milk $\})=$
$\frac{|(\{\mbox{Cakes, M\"usli}\}\cup \{\mbox{Milk}\})'| }{|G|}=\frac{|\{C2,C5\}|}{5}=2/5$

\item $conf(\{$Cakes, M\"usli $\} \rightarrow \{$ Milk $\})=$
$\frac{|(\{\mbox{Cakes, M\"usli}\}\cup \{\mbox{Milk}\})'| }{|\{\mbox{Cakes, M\"usli}\}'|}=$$\frac{|\{c_2,c_5\}|}{|\{c_2,c_5\}|}=1$
    
\ei
$\square$
\end{example}

The main task of association rules mining is formulated as follows: Find all association rules of a context, where support and confidence of the rules are greater than predefined thresholds, min-confidence and min-support, denoted as $min\_conf$ and $min\_supp$, respectively \cite{agrawal:1993}

\begin{proposition}(Association rules and  implications)

Let $\context$ be a context, then its associations rules under condition $min\_supp=0\%$ and $min\_conf=100\%$ are implications of the same context.
\end{proposition}

Sometimes an association rule can be written as $A\xrightarrow[s]{c} B$, where $c$ and $s$ are confidence and support of the given rule.

Two main steps of association rules mining are given below:

\be
\item  Find frequent sets of attributes (frequent itemsets), i.e. sets of attributes (items) that have support greater than $min\_supp$.
\item Building association rules based on found frequent itemsets.
\ee

The first step is the most expensive, the second one is rather trivial.

The well-known algorithm for frequent itemset mining is Apriori \cite{Agrawal:94} uses the antimonotony property to ease frequent itemsets enumeration. 

\begin{property}(Antimonotony property) For $ \forall A,B \subseteq M \mbox{ and } A\subseteq B  \Rightarrow supp(B) \leq supp(A).$
\end{property}

This property implies the following facts:
\bi
 \item The larger set, the smaller support it has or its support remains the same;
 \item Support of any itemset is not greater than a minimal support of any its subset;
 \item Aa itemset of size $n$ is frequent if and only if all its $(n-1)$-subsets are frequent.
\ei

The Apriori algorithm finds all frequent itemsets.

It is check iteratively the set of all itemsets in a levelwise manner. At each iteration one level is considered, i.e.
a subset of candidate itemsets $C_i$ is composed by collecting the frequent itemsets
discovered during the previous iteration (AprioriGen procedure). Then supports of all candidate itemsets
are counted, and the infrequent ones are discarded.

\begin{algorithm}
\caption{Apriori($Context$, $min\_supp$)}\label{alg:apriori}
    \begin{algorithmic}[1]
      	\REQUIRE $Context$, $min\_supp$ is a minimal support
    	 	\ENSURE all frequent itemsets $I_F$
    	\STATE $C_1 \gets$ 1-itemsets
			\STATE $i \gets 1$
    	\WHILE{$C_i \neq \emptyset$}
			\STATE $SupportCount(C_i)$
    \STATE $F_i \gets \{ f\in C_i\ | f.support \geq min\_supp\}$
    \STATE \COMMENT{$F_i$ is a set of frequent $i$-itemsets}
    \STATE $C_{i+1}\gets AprioriGen(F_i)$
		\STATE \COMMENT{$C_i$ is a set of $(i+1)$-candidates}
    \STATE $i{++}$
			\ENDWHILE
			\STATE $I_F \gets \bigcup F_i$
  \RETURN{$I_F$}
    \end{algorithmic}
\end{algorithm}

For frequent itemsets of size $i$, procedure AprioriGen  finds $(i+1)$-supersets and returns only the set of potentially frequent candidates.

\begin{algorithm}
\caption{AprioriGen($F_i$)}\label{alg:aprgen}
    \begin{algorithmic}[1]
      	\REQUIRE $F_i$ is a set of frequent $i$-itemsets
    	 	\ENSURE $C_{i+1}$ is a set of $(i+1)$-itemsets candidates 
				\STATE{insert into  $C_{i+1}$} 
				\COMMENT{union}
				\STATE{select  $p[1], p[2],\ldots, p[i], q[i]$}
				\STATE{from  $F_i.p$, $F_i.q$}
				\STATE{where  $p[1]=q[1],\ldots, p[i-1]=q[i-1], p[i]<q[i]$}
				\FORALL{$c \in C_{i+1}$}  
				\STATE{}\COMMENT{elimination}
				\STATE{$S \gets (i-1)$-itemset $c$}
				\FORALL{$s \in S$}
				\IF{$s \not\in F_i$} 
				\STATE{$C_{i+1}\gets C_{i+1}\setminus c$}
				\ENDIF
				\ENDFOR
				\ENDFOR
				\RETURN{$C_{i+1}$}
    \end{algorithmic}
\end{algorithm}

\begin{example} Union and elimination steps of AprioriGen for a certain context.

\bi
\item The set of frequent 3-itemsets: $F_3=\{\{a,b,c\},\{a,b,d\},\{a,c,d\},\{a,c,e\},\{b,c,d\}\}.$
\item The set of candidate 4-itemsets (union step): $C_4 = \{\{a,b,c,d\}, \{a,c,d,e\}\}.$
\item The remaining candidate is $C_4 = \{\{a,b,c,d\}\}$, since is eliminated $\{a,c,d,e\}$ because $\{c,d,e\} \not\in F_3$ (elimination step).
\ei 
$\square$
\end{example}

The worst-case computational complexity of the Apriori algorithm is $O(|G||M|^22^{|M|}$ since all the itemsets may be frequent. However, it takes only $O(|M|)$ datatable scans compared to $O(2^|M|)$ for brute-force method. 

Rules extraction is based on frequent itemsets.

Let $F$ be a frequent 2-itemset. We compose a rule $f\rightarrow F\setminus f$ if

 $$conf(f\rightarrow F\setminus f)=\frac{supp(F)}{supp(f)} \geq min\_conf, \mbox{where} f \subset F.$$

\begin{property}

Confidence $conf(f\rightarrow F\setminus f)=\frac{supp(F)}{supp(f)}$ is minimal, when $supp(f)$ is maximal.

\end{property}

\bi
\item Confidence is maximal when rule consequent $F\setminus f$ consists of one attribute (1-itemset). The subsets of such an consequent have greater support and it turn smaller confidence.
\item Recursive procedure of rules extraction starts with $(|F|-1)$-itemset $f$ fulfilling $min\_conf$ and $min\_sup$; then, it forms the rule $f\rightarrow F\setminus f$ and checks all its subsets $(|F|-2)$-itemset (if any) and so on.
\ei

\begin{exercise}

Find all frequent itemsets for the customers context with with Apriori algorithm and $min\_sup=1/3$. $\square$

\end{exercise}

\paragraph{Condensed representation of frequent itemsets}

According to basic results from Formal Concept Analysis, it is not necessary count the support of all frequent itemsets. Thus, it is possible to derive from some known supports the supports of all other itemsets: it is enough to know the support of all frequent concept intents.
And it is also not necessary to compute all frequent itemsets for solving the association rule problem: it is sufficient to consider
the frequent concept intents that also called closed itemsets in Data Mining.
In fact, closed itemsets was independently discovered by three groups of researches in the late 90s \cite{Pasquier:1999,Zaki:99,Stumme:1999}.
	
Let $\mathbb{K}=(G, M, I)$ be a formal context.

\begin{definition}
 A set of attributes $FC\subseteq M$ is called \textbf{frequent closed itemset}, if $supp(FC)\geq min\_supp$ and there is no any $F$ such that $F \supset FC$ and $supp(F)=supp(FC)$.
\end{definition}

\begin{definition}
A set of attributes $MFC\subseteq M$ is called \textbf{maximal frequent itemset} if it is frequent and there is no any $F$ such that$F \supset FMC$ and $supp(F)\geq min\_supp$.
\end{definition}

\begin{proposition}
In a formal context $\context$, $\mathcal{MFC}\subseteq \mathcal{FC} \subseteq \mathcal{F}$, where $\mathcal{MFC}$ is  the set of maximal frequent itemset, $\mathcal{FC}$ is  the set of frequent closed itemsets, and $\mathcal{F}$ is the set of frequent itemsets of $\mathbb{K}$ with a minimal support $min\_supp$.
\end{proposition}

\begin{proposition}
 The concept lattice of a formal context $\mathbb{K}$ is (isomorphic to) its lattice of frequent closed itemsets with $min\_supp=0$.
\end{proposition}
	
\begin{figure}[h]
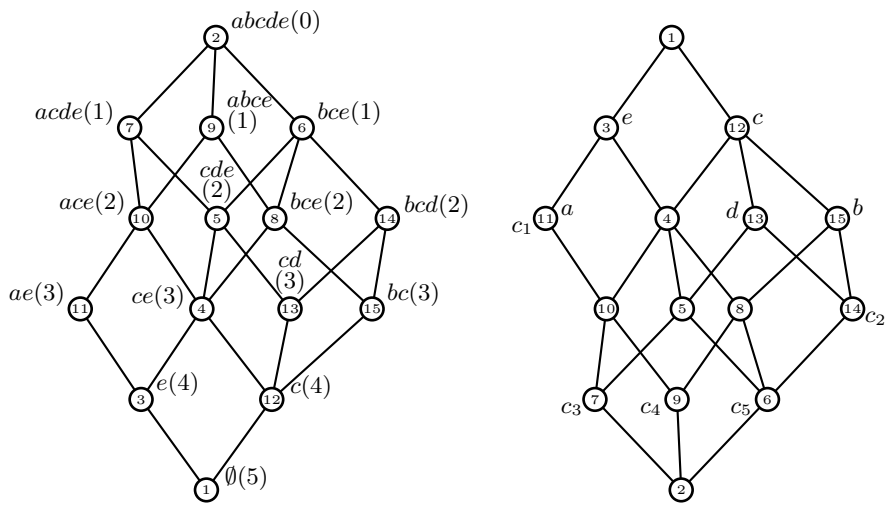

\begin{minipage}[h]{0.50\linewidth}  
  \centering
\unitlength 0.20mm
\begin{diagram}{300}{310}
\Node{1}{166.0}{0.0}
\Node{2}{172.25}{300.0}
\Node{3}{123}{60}
\Node{4}{163}{120}
\Node{5}{173}{180}
\Node{6}{229}{240}
\Node{7}{115.5}{240}
\Node{8}{211}{180}
\Node{9}{169.5}{240}
\Node{10}{123}{180}
\Node{11}{83}{120}
\Node{12}{209}{60}
\Node{13}{221}{120}
\Node{14}{285}{180}
\Node{15}{275}{120}
\Edge{14}{15}
\Edge{5}{4}
\Edge{2}{9}
\Edge{7}{10}
\Edge{6}{8}
\Edge{13}{12}
\Edge{4}{3}
\Edge{10}{11}
\Edge{12}{1}
\Edge{6}{5}
\Edge{9}{10}
\Edge{2}{7}
\Edge{8}{4}
\Edge{14}{13}
\Edge{15}{12}
\Edge{7}{5}
\Edge{10}{4}
\Edge{11}{3}
\Edge{2}{6}
\Edge{9}{8}
\Edge{3}{1}
\Edge{4}{12}
\Edge{5}{13}
\Edge{6}{14}
\Edge{8}{15}
\NoDots
\rightAttbox{1}{12}{0}{$\emptyset(5)$}
\rightAttbox{2}{10}{2}{$abcde(0)$}
\rightAttbox{3}{10}{2}{$e(4)$}
\leftAttbox{4}{11}{0}{$ce(3)$}
\centerAttbox{5}{0}{10}{$cde$\\$(2)$}
\rightAttbox{6}{10}{2}{$bce(1)$}
\leftAttbox{7}{10}{2}{$acde(1)$}
\rightAttbox{8}{10}{2}{$bce(2)$}
\rightAttbox{9}{10}{-4}{$abce$\\$(1)$}
\leftAttbox{10}{10}{2}{$ace(2)$}
\leftAttbox{11}{10}{2}{$ae(3)$}
\rightAttbox{12}{12}{0}{$c(4)$}
\centerAttbox{13}{0}{10}{$cd$\\$(3)$}
\rightAttbox{14}{11}{0}{$bcd(2)$}
\rightAttbox{15}{10}{2}{$bc(3)$}
\Numbers
\CircleSize{15}
\end{diagram}
 \end{minipage} 
\hfill
\begin{minipage}[h]{0.50\linewidth}
  \unitlength 0.20mm
\begin{diagram}{300}{310}
\Node{1}{166.0}{300.0}
\Node{2}{172.25}{0.0}
\Node{3}{123}{240}
\Node{4}{163}{180}
\Node{5}{173}{120}
\Node{6}{229}{60}
\Node{7}{115.5}{60}
\Node{8}{211}{120}
\Node{9}{169.5}{60}
\Node{10}{123}{120}
\Node{11}{83}{180}
\Node{12}{209}{240}
\Node{13}{221}{180}
\Node{14}{285}{120}
\Node{15}{275}{180}
\Edge{14}{15}
\Edge{5}{4}
\Edge{2}{9}
\Edge{7}{10}
\Edge{6}{8}
\Edge{13}{12}
\Edge{4}{3}
\Edge{10}{11}
\Edge{12}{1}
\Edge{6}{5}
\Edge{9}{10}
\Edge{2}{7}
\Edge{8}{4}
\Edge{14}{13}
\Edge{15}{12}
\Edge{7}{5}
\Edge{10}{4}
\Edge{11}{3}
\Edge{2}{6}
\Edge{9}{8}
\Edge{3}{1}
\Edge{4}{12}
\Edge{5}{13}
\Edge{6}{14}
\Edge{8}{15}
\NoDots
\leftObjbox{6}{10}{2}{$c_5$}
\leftObjbox{7}{8}{2}{$c_3$}
\leftObjbox{9}{10}{2}{$c_4$}
\leftObjbox{11}{8}{2}{$c_1$}
\rightObjbox{14}{8}{2}{$c_2$}
\rightAttbox{3}{10}{2}{$e$}
\rightAttbox{11}{10}{2}{$a$}
\rightAttbox{12}{10}{2}{$c$}
\leftAttbox{13}{11}{0}{$d$}
\rightAttbox{15}{10}{2}{$b$}
\Numbers
\CircleSize{15}
\end{diagram}
  \end{minipage} 
 \caption{The line diagrams of the lattice of closed itemsets (left, with support given in parentheses) and the concept lattice for the customers context (right, with reduced labeling)}
  \label{fig:custlatts}
\end{figure}

One may check that the lattices, whose diagrams are depicted in Fig.~\ref{fig:custlatts}, are isomorphic.

The set of all frequent concepts of the context $\context$ for the threshold $min\_sup$ is also known as the ``iceberg concept lattice'' \cite{Stumme:02}, mathematically it corresponds to the order filter of the concept lattice. However, the idea of usage the size of concept's extent, intent (or even their different combinations) as a concept quality measure is not new in FCA \cite{Kuznetsov:1996}.

Of course, the application domain is not restricted to market basket analysis; thus, the line diagram built in ConExp shows 25 largest concepts of visitors of HSE website in terms of news websites in 2006.

\begin{figure}
	\centering
		\includegraphics[scale=.7]{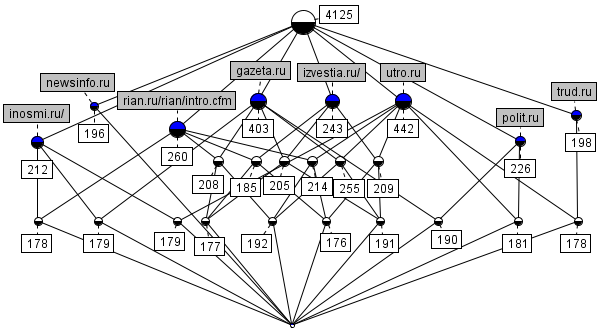}
	\caption{The line diagram of 25 largest concepts for the context of HSE web users}
	\label{fig:25iceberg}
\end{figure}

For real datasets, association rules mining usually results in the large number of rules. However, not all rules are necessary to present the information. Similar compact representation can be used here; thus, one can represent all valid association rules by their subsets that called
bases. For example, the Luxenburger base is a set of association rules in the form 
$$\{B_1\to B_2| (B_1^\prime, B_1) \mbox{ is an upper neighbour of concept } (B_2^\prime,B_2)\}.$$

The rest rules and their support and confidence can be derived by some calculus, which is not usually clear from the base definition.

\begin{exercise}
1. Find the Luxenburger base for the customers context with  $min\_sup=1/3$ and $min\_conf=1/2$.
2. Check whether Concept Explorer generates the association rule base (for the same context) that consists of the Duquenne-Guigues base and  the Luxenburger base. $\square$
\end{exercise}

One of the first algorithms that were explicitly designed to compute frequent
closed itemsets is Close \cite{Pasquier:1999}. Inspired by Apriori it traverses the database in a level-wise manner and generates requent closed itemsets by computing the closures of all minimal generators. See more detailed survey in \cite{Lakhal:05}.

It is interesting that after the years of co-existence, one of the Apriori's authors has started to apply FCA in text mining~\cite{Agrawal:2014}. FCA is also included into textbooks on data mining, see Chapter 8\&9 in~\cite{zaki:2014}.

Another interesting subdomain of frequent pattern mining, where lattice-based methods are successfully used, is so called sequential pattern mining \cite{Zaki:2001,Buzmakov:2013}.

\subsubsection{Multimodal clustering (biclustering and triclustering)}\label{ssec:tric}

Clustering is an activity for finding homogeneous groups of instances in data. In machine learning, clustering is a part of so called unsupervised learning.
The widely adopted idea of cluster relates to instances in a feature space. A cluster in this space
is a subset of data instances (points) that are relatively close to each other but relatively far from other data points. 
Such feature space clustering algorithms are a popular tool in marketing research, bioinformatics, finance, image analysis, web mining, etc. 
With the growing popularity of recent data sources such as biomolecular techniques and Internet, other than instance-to-feature data appear for analysis.

One example is gene expression matrices, entries of which show expression levels of gene material captured in a polymerase reaction. Another example would be $n$-ary relations among several sets of entities such as:
\begin{itemize}
\item Folksonomy data~\cite{Wal:2007} capturing a ternary relation among three sets: users, tags, and resources; 
\item Movies database IMDb (\footnote{\url{www.imdb.com}}) describing a binary relation of ``relevance'' between a set of movies and a set keywords or a ternary relation between sets of movies, keywords and genres;
\item product review websites featuring at least three itemsets (product, product features, product-competitor);
\item job banks comprising at least four sets (jobs, job descriptions, job seekers, seeker skills).
\end{itemize}
 
For two-mode case other cluster approaches demonstrates growing popularity. Thus the notion of bicluster in a data matrix (coined by B.~Mirkin~in \cite{Mirkin:1996}, p. 296) represents a relation between two itemsets. Rather than a single subset of entities, a bicluster features two subsets of different entities. 

In general, the larger the values in the submatrix, the higher interconnection between the subsets, the more relevant is the corresponding bicluster. In the relational data, presence-absence facts represented by binary 1/0 values and this condition expresses the proportion of unities in the submatrix, its ``density'': the larger, the better. It is interesting that a bicluster of the density 1 is a formal concept if its constituent subsets cannot be increased without a drop in the  density value, i.e. a maximal rectangle of 1s in the input matrix w.r.t. permutations of its rows and columns \cite{Ganter:1999}. Usually one of the related sets of entities is a set of objects, the other one is a set of attributes. So, in contrary to ordinary clustering, bicluster $(A,B)$ captures similarity (homogeneity) of objects from $A$ expressed in terms of their common (or having close values) attributes $B$, which usually embrace only a subset of the whole attribute space.

Obviously, biclusters form a set of homogeneous chunks in the data so that further learning can be organized within them. The biclustering techniques and FCA machinery are being developed independently in independent communities using different mathematical frameworks. Specifically, the mainstream in Formal Concept Analysis is based on ordered structures, whereas biclustering relies on conventional optimisation approaches, probabilistic and matrix algebra frameworks \cite{Madeira:2004,Eren:2012}. However, in fact these different frameworks considerably overlap in applications, for example: finding co-regulated genes over gene expression data \cite{Madeira:2004,Besson:2005,Barkow:2006,Tarca:2007,Hanczar:2010,Kaytoue:2011,Eren:2012}, prediction of biological activity of chemical compounds \cite{Blinova:2003,Kuznetsov:2005,DiMaggio:2010b,Asses:2012}, text summarisation and classification \cite{Dhillon:2001,Cimiano:2005,Banerjee:2007,Ignatov:2009,Carpineto:2009}, structuring websearch results and browsing navigation in Information Retrieval \cite{Carpineto:2005,Koester:2006,Eklund:2012,Poelmans:2012}, finding communities in two-mode networks in Social Network Analysis \cite{Duquenne:1996,Freeman:1996,Latapy:2008,Roth:2008,Gnatyshak:2012} and Recommender Systems \cite{duBoucherRyan:2006,Symeonidis:2008a,Ignatov:2008,Nanopoulos:2010,Ignatov:2014}.

For example, consider a bicluster definition from paper~\cite{Barkow:2006}. Bi-Max algorithm described in~\cite{Barkow:2006} constructs  \textbf{inclusion-maximal biclusters} defined as follows:

\bd Given $m$ genes, $n$ situations and a binary table $e$ such that $e_{ij} = 1$ (gene $i$ is active in situation $j$) or $e_{ij} = 0$ (gene $i$ is not active in situation $j$) for all $i \in [1,m]$ and $j \in [1,n]$, the pair $(G, C) \in 2^{\{1,\dots, n\}} \times 2^{\{1,\dots, m\}}$ is called an \textbf{inclusion-maximal bicluster} if and only if (1) $\forall i \in G, j \in C : e_{ij} = 1$ and (2) $\nexists (G_1 , C_1 ) \in 2^{\{1,\dots, n\}} \times 2^{\{1,\dots, m\}}$ with (a) $\forall i_1 \in G_1, \forall j_1 \in C_1$: $e_{{i_1}{j_1}} = 1$ and (b) $G \subseteq G_1 \wedge C \subseteq C_1 \wedge (G_1, C_1 ) \ne (G, C)$.
\ed

Let us denote by $H$ the set of genes (objects in general), by $S$ the set of situations (attributes in general), and by $E\subseteq H\times S$ the binary relation given by the binary table $e$, $|H| = m$, $|S| = n$. Then one has the following proposition:

\begin{proposition}\label{prop:bimax_bicl=concept}
For every pair $(G,C)$, $G\subseteq H$, $C\subseteq S$ the following two statements are equivalent.

1. $(G,C)$ is an inclusion-maximal bicluster of the table $e$;

2. $(G,C)$ is a formal concept of the context $(H,S,E)$.

\end{proposition}

\begin{exercise} Prove Proposition~\ref{prop:bimax_bicl=concept}.$\square$
\end{exercise}

\subsubsection{Object-Attribute-biclustering}\label{sssec:oabic}

Another example is OA-biclustering proposed in \cite{Ignatov:2010,Ignatov:2012a} as a reliable relaxation of formal concept.

\begin{definition}\label{def:bicl}
If $(g,m)\in I$, then $(m',g')$ is called an object-attribute or \emph{OA-bicluster}
with density $\rho(m',g')=\frac{|I\cap (m'\times g')|}{|m'|\cdot|g'|}$.
\end{definition}

Here are some basic properties of OA-biclusters.

\begin{proposition}\label{proposition:bicl}

\noindent 1. $0\leq \rho\leq 1$.

\noindent 2. OA-bicluster $(m',g')$ is a formal concept iff $\rho = 1$.

\noindent 3. if $(m',g')$ is a OA-bicluster, then $(g'',g')\leq (m',m'')$.

\end{proposition}

\begin{exercise} a. Check that properties 1. and 2. from  Proposition~\ref{proposition:bicl} follow directly by definitions.
\noindent b. Use antimonotonicity of $(\cdot)'$ to prove 3. $\square$
\end{exercise}

\begin{figure}[h]
 \begin{center}
  \includegraphics[scale=0.5]{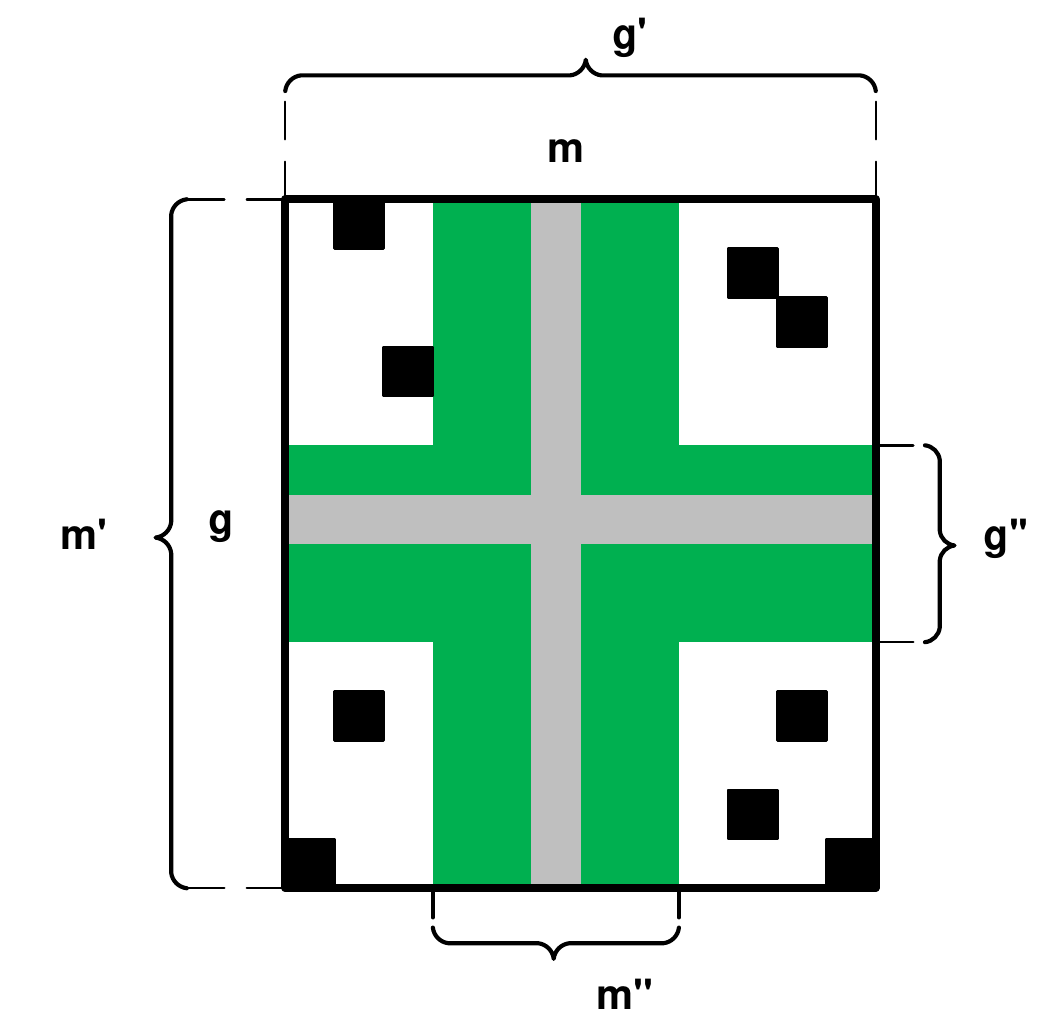}
   \caption{Bicluster based on object and attribute closures}\label{fig:Bicl}
 \end{center}
\end{figure}

In figure \ref{fig:Bicl} one can see the structure of the OA-bicluster for a particular pair $(g,m) \in I$ of a certain context $(G,M,I)$. In general, only the regions $(g'',g')$ and $(m',m'')$ are full of non-empty pairs, i.e. have maximal density $\rho=1$, since they are object and attribute formal concepts respectively. Several black cells indicate non-empty pairs which one may found in such a bicluster. It is quite clear, the density parameter $\rho$ would be a bicluster quality measure which shows how many non-empty pairs the bicluster contains.

\begin{definition}\label{def:dense}
Let  $(A,B) \in 2^G \times 2^M$ be an OA-bicluster and  $\rho_{min}$ be a nonnegative real number, such that $0 \leq \rho_{min} \leq 1$, then $(A,B)$ is called  \emph{dense} if it satisfies the constraint $\rho(A,B)\geq \rho_{min}$.
\end{definition}

Order relation $\sqsubseteq$ on OA-biclusters is defined component-wise: $(A,B) \sqsubseteq (C,D)$ iff $A \subseteq C$ and $B \subseteq D$.

Monotonicity (antimonotonicity) of constraints is often used in mining association rules for effective algorithmic solutions.

\begin{proposition}\label{proposition:dense}
The constraint  $\rho(A,B)\geq \rho_{min}$ is neither monotonic nor anti-monotonic w.r.t. $\sqsubseteq$ relation.

\end{proposition}

\begin{exercise} 1. To prove Proposition~\ref{proposition:dense} for context $\K$ consider OA-biclusters $b_1=(\{g_1,g_3,g_4,g_5\}, \{m_1,m_4,m_5\})$, $b_3=(G, \{m_1,m_2,m_3\})$ and $b_2=(G,M)$.


\begin{center}
\begin{cxt}%
\cxtName{}%
\atr{$m_1$}%
\atr{$m_2$}%
\atr{$m_3$}%
\atr{$m_4$}%
\atr{$m_5$}%
\obj{xxxxx}{$g_1$}
\obj{xxxx.}{$g_2$}
\obj{x..xx}{$g_3$}
\obj{x..xx}{$g_4$}
\obj{x..xx}{$g_5$}
\end{cxt}
\end{center}

\noindent 2. Find generating pairs $(g,m)$ for all these three biclusters. $\square$

\end{exercise}

However, the constraint on $\rho_{min}$  has other useful properties.

If $\rho=0$, this means that we consider the set of all OA-biclusters of the context $\K$.
For $\rho_{min}=0$ every formal concept is ``contained'' in a OA-bicluster of the context $\K$, i.e., the following proposition holds.

\begin{proposition}\label{prop:concept_in_bicl}
For each $(A_c,B_c) \in \BGMI$ there exists a OA-bicluster $(A_b,B_b) \in \mathbf{B}$ such that $(A_c,B_c) \sqsubseteq (A_b,B_b)$.
\end{proposition}

\textbf{Proof.} Let $g \in A_c$, then by antimonotonicity of $(\cdot)'$ we obtain $g' \supseteq B_c$. Similarly, for $m \in B_c$ we have $m' \supseteq A_c$. Hence, $(A_b,B_b)\sqsubseteq (m',g')$. $\square$

The number of OA-biclusters of a context can be much less than the number of formal concepts (which may be exponential in $|G|+|M|$), as stated by the following proposition.

\begin{proposition}\label{prop:improvement}
For a given formal context $\K=(G,M,I)$ and $\rho_{min}=0$ the largest number of OA-biclusters  is equal to $|I|$, all OA-biclusters can be generated in time $O(|I|\cdot(|G|+ |M|))$.
\end{proposition}

\begin{proposition}\label{prop:improvement2}
For a given formal context $\K=(G,M,I)$ and $\rho_{min}>0$ the largest number of OA-biclusters  is equal to $|I|$, all OA-biclusters can be generated in time $O(|I|\cdot |G|\cdot|M|)$.
\end{proposition}

\begin{algorithm}
\caption{OA-bicluster computation\label{BiA}}
    \begin{algorithmic}[1]
      	\REQUIRE $\K=(G,M,I)$ is a formal context, $\rho_{min}$ is a threshold density value of bicluster density
    	 	\ENSURE $B=\{(A_k,B_k)| (A_k,B_k)$ is a bicluster$\}$ 
				\STATE{$B \gets \emptyset$} 
				\IF{$\rho_{min}=0$}
					\FORALL{$(m,g) \in I$}
						\STATE{$B.Add(m',g')$}
					\ENDFOR
				\ELSE
					\FORALL{$(m,g) \in I$}
						\IF{$\rho(m',g')\geq\rho_{min}$}
							\STATE{$B.Add(m',g')$}
						\ENDIF
					\ENDFOR
				\ENDIF
				\STATE $B.RemoveDuplicates()$
				\RETURN{$B$}
    \end{algorithmic}
\end{algorithm}

Algorithm \ref{BiA} is a rather straightforward implementation by definition, which takes initial formal context and minimal density threshold as parameters and computes biclusters for each (object, attribute) pair in relation $I$. However, in its latest implementations we effectively use hashing for duplicates elimination. In our experiments on web advertising data, the algorithm produces 100 times less patterns than the number of formal concepts. In general, for the worst case these values are $2^{\min(|G|,|M|)}$ vs $|I|$. The time complexity of our algorithm is polinomial ($O(|I||G||M|)$) vs exponential in the worst case for Bi-Max ($O(|I||G||L|\log|L|)$) or CbO ($O(|G|^2|M||L|)$), where $|L|$ is a number of generated concepts which is exponential in the worst case ($|L|=2^{\min(|G|,|M|)}$).

\subsubsection{Triadic FCA and triclustering}

\begin{figure}
	\centering
		\includegraphics[width=0.8\textwidth]{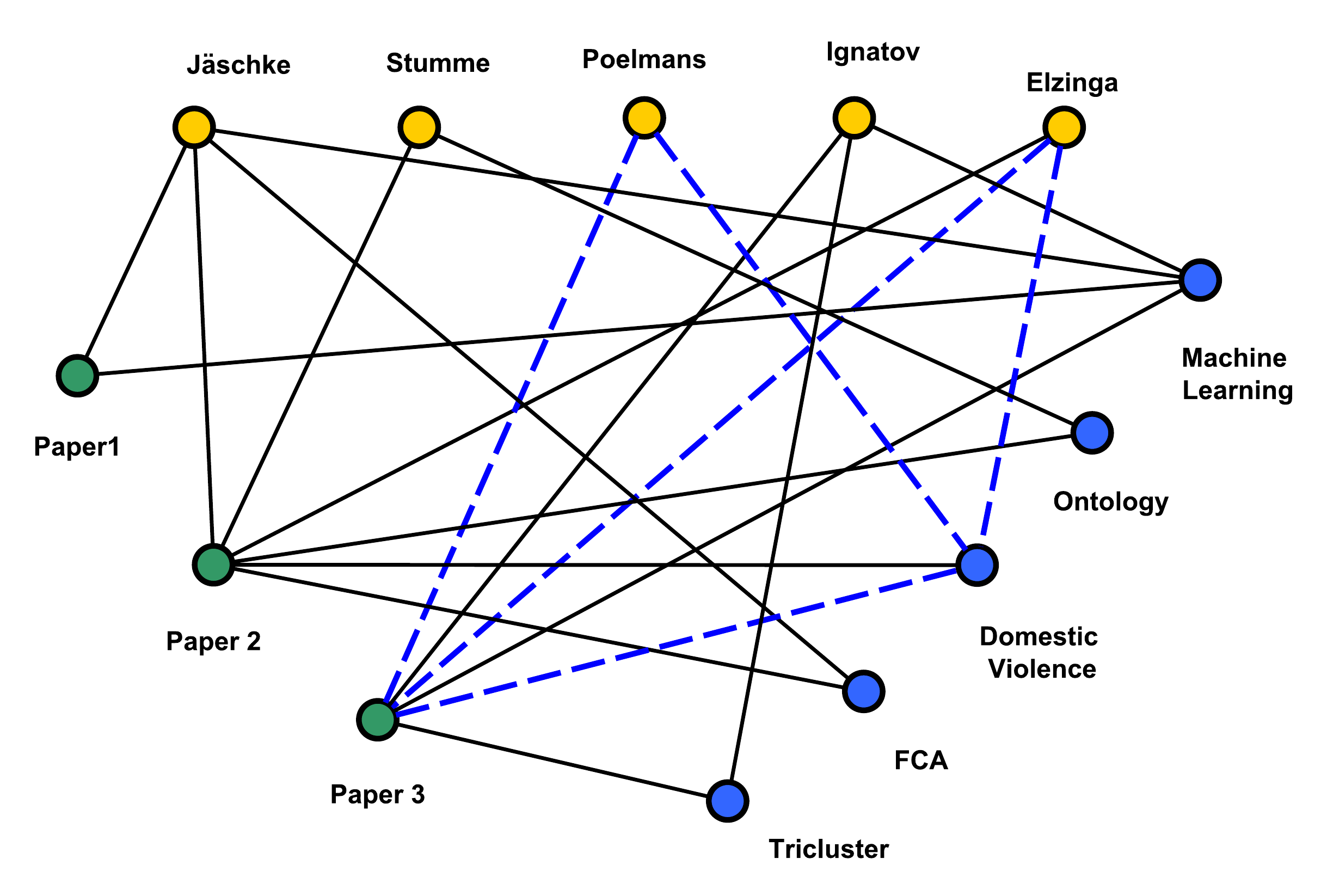}
	\caption{An example of Bibsonomy relation for three paper, five authors and five tags.}
	\label{fig:bibs}
\end{figure}

As we have mentioned, there are such data sources as folksonomies, for example, a bookmarking website for scientific literature Bibsonomy \footnote{\url{bibsonomy.org}} \cite{Benz:2010}; the underlying structure includes triples (user, tag, bookmark) like one in Fig.~\ref{fig:bibs}.

Therefore, it can be useful to extend the biclustering and Formal Concept Analysis to process relations among more than two datasets. A few attempts in this direction have been published in the literature. For example, 
Zaki et al.~\cite{Zaki:2005} proposed Tricluster algorithm for mining biclusters extended by time dimension to real-valued gene expression data.  A triclustering method was designed in \cite{Li:2009} to mine gene expression data using black-box functions and parameters coming from the domain. In the Formal Concept Analysis framework,  theoretic papers  \cite{Wille:1995,Lehmann:1995} introduced the so-called  Triadic Formal Concept Analysis. In~\cite{Ganter:1994},  triadic formal concepts apply to analyse small datasets in a psychological domain.  Paper \cite{Jaschke:2006}  proposed rather scalable method \textsc{Trias} for mining frequent triconcepts in Folksonomies. Simultaneously, a less efficient method on mining closed cubes in ternary relations was proposed by Ji et al.~\cite{Ji:2006}. There are several recent efficient algorithms for mining closed ternary sets (triconcepts) and even more general algorithms than \textsc{Trias}. Thus, Data-Peeler  \cite{Cerf:2009} is able to mine $n$-ary formal concepts and its descendant mines fault-tolerant $n$-sets \cite{Cerf:2013}; the latter was compared with DCE algorithm for fault-tolerant $n$-sets mining from \cite{Georgii:2011}. The paper \cite{Spyropoulou:2014} generalises $n$-ary relation mining to multi-relational setting in databases using the notion of algebraic closure.

In triadic setting, in addition to set of objects, $G$, and set of attributes, $M$, we have $B$, a set of conditions.
Let $\mathbb{K}=(G,M,B,I)$ be a \textbf{triadic context}, where $G$, $M$, and $B$ are sets, and $I$ is a ternary relation: $I\subseteq G\times M\times B$.
The \textbf{ triadic concepts} of an triadic context $(G, M, B, Y \subseteq G \times M \times B)$ are exactly the maximal $3$-tuples $(A_1, A_2 , A_3)$ in $2^{G} \times 2^{M} \times 2^{B}$ with $A_1 \times A_2 \times A_3 \subseteq Y$ with respect to component-wise set inclusion \cite{Wille:1995,Lehmann:1995}. The notion of $n$-adic concepts can be introduced in the similar way to the triadic case \cite{Voutsadakis:02}.

\begin{example} For the bibsonomy example, one of the triadic concepts is 

$$(\{Poelmans, Elzinga\}, \{Domestic \ Violence\}, \{paper 3\})$$ (see dotted edges on the graph in Fig.\ref{fig:bibs}). It means that both users Poelmans and Elzinga marked paper~3 by the tag ``Domestic Violence''.
$\square$
\end{example}

Guided by the idea of finding scalable and noise-tolerant triconcepts, we had a look at triclustering paradigm in general for a triadic binary data, i.e. for tricontexts as input datasets.

Suppose $X$, $Y$, and $Z$ are some subsets of $G$, $M$, and $B$ respectively.

\begin{definition}
    Suppose $\mathbb{K}=(G,M,B,I)$ is a triadic context and $Z \subseteq G$, $Y \subseteq M$, $Z \subseteq B$.
    A triple $T=(X,Y,Z)$ is called an \emph{OAC-tricluster}.
    Traditionally, its components are called \emph{(tricluster) extent, (tricluster) intent, and (tricluster) modus}, respectively.
\end{definition}

The \emph{density} of a tricluster $T=(X,Y,Z)$ is defined as the fraction of all triples of $I$ in $X\times Y\times Z$:

$$    \rho(T):=\frac{|I\cap(X\times Y\times Z)|}{|X||Y||Z|}.$$

\begin{definition}
    The tricluster $T$ is called \emph{dense} iff its density is not less than some predefined threshold, i.e. $\rho(T)\ge\rho_{min}$.
\end{definition}

The collection of all triclusters for a given tricontext $\K$ is denoted by $\mathcal{T}$.

Since we deal with all possible cuboids in Cartesian product $G\times M\times B$, it is evident that the number of all OAC-triclusters, $|\mathcal{T}|$, is equal to $2^{|G|\cdot |M|\cdot |B|}$. However not all of them are supposed to be dense, especially for real data which are frequently quite sparse. Thus we have proposed two possible OAC-tricluster definitions, which give us an efficient way to find within polynomial time a number of (dense) triclusters not greater than the number $|I|$ of triples in the initial data.

In \cite{Ignatov:2015}, we have compared a  set of  triclustering techniques  proposed within Formal Concept Analysis and/or bicluster analysis perspectives: \textsc{OAC-box} \cite{Ignatov:2011},  \textsc{Tribox} \cite{Mirkin:2011},  \textsc{SpecTric} \cite{Ignatov:2013b} and a recent \textsc{OAC-prime} algorithm. This novel algorithm, \textsc{OAC-prime}, overcomes computational and substantive  drawbacks of the earlier formal-concept-like algorithms. 
 In our spectral  approach (SpecTric algorithm)  we rely on an extension of the well-known
reformulation of a bipartite graph partitioning problem to the spectral partitioning of a graph (see,
e.g. \cite{Dhillon:2001}).
For comparison purposes, we have proposed new developments in the following components of the experiment setting:
\begin{enumerate}
\item  Evaluation criteria: The average density, the coverage, the diversity and the number of triclusters, and the computation time and noise tolerance for the algorithms.
\item Benchmark datasets: We use triadic datasets from publicly available internet data as well as synthetic datasets with various noise models.
\end{enumerate}
A preceding work was done in \cite{Gnatyshak:2013}.

As a result we have not defined an absolute winning methods, but the multicriteria choice allows an expert to decide which of the criteria are most important in a specific case and make a choice. Thus our experiments show that our Tribox and OAC-prime algorithms
can be reasonable alternatives to triadic formal concepts and lead to Pareto-effective solutions.
In fact \textsc{TriBox} is better with respect to noise-tolerance and the number of clusters, OAC-prime is the best on scalability to large real-world datasets. In paper~\cite{Gnatyshak:2014}, an efficient version of online OAC-prime has been proposed.

In our experiments we have used a context of top~250 popular movies from \url{www.imdb.com}, objects are movie titles, attributes are tags, whereas conditions are genres. Prime OAC-triclustering showed rather good results being one the fastest algorithm under comparison.

\begin{example}
Examples of Prime OAC triclusters with their density indication for the IMDB context are given below:
\begin{enumerate}
    \item $36\%$, \textcolor{obj-red}{\{The Shawshank Redemption  (1994), Cool Hand Luke (1967), American History X (1998), A Clockwork Orange (1971), The Green Mile (1999)\}}, \textcolor{obj-blue}{\{Prison, Murder, Friend, Shawshank, Banker\}}, \textcolor{obj-green}{\{Crime, Drama\}}
    \item $56,67\%$, \textcolor{obj-red}{\{The Godfather: Part II (1974), The Usual Suspects (1995)\}}, \textcolor{obj-blue}{\{Cuba, New York, Business, 1920s, 1950s\}}, \textcolor{obj-green}{\{Crime, Drama, Thriller\}}
    \item $60\%$, \textcolor{obj-red}{\{Toy Story (1995), Toy Story 2 (1999)\}}, \textcolor{obj-blue}{\{Jealousy, Toy, Spaceman, Little Boy, Fight\}}, \textcolor{obj-green}{\{Fantasy, Comedy, Animation, Family, Adventure\}}
\end{enumerate}$\square$
\end{example}

\subsection{FCA in Classification}

It is a matter of fact that Formal Concept Analysis helped
to algebraically rethink several models and methods in Machine
Learning such as version spaces~\cite{Ganter:2003}, learning from
positive and negative examples~\cite{Blinova:2003,Kuznetsov:2004a}, and decision trees \cite{Kuznetsov:2004a}.  It was also shown that concept lattice is a perfect search space for learning globally optimal decision trees \cite{Belohlavek:2009}. Already in early 90s both supervised and unsupervised machine learning techniques and applications based on Formal Concept Analysis were introduced in the machine learning community. E.g., in ML-related venues there were reported results on the concept lattice based clustering in GALOIS system that suited for information retrieval via browsing~\cite{Carpineto:1993,Carpineto:1996}. 
\cite{Fu:2004} performed a comparison of seven FCA-based classification algorithms. 
\cite{Rudolph:2007} and \cite{Tsopze:2007} propose independently to use FCA to design a neural network architecture. 
In \cite{Outrata:2010,Belohlavek:2014} FCA was used as a data preprocessing technique to transform the attribute space to improve the results of  decision tree induction. Note that FCA helps to perform feature selection via conceptual scaling and has quite evident relations with Rough Sets theory, a popular tool for feature selection in classification~\cite{Ganter:2008}. \cite{Visani:2011} proposed Navigala, a navigation-based approach for supervised classification, and applied it to noisy symbol recognition. Lattice-based approaches were also successfully used for classification of data with complex descriptions such as graphs or trees \cite{Kuznetsov:2005,Zaki:2006}. Moreover, in~\cite{Flach:2012} (Chapter 4, ``Concept Learning'') FCA is suggested as an alternative learning framework.

\subsubsection{JSM-method of hypothesis generation}

The JSM-method proposed by Viktor K. Finn in late 1970s was proposed as attempt to describe induction in purely deductive
form and thus to give at least partial justification of induction \cite{Finn:1983}. The method is named
to pay respect to the English philosopher John Stuart Mill, who proposed several schemes
of inductive reasoning in the 19th century. For example, his Method of Agreement, is formulated as follows:
``If two or more instances of the phenomenon under investigation have only one
circumstance in common, ... [it] is the cause (or effect) of the given phenomenon.''

The method proved its ability to enable learning from
positive and negative examples in various domains~\cite{Kuznetsov:1991}, e.g., in life sciences \cite{Blinova:2003}.

For RuSSIR audience, the example of the JSM-method application in paleography might be especially interesting \cite{Gusakova:2001}:
JSM was used for dating birch-bark documents of 10--16 centuries of the Novgorod republic. There were five types of attributes: individual letter features, features
common to several letters, handwriting, language features (morphology, syntax,
and typical errors), style (letter format, addressing formulas and their key words).

Even though, the JSM-method was formulated in a mathematical logic setting, later on the equivalence between JSM-hypotheses and formal concepts was recognized \cite{Kuznetsov:1996}.

The following definition of a hypothesis (``no counterexample-hypothesis'') in
FCA terms was given in \cite{Ganter:2000}.

Let  $\K = (G,M, I)$ be a context. There are a \textbf{target attribute} $w\notin M$,
\bi
\item \textbf{positive examples}, i.e. set $G_+\subseteq G$ of objects known to have $w$,
\item \textbf{negative examples}, i.e. set $G_-\subseteq G$ of objects known not to have $w$,
\item \textbf{undetermined examples}, i.e. set $G_{\tau}\subseteq G$  of objects for which it
is unknown whether they have the target attribute or do not have it.
\ei

\noindent There are three subcontexts of $\K=(G,M,I)$, the first two are used for
the training sample:
$\K_\varepsilon: =(G_\varepsilon,M,I_\varepsilon),$ $\varepsilon\in
\{-, +, \tau\}$ with respective derivation operators $(\cdot)^+$, $(\cdot)^-$, and $(\cdot)^\tau$.

\bd
A \textbf{positive hypothesis} $H\subseteq M$
is an intent of $\K_+$ not contained in the intent $g^-$ of any
negative example $g\in G_-$: $\forall g\in G_-\quad  H\not\subseteq g^-$.
Equivalently,
$$H^{++} = H,\quad H'\subseteq G_+\cup G_{\tau}.$$

\ed

Negative hypotheses are defined similarly. An intent of $\K_+$ that
is contained in the intent of a negative example is called a  \textbf{falsified (+)-generalisation}.

\begin{example} In Table~\ref{tbl:scor}, there is a many-valued context representing credit scoring data.

$G_+=\{1,2,3,4\}$, $G_-=\{5,6,7\}$, and $G_{\tau}=\{8,9,10\}$. The target attribute takes values $+$ and $-$ meaning ``low risk'' and ``high risk'' client, respectively.

\begin{table}
	\centering
	\caption{Many-valued classification context for credit scoring}\label{tbl:scor}
		\begin{tabular}{|c||c|c|c|c|c|}
		\hline
		G / M & Gender & Age & Education & Salary & Target\\
		\hline
		\hline
1 & M & young & higher & high & $+$\\
2 & F & middle & special & high & $+$\\
3 & F & middle & higher & average & $+$\\
4 & M & old & higher & high & $+$\\
		\hline
5 & M & young & higher & low & $-$\\
6 & F & middle & secondary & average &$-$\\
7 & F & old & special & average &$-$\\
		\hline
8 & F & young & special & high & $\tau$\\
9 & F & old & higher & average & $\tau$\\
10 & M & middle & special & average & $\tau$\\
		\hline
		\end{tabular}
\end{table}

To apply JSM-method in FCA terms we need to scale the given data. One may use nominal scaling as below.

\begin{center}
\begin{cxt}%
\cxtName{}%
\att{$M$}%
\att{$F$}%
\att{$Y$}%
\att{$Mi$}%
\att{$O$}%
\att{$HE$}%
\att{$Sp$}%
\att{$Se$}%
\att{$HS$}%
\att{$A$}%
\att{$L$}%
\att{$w$}%
\att{$\bar{w}$}%
\obj{x.x..x..x..x.}{$g_1$}
\obj{.x.x..x.x..x.}{$g_2$}
\obj{.x.x.x...x.x.}{$g_3$}
\obj{x...xx..x..x.}{$g_4$}
\obj{x.x..x....x.x}{$g_5$}
\obj{.x.x...x.x..x}{$g_6$}
\obj{.x..x.x..x..x}{$g_7$}
\end{cxt}
\end{center}

Then we need to find positive and negative non-falsified hypotheses. If Fig.~\ref{fig:posneglatts} there are two lattices of positive and negative examples for the input context, respectively.

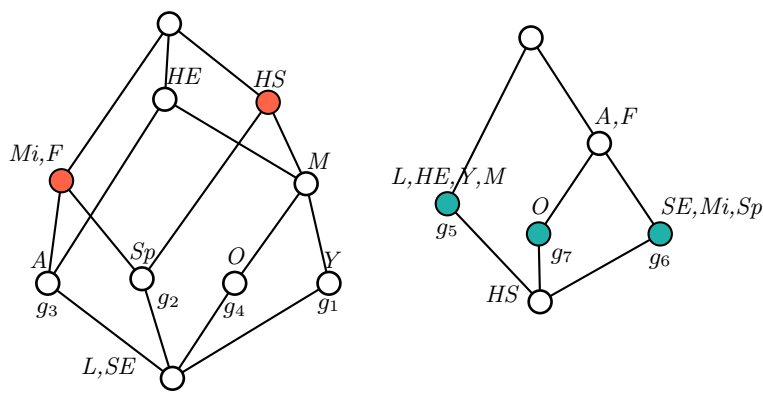
\begin{figure}[h]
\begin{minipage}[h]{0.50\linewidth}  
  \centering
	\begin{picture}(150,160)
\unitlength 0.20mm
\begin{diagram}{270}{260}
\Node{1}{155}{252}
\Node{2}{157}{17}
\Node{3}{84}{148}
\Node{4}{75}{80}
\Node{5}{137}{83}
\Node{6}{152}{202}
\Node{7}{245}{146}
\Node{8}{260.0}{80.0}
\Node{9}{198}{80.0}
\Node{10}{220.0}{200.0}
\Edge{6}{1}
\Edge{7}{10}
\Edge{3}{1}
\Edge{4}{3}
\Edge{5}{10}
\Edge{2}{4}
\Edge{4}{6}
\Edge{2}{9}
\Edge{8}{7}
\Edge{9}{7}
\Edge{2}{5}
\Edge{5}{3}
\Edge{10}{1}
\Edge{7}{6}
\Edge{2}{8}
\NoDots
\centerObjbox{4}{0}{13}{$g_3$}
\rightObjbox{5}{10}{10}{$g_2$}
\centerObjbox{8}{0}{10}{$g_1$}
\centerObjbox{9}{0}{13}{$g_4$}
\leftAttbox{2}{25}{0}{L,SE}
\leftAttbox{3}{0}{10}{Mi,F}
\leftAttbox{4}{0}{10}{A}
\centerAttbox{5}{0}{10}{Sp}
\rightAttbox{6}{0}{10}{HE}
\rightAttbox{7}{0}{10}{M}
\centerAttbox{8}{0}{10}{Y}
\centerAttbox{9}{0}{13}{O}
\centerAttbox{10}{0}{10}{HS}
\CircleSize{15}
\end{diagram}
\put(-186,148){\color{tomato}{\circle*{14}}}%
\put(-50,200){\color{tomato}{\circle*{14}}}%
\end{picture}
 \end{minipage} 
\hfill
\begin{minipage}[h]{0.50\linewidth}
\begin{picture}(100,100)
  \unitlength 0.20mm
\begin{diagram}{210}{200}
\Node{1}{105}{190}
\Node{2}{111}{15}
\Node{3}{50.0}{80}
\Node{4}{150.0}{120}
\Node{5}{190.0}{60}
\Node{6}{110.0}{60}
\Edge{2}{3}
\Edge{5}{4}
\Edge{6}{4}
\Edge{4}{1}
\Edge{2}{6}
\Edge{3}{1}
\Edge{2}{5}
\NoDots
\centerObjbox{3}{0}{13}{$g_5$}
\centerObjbox{5}{0}{13}{$g_6$}
\rightObjbox{6}{8}{10}{$g_7$}
\leftAttbox{2}{15}{0}{HS}
\centerAttbox{3}{0}{10}{L,HE,Y,M}
\centerAttbox{4}{10}{10}{A,F}
\rightAttbox{5}{0}{10}{SE,Mi,Sp}
\centerAttbox{6}{0}{13}{O}
\CircleSize{15}
\end{diagram}
\put(-20,60){\color{lightseagreen}{\circle*{14}}}%
\put(-160,80){\color{lightseagreen}{\circle*{14}}}%
\put(-100,60){\color{lightseagreen}{\circle*{14}}}%
\end{picture}
  \end{minipage} 
 \caption{The line diagrams of the lattice of positive hypotheses (left) and the lattice of negative hypotheses (right).}
  \label{fig:posneglatts}
\end{figure}

Shaded nodes correspond to maximal non-falsified hypotheses, i.e. they have no upper neighbors being non-falsified hypotheses.

For $\K_+$ hypothesis $\{HE\}$ is falsified since object $g_5$ provides a counterexample, i.e. $\{HE\} \subseteq g_5^-=\{M,Y,HE,L\}$.

For $\K_-$ hypothesis $\{A,F\}$ is falsified since there is a positive counterexample, namely $\{A,F\} \subseteq g_3^+=\{F,M,HE,A\}$.$\square$

\end{example}

Undetermined examples $g_\tau$ from $G_{\tau}$ are classified as follows:

\bi
\item If $g^{\tau}_\tau$ contains a positive, but no negative hypothesis, then $g_\tau$ is \textbf{classified positively} (presence of target attribute $w$ predicted).

\item If $g^{\tau}_\tau$ contains a negative, but no positive hypothesis, then $g_\tau$ \textbf{classified negatively} (absence of target attribute $w$ predicted).

\item If $g^{\tau}_\tau$ contains both negative and positive hypotheses, or if $g^{\tau}_\tau$
 does not contain any hypothesis, then object classification is \textbf{contradictory} or \textbf{undetermined}, respectively.
\ei

It is clear,  for performing classification it is enough to have only minimal hypotheses
(w.r.t. $\subseteq$), negative and positive ones.

\begin{exercise}
For the credit scoring context, classify all undetermined examples. $\square$

\end{exercise}

There is a strong connection between hypotheses and implications.

\begin{proposition}
A positive hypothesis $h$ corresponds to an implication
$h\to \{w\}$ in the context $K_+=(G_+,M\cup\{w\},I_+\cup G_+\times \{w\})$.

A negative hypothesis $h$ corresponds to an implication
$h\to \{\bar{w}\}$ in the context $K_-=(G_-,M\cup\{\bar{w}\},I_-\cup G_-\times \{\bar{w}\})$.

Hypotheses are  implications which premises are closed (in $K_+$ or in $K_-$).
\end{proposition}

A detailed yet retrospective survey on JSM-method (in FCA-based and original formulation) and its applications can be found in \cite{Kuznetsov:2005a}. A further extension of JSM-method to triadic data with target attribute in FCA-based formulation can be found in \cite{Zhuk:2014,Ignatov:2014a}; there, the triadic extension of JSM-method used CbO-like algorithm for classification in Bibsonomy data.

However, we saw that original data often need scaling, but, for example, it is not evident what to do in case of learning with labeled graphs. To name a few problems of this kind we would mention structure-activity relationship problems for chemicals given by molecular graphs and learning semantics from graph-based (XML, syntactic tree) text representations. Motivated by search of possible extensions of original FCA machinery to analyse data with complex structure, Ganter and Kuznetsov proposed so called Pattern Structures~\cite{Ganter:2001}.

\subsection{Pattern Structures for data with complex descriptions}

The basic definitions of Pattern Structures were proposed in \cite{Ganter:2001}.

 Let  $G$ be a set of objects and $D$ be a set of all possible object descriptions. Let $\sqcap$ be a similarity operator. It helps to work with objects that have non-binary attributes like in traditional FCA setting, but those that have complex descriptions like intervals~\cite{Kaytoue:2011}, sequences~\cite{Buzmakov:2013s} or (molecular) graphs~\cite{Kuznetsov:2005}. Then $(D,\sqcap)$  is a meet-semi-lattice of object descriptions. Mapping  $\delta: G \to D$ assigns an object $g$ the description $d \in (D, \sqcap)$.    

A triple  $(G,(D,\sqcap),\delta)$ is a pattern structure. Two operators $( \cdot)^\square$ define Galois connection between $(2^G, \subseteq)$ and $(D,\sqcap)$:

\begin{eqnarray}
A^\square =\bigsqcap\limits_{g \in A} \delta(g)    \mbox{ for } A \subseteq G	\label{op1}\\
d^\square =\{g \in G | d \sqsubseteq \delta(g) \}   \mbox{ for }  d \in (D,\sqcap), \mbox{ where }\label{op2} \\
d \sqsubseteq \delta(g) \iff d \sqcap \delta(g)=d	.\nonumber
\end{eqnarray}

For a set of objects $A$ operator~\ref{op1} returns the common description (pattern) of all objects from $A$. For a description $d$ operator~\ref{op2} returns the set of all objects that contain $d$.

A pair $(A,d)$ such that $A\subseteq G$ and $d \in (D,\sqcap)$ is called a pattern concept of the pattern structure $(G,(D,\sqcap),\delta)$ iff $A^\square=d$ and $d^\square=A$. In this case $A$ is called a pattern extent and $d$ is called a pattern intent of a pattern concept  $(A, d)$. 
Pattern concepts are partially ordered by $(A_1,d_1) \leq (A_2,d_2 ) \iff A_1 \subseteq A_2 (\iff d_2 \sqsubseteq d_1)$. The set of all pattern concepts forms a complete lattice called a pattern concept lattice.

\paragraph{Intervals as patterns.}

It is obvious that similarity operator on intervals should fulfill the following condition:  two intervals should belong to an interval that contains them. Let this new interval be minimal one that contains two original intervals. Let  $[a_1,b_1]$ and $[a_2,b_2]$ be two intervals such that $a_1,b_1, a_2, b_2 \in \mathbb R$, $a_1\leq b_1$ and $a_2 \leq b_2$, then their similarity is defined as follows:
 
$$[a_1,b_1 ] \sqcap [a_2,b_2 ]=[\min(a_1,a_2 ), \max(b_1,b_2 )].$$

Therefore 
\begin{eqnarray*}
[a_1,b_1 ] \sqsubseteq [a_2,b_2 ] \iff [a_1,b_1 ] \sqcap [a_2,b_2 ]=[a_1,b_1 ]  \\
\iff \big[ \min (a_1, a_2) , \max (b_1, b_2) \big]  =[ a_1, b_1]\\
\iff a_1 \leq  a_2  \mbox{ and }  b_1 \geq b_2 \iff [a_1,b_1 ] \supseteq [a_2,b_2] \\
\end{eqnarray*}

Note that $a \in \mathbb R$ can be represented by $[a,a]$.

\paragraph{Interval vectors as patterns.}

Let us call p-adic vectors of intervals as interval vectors. In this case for two interval vectors of the same dimension $e=\langle [a_i,b_i] \rangle_{i \in [1,p]}$ and $f= \langle [c_i,d_i] \rangle _{i\in [1,p]}$ we define similarity operation via the intersection of the corresponding components of interval vectors, i.e.:

$$e \sqcap f=\langle [a_i,b_i ]\rangle_{i \in [1,p]} \sqcap \langle [c_i,d_i ] \rangle_{i\in [1,p]} \iff e \sqcap f=\langle [a_i,b_i] \sqcap [c_i,d_i] \rangle_{i \in [1,p] }$$

Note that interval vectors are also partially ordered:
$$e \sqsubseteq f \iff\langle [a_i,b_i ] \rangle _{i\in [1,p] } \sqsubseteq \langle [c_i,d_i ] \rangle_{i \in [1,p]} \iff  [a_i,b_i ] \sqsubseteq [c_i,d_i ]$$  for all  $i \in [1,p]$.

\begin{example}
Consider as an example the following table of  movie ratings:

\begin{table}[ht]
\caption{Movie rates}\label{tbl-data}
\begin{center}
\begin{small}
            \begin{tabular}{|c|c|c|c|c|c|c|c|c|}
              \hline
               & The Artist  &Ghost & Casablanca &Mamma Mia!&  Dogma& Die Hard &Leon   \\
           \hline
            User1& 4 & 4 & 5 & 0 & 0 & 0 & 0 \\
            \hline
            User2 &5 & 5 & 3 & 4 & 3 & 0 & 0 \\
            \hline
            User3 &0 & 0 & 0 & 4 & 4 & 0 & 0 \\
            \hline
            User4 &0 & 0 & 0 & 5& 4 & 5 & 3 \\
             \hline
             User5&0 & 0 & 0 & 0 & 0 & 5 &  5\\
             \hline
            User6&0 & 0 & 0 & 0 & 0 & 4 & 4 \\
              \hline
            \end{tabular}
\end{small}
\end{center}
\end{table}

Each user of this table can be described by vector of ratings' intervals.
For example, $\delta(u_1)=\langle [4,4],[4,4],[5,5],[0,0],[0,0],[0,0],[0,0] \rangle$.
If some new user $u$ likes movie Leon, a movie recommender system would reply who else like this movie by applying operator~\ref{op2}:
$[4,5]_{Leon}^\square=\{u_5,u_6\}$. Moreover, the system would retrieve the movies that users 5 and 6 liked, hypothesizing that they have similar tastes with $u$. Thus, operator~\ref{op1} results in $d=\{u_5,u_6\}^\square=\langle [0,0],[0,0],[0,0],[0,0],[0,0],[4,5],[4,5] \rangle$, suggesting that Die Hard is worth watching for the target user $u$.

Obviously, the pattern concept $(\{u_5,u_6\},d)$ describes a small group of like-minded users and their shared preferences are stored in the vector $d$ (cf. bicluster). $\square$

Taking into account constant pressing of industry requests for Big Data tools, several ways of their fitting to this context were proposed in~\cite{Kuznetsov:2013p,Kuznetsov:2013}; thus, for Pattern Structures in classification setting, combination of lazy evaluation with projection approximations of initial data, randomisation and parallelisation, results in reduction of algorithmic complexity to low degree polynomial.
This observations make it possible to apply pattern structures in text mining and learning from large text collections~\cite{Strok:2014}.
Implementations of basic Pattern Structures algorithms are available in FCART. $\square$

\end{example} 

\begin{exercise} 1. Compose a small program, e.g. in Python, that enumerates all pattern concepts from the movie recommender example directly by definition or adapt CbO this end.
2. In case there is no possibility to perform 1., consider the subtable of the first four users and the first four movies from the movie recommender example. Find all pattern concepts by the definition. Build the line diagram of the pattern concept lattice. $\square$
\end{exercise}

However, Pattern Structures is not the only attempt to fit FCA to data with more complex description than Boolean one. 
Thus, during the past years, the research on extending FCA theory to cope with imprecise and incomplete information made significant progress.
The underlying model is a so called fuzzy concepts lattice; there are several definitions of such a lattice, but the basic assumption usually is that an object may posses attributes to some degree~\cite{Belohlavek:2011}. For example, in sociological studies age representation requires a special care:  a person being a teenager cannot be treated as a truly adult one on the first day when his/her age exceeds a threshold of 18 years old (moreover, for formal reasons this age may differ in different countries). However, it is usually the case when we deal with nominal scaling; even ordinal scaling may lead to information loss because of the chosen granularity level. So, we need a flexible measure of being an adult and a teenage person at the same and it might be a degree lying in [0,1] interval for each such attribute. Another way to characterise this imprecision or roughness can be done in rough sets terms~\cite{Kent:1996}. An interested reader is invited to follow a survey on Fuzzy and Rough FCA in~\cite{Poelmans:2014}. The correspondence between Pattern Structures and Fuzzy FCA can be found in~\cite{Pankratieva:2012}.

\subsection{FCA-based Boolean Matrix Factorisation}

Matrix Factorisation (MF) techniques are in the typical inventory of Machine Learning (\cite{Flach:2012}, chapter Features), Data Mining (\cite{zaki:2014}, chapter Dimensionality Reduction) and Information Retrieval (\cite{Manning:2008}, chapter Matrix decompositions and latent semantic indexing).  Thus MF used for dimensionality reduction and feature extraction, and, for example, in Collaborative filtering recommender MF techniques are now considered industry standard \cite{Koren:2009}.

Among the most popular types of MF we should definitely mention Singular Value Decomposition (SVD) \cite{Elden:2007} and its various modifications like Probabilistic Latent Semantic Analysis (PLSA) \cite{Hofmann:2001} and SVD++ \cite{Koren:2008}. However, several existing factorisation techniques, for example, non-negative matrix factorisation (NMF) \cite{Lin:2007} and Boolean matrix factorisation (BMF) \cite{Belohlavek:2010}, seem to be less studied in the context of modern Data Analysis and Information Retrieval.

Boolean matrix factorisation (BMF) is a decomposition of the original matrix   $ I\in \{0,1 \}^{n\times m} $, where $I_ {ij}\in\{0,1 \}, $  into a Boolean matrix product $ P \circ Q $ of binary matrices $ P \in \{0,1 \}^{n \times k} $ and $ Q \in \{0,1 \} ^ {k \times m} $ for the smallest possible number of $k. $
Let us define Boolean matrix product as follows:
\begin{equation}
(P\circ Q)_{ij}=\bigvee_{l=1}^k P_{il}\cdot Q_{lj},
\label{def:bmf}
\end{equation}
where $ \bigvee $ denotes disjunction, and $ \cdot $ conjunction.

Matrix $ I $ can be considered a matrix of binary relations between set $ X $ of objects (users), and a set $ Y $ of attributes (items that users have evaluated). We assume that $ xIy $  iff the user $x$  evaluated object $y$. The triple $ (X, Y, I) $ clearly forms a formal context.

Consider a set $ \mathcal {F} \subseteq \mathcal {B} (X, Y, I) $, a subset of all formal concepts of context $ (X, Y, I) $, and introduce matrices $ P_ {\mathcal {F}} $ and $ Q_ {\mathcal {F}}: $
$$(P_{\mathcal{F}})_{il}=\left\{
                           \begin{array}{ll}
                             1, i\in A_l,\\
                             0, i\notin A_l,
                           \end{array}
                         \right.
  \ \ (Q_{\mathcal{F}})_{lj}=\left\{
                               \begin{array}{ll}
                                 1, j\in B_l, \\
                                 0, j\notin B_l.
                               \end{array}
                             \right.
  ,$$
where $(A_l, B_l) $ is a formal concept from $F$.

We can consider decomposition of the matrix $I$ into binary matrix product $ P_ \mathcal {F} $ and $ Q_ \mathcal {F} $ as described above. The following theorems are proved in \cite{Belohlavek:2010}:

\begin{theorem}(Universality of formal concepts as factors). For every $I$ there is  $\mathcal{F}\subseteq \mathcal{B}(X,Y,I)$,  such that  $I=P_\mathcal{F}\circ Q_\mathcal{F}.$
\end{theorem} 

\begin{theorem} (Optimality of formal concepts as factors). Let $I=P\circ Q$  for $n\times k$ and $k\times m$ binary matrices $P$ and $Q$. Then there exists a set $\mathcal{F}\subseteq \BV(X,Y,I)$ of formal concepts of $I$  such that $|\mathcal{F}|\leq k$ and  for the $n \times |\mathcal{F}|$ and $|\mathcal{F}| \times m$ binary matrices $P_\mathcal{F}$ and $Q_\mathcal{F}$ we have $I=P_\mathcal{F}\circ Q_\mathcal{F}.$
\end{theorem} 

\begin{example}
Transform the  matrix of ratings described above by thresholding ($geq3$), to a Boolean matrix, as follows:
 $$\left(
    \begin{array}{ccccccc}
         1 & 1 & 1 & 0 & 0 & 0 & 0 \\
        1 & 1 & 1 & 1 & 1 & 0 & 0 \\
        0 & 0 & 0 & 1 & 1 & 0 & 0 \\
        0 & 0 & 0 & 1& 1 & 1 & 1 \\
        0 & 0 & 0 & 0 & 0 & 1 &  1\\
        0 & 0 & 0 & 0 & 0 & 1 & 1 \\
    \end{array}
  \right)=I.
$$
The decomposition of the matrix $ I $ into the Boolean product of $ I = A_ {\mathcal {F}} \circ B_ {\mathcal {F}} $ is the following:
$$\left(
    \begin{array}{ccccccc}
         1 & 1 & 1 & 0 & 0 & 0 & 0 \\
        1 & 1 & 1 & 1 & 1 & 0 & 0 \\
        0 & 0 & 0 & 1 & 1 & 0 & 0 \\
        0 & 0 & 0 & 1& 1 & 1 & 1 \\
        0 & 0 & 0 & 0 & 0 & 1 &  1\\
        0 & 0 & 0 & 0 & 0 & 1 & 1 \\
    \end{array}
  \right)= \left( \begin{array}{ccc}
         1 & 0 & 0 \\
        1 & 1 & 0 \\
        0 & 1 & 0 \\
        0 & 1& 1 \\
        0 & 0 & 1 \\
        0 & 0 & 1 \\
    \end{array}
  \right)
  \circ  \left( \begin{array}{ccccccc}
         1 & 1 & 1 & 0 & 0 & 0 & 0 \\
        0 & 0 & 0 & 1 & 1 & 0 & 0 \\
        0 & 0 & 0 & 0 & 0 & 1 &1 \\
          \end{array}
  \right).$$

Even this tiny example shows that the algorithm has identified three factors that significantly reduces the dimensionality of the data. $\square$

\end{example}

There are  several algorithms for finding $P_ \mathcal {F}$ and $Q_ \mathcal {F}$ by calculating formal concepts based on these theorems \cite{Belohlavek:2010}. Thus, the approximate algorithm (Algorithm 2 from \cite{Belohlavek:2010}) avoids computation of all possible formal concepts and therefore works much faster than direct approach by all concepts generation. Its running time complexity in the worst case yields $O(k|G||M|^3)$, where $k$ is the number of found factors, $|G|$ is the number of objects, $|M|$  is the number of attributes.

As for applications, in \cite{Outrata:2010,Belohlavek:2014}, FCA-based BMF was used as a feature extraction technique for improving the results of  classification. Another example closely relates to IR; thus, in \cite{Nenova:2013,Ignatov:2014} BMF demonstrated comparable results to SVD-based collaborative filtering in terms of MAE and precision-recall metrics.  

Further extensions of BMF to triadic and n-ary data were proposed in~\cite{Belohlavek:2013} and~\cite{Miettinen:2011}, respectively (the last one in not FCA-based)

\subsection{Case study: admission process to HSE university}

in this case study we reproduce results of our paper from~\cite{Romashkin:2011}. Assuming probable confusion of the Russian educational system, we must say a few words about the National Research University Higher School of Economics\footnote{\url{http://www.hse.ru/en/}} and its admission process.

Nowadays HSE is acknowledged as a leading university in the field of economics, management, sociology, business informatics, public policy and political sciences among Russian universities. Recently a number of bachelor programmes offered by HSE has been increased. In the year 2010 HSE offered 20 bachelor programmes. We consider only bachelor programmes in our investigation.

In order to graduate from school and  enter a university or a college every Russian student must pass a Unified State Exam (Russian transcription: EGE), similar to US SAT--ACT or UK A-Level tests. During 2010 admission to U-HSE, entrants were able to send their applications to up to three programmes simultaneously. Some school leavers (major entrants of HSE bachelor programmes) chose only one programme, some chose two or three. Then entrants had to choose only one programme to study among successful applications.

We used data representing admission to HSE in 2010. It consists of information about 7516 entrants. We used mainly information about programmes (up to three) to which entrants apply\footnote{HSE is a state university, thus most of student places are financed by government. In this paper we consider only such places.}. Exactly 3308 entrants successfully applied at least to one programme, but just 1504 become students. Along with this data we also used the data of entrants' survey (76\% of entire assembly).

Further in the paper we mostly used data for the Applied Mathematics and Informatics programme to demonstrate some results. The total number of applications to the Applied Mathematics and Informatics programme was 843, of which 398 were successful but only 72 of them were actually accepted into the program. It might seem confusing only 72 out of 398 eligible prospective students decided to enroll, but since the admission process was set up in two stages, and at each stage only 72 entrants were eligible to attend the program, some of them decided to go for a different programme or university. As a result, the number of entrants whose applications were successful in any way came down to 398. Such situation is typical for all the bachelor programmes at HSE.

FCA requires object-attribute data. In our case objects are entrants and programmes they apply to are attributes. Together they are treated as a context. A series of contexts were constructed. Namely, we built a context for every programme where objects were entrants applying to that programme and attributes were other programmes they applied to. We built a separate context for every programme because it is meaningless to consider all programmes at once as programmes are very different in size and the resulting lattice would represent only the largest of them.

Likewise, we built a context for every programme where objects were entrants and attributes were
programmes to which entrants successfully applied as well as the programmes that the entrants decided to enroll into, including those at other universities.

These contexts were then used to build concept lattices. Since the resulting lattices had too complicated a
structure to interpret, we filtered concepts by their extent size (extent size is the number of objects, in our
case it is the number of entrants), thus remaining concepts express only some of the more common patterns in entrants decisions.

\begin{figure}[tp]
\centering
\includegraphics[width=\textwidth]{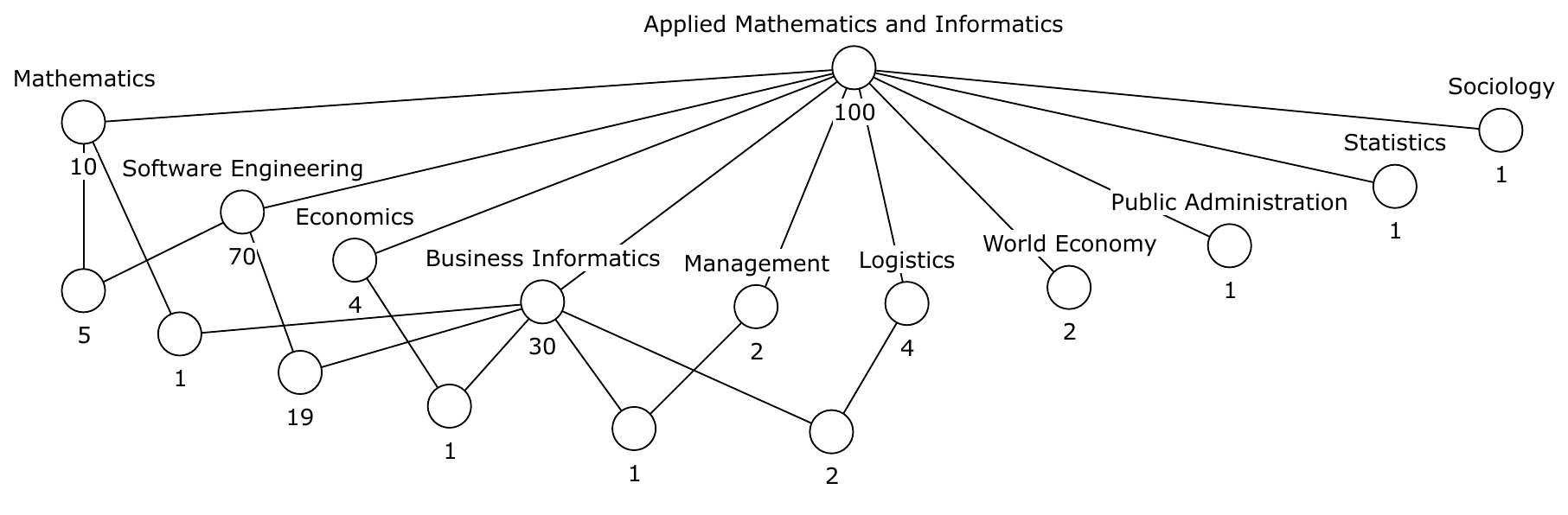}
\caption{Other programmes which entrants of Applied Mathematics and Informatics programme also applied.}
\label{pm.17}
\end{figure}

To which programmes entrants often apply simultaneously? Trying to answer this question for every programme, we built diagrams\footnote{
As any other data mining technique FCA implies an intensive use of software. All diagrams mentioned in this paper have been produced with meud (\url{https://github.com/jupp/meud-wx}).} similar to figure~\ref{pm.17}. Such diagrams help us to reveal common patterns in entrants choices. Typical applications of FCA imply building formal concept lattices discussed earlier, but here we filter concepts by extent size to avoid complexity caused by noise in the data. Thus the order on remaining concepts is no longer a lattice, it is a partial order. Meaning of the labels on the diagram is obvious. A label above a node is a programme, a label below a node is a percent of entrants to Applied Mathematics and Informatics programme who also applied to programmes connected to a node from above. For example, the most left and bottom node on the diagram means that five percent of applied math's entrants also apply to Mathematics and Software Engineering. Then if we look at the nodes above the current node we may notice that ten percent Applied Mathematics and Informatics applicants also apply to Mathematics programme, and 70 percent also applied to Software Engineering.

Now let us try to interpret some knowledge unfolded by the diagram in figure~\ref{pm.17}. 70 percent of entrants who applied to Applied Mathematics and Informatics also apply to Software Engineering. The same diagram for Software Engineering states that 80 percent of Software Engineering applicants also apply to Applied Mathematics and Informatics. How this fact can be explained? Firstly it can easily be explained by the fact that these two programmes require to pass the same exams. Therefore there were not any additional obstacles to apply to both programmes simultaneously. Another possible explanation is that it is uneasy for entrants to distinguish these two programmes and successful application to any of them would be satisfactory result.

Analysing diagrams of other programmes' applications we found that equivalence of required exams is probably the most significant reason to apply to more than one programme.

\paragraph{Entrants' "Efficient" choice.}

If an entrant successfully applied to more than one bachelor programme
he or she must select a programme to study. Unlike the previous case, entrants have to select exactly
one programme which gives us more precise information about entrants preferences. For that reason we
define this situation as an efficient choice, efficient in the sense of more expressive about true entrants
preferences.

\begin{figure}[tp]
\centering
\includegraphics[width=\textwidth]{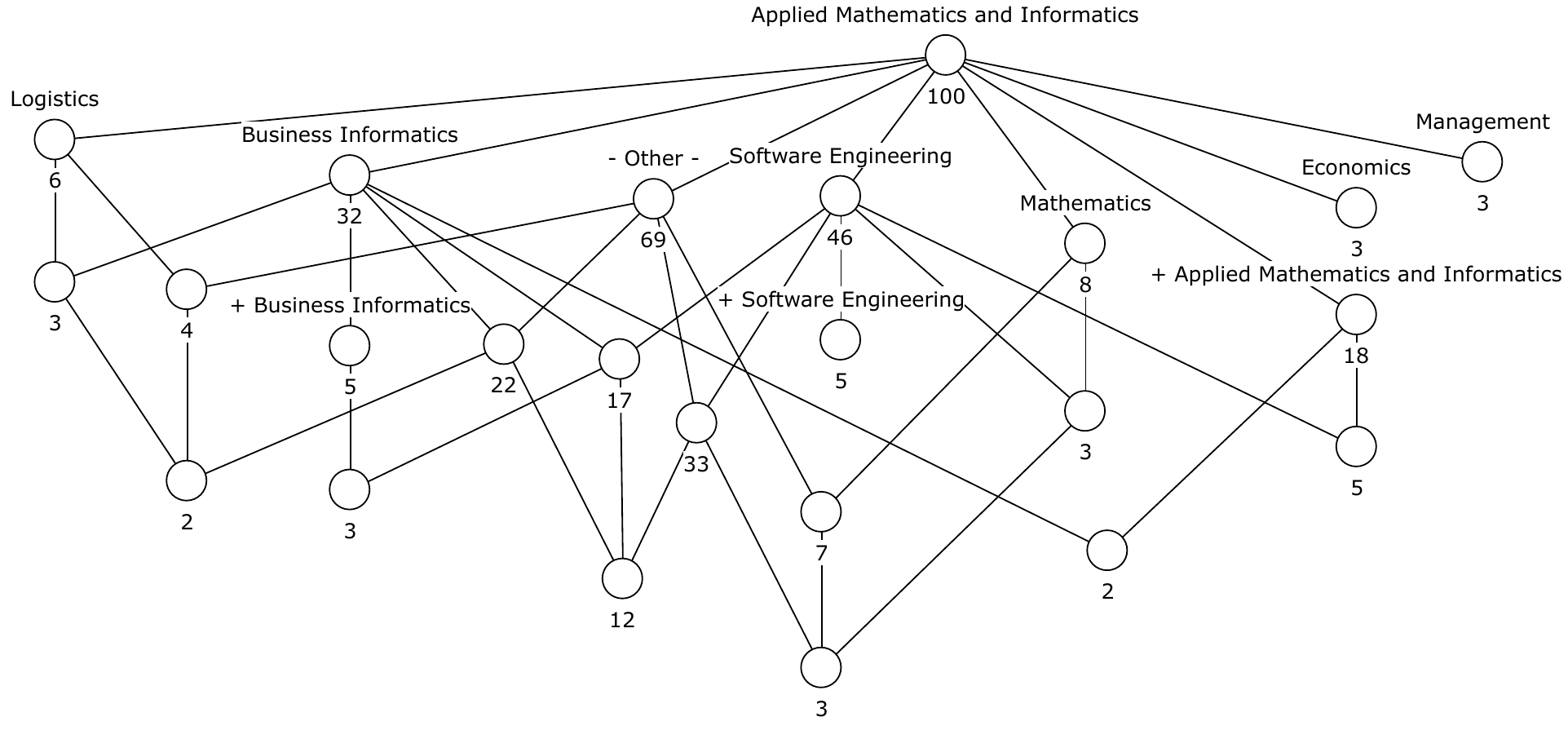}
\caption{``Efficient'' choice of entrants to Applied Mathematics and Informatics programme.}
\label{pm.24}
\end{figure}

 Figure~\ref{pm.24} presents the efficient choice of entrants to Applied Mathematics and Informatics
programme. The meaning of diagram labels is almost the same as in Fig.~\ref{pm.17}. Programmes without plus sign
(+) are successful applications, programmes with preceding plus sign are programmes chosen to study by
entrants. Label ``- Other -'' means that the entrant canceled his application preferring another university or
not to study this year altogether.

Together with diagram in Fig.~\ref{pm.17} 1 this diagram provides us with more precise knowledge about preferences
of entrants to the Applied Mathematics and Informatics programme. More than two thirds of entrants
who successfully apply to the Applied Math programme nevertheless prefer to study at another university.
Whereas just 18 percent of successful applicants then become students on the Applied Mathematics and
Informatics programme. Exactly 5 percent prefer to study Software Engineering and 5 percent of entrants who choose Applied Mathematics and Informatics also successfully applied to Software Engineering.
It can be interpreted as equality of entrants preferences concerning these two programmes. Additionally, 5
percent prefer Business Informatics and only two percent of entrants who prefer Applied Mathematics and
Informatics also successfully apply to Business Informatics, therefore in the pair Business Informatics
and Applied Mathematics and Informatics the latter one is less preferable by entrants.

Here we should note that the sum of nodes percents with labels containing plus sign and node ``- Other -'' must equal to 100\%, however here it does not because we excluded some nodes during filtering.

We built diagrams of ``efficient'' choice for every programme. Analysis of these diagrams helps us to recognise some relations between programmes in terms of entrants preferences. For example, some programmes in most cases is rather backup than actual entrants preference. Some programmes are close to each other by subject of study, these relations are also expressed by diagrams. With help of formalised survey data we found some possible factors of entrants' choice among some particular programmes. These knowledge can help our university to understand entrants' attitude to its undergraduate programmes and thus correct the structure and positioning of them.

Another Educational data mining case includes analysis of student achievements in two subsequent year for the same group by means of grading data~\cite{Ignatov:2011edu}.
	
\subsection{Machine learning exercises with JSM-method in QuDA}

QuDA was developed in early 2000s as ``a software environment for those who want to learn Data Mining by doing'' at the Intellectics group of the Darmstadt Technical University of Technology ~\cite{Grigoriev:2003a,Grigoriev:2003b,Grigoriev:2004}. It
includes various techniques, such as association rule mining, decision trees and rule-based learning, JSM-reasoning (including various reasoning scheme~\cite{Grigoriev:2002}), Bayesian learning, and interesting subgroup discovery. It also provides the experimenter with error estimation and model selection tools as well several preprocessing and postprocessing utilities, including data cleansing tools,
line diagrams, visualisation of attribute distributions, and a convenient rule navigator, etc.
It was mostly aimed to support scientific and teaching activities in the field
of Machine Learning and Data Mining. However, since QuDA has open architecture and support the most common data formats as well as the Predictive Model Markup Language (PMML)\footnote{\url{http://www.dmg.org/}}, it can be easily integrated into a working Data Mining circle.
Originally, it was an acronym for ``Qualitative Data Analysis''. Now, since QuDA finally includes many quantitative methods integrated into it from WEKA\footnote{\url{http://www.cs.waikato.ac.nz/ml/weka/}}, this name is a backronym \footnote{\url{http://en.wikipedia.org/wiki/Backronym}} since it has lost its original meaning.

\begin{exercise} Download QuDa~\footnote{\url{http://sourceforge.net/projects/quda/}; its alternative compilation for RuSSIR 2014 is vaialable at \url{http://bit.ly/QuDA4RuSSIR2014
}}. Refer to QuDa's manual~\cite{Grigoriev:2003a} for details and prepare the credit scoring context in csv format for opening in the QuDA environment. Perform nominal scaling of attributes and apply JSM classifier with basic setup. Compare the obtained rules with the hypotheses obtained manually. $\square$
\end{exercise}

\begin{exercise}  For zoo dataset available with QuDa (or some other dataset that suitable for classification from UCI ML repository\footnote{\url{http://archive.ics.uci.edu/ml/datasets.html}}), perform nominal scaling and comparison of JSM-classification against all available methods  1) by splitting data into 80:20 training-to-test sample size ration 2) by 10-fold cross-validation. Compare learning curves and confusion matrices. Identify all non-covered examples by JSM-method. Change scaling type for attribute ``number of legs''.
Reiterate comparison and check which methods have improved their classification quality. $\square$
\end{exercise}

\section{FCA in Information Retrieval and Text Mining}

Lattice-based models and FCA itself are not mainstream directions of modern IR; they attracted numerous researchers because of their interpretability and human-centerdness, but their intrinsic complexity is a serious challenge to make them working on a Web scale.

Thus, from  early works on Information Retrieval it is known that usage of a lattice as a search space requires treatment of the enormous number of subsets of documents: $10^{310,100}$ for a collection of one million documents \cite{Mooers:1958}. At that time it was rather natural in library classification domain to consider documents and their categories, which may form requests as a combination of simple logical operations like AND, NOT, and OR~\cite{Fairthorne:1956}.
Thus, Mooers considered transformations $T:P \to L$, where $P$ is the space of all possible document descriptors and $L$ is the space of all possible document subsets \cite{Mooers:1958}. Thus, $T$ retrieves the largest set of documents from $L$ according to a query (prescription) from $P$.

At that period in Soviet Russia, All-Soviet Institute for Scientific and Technical Information (VINITI) was organised to facilitate information interchange and fulfill growing scientific needs in cataloging and processing of scientific publications.
Around the mid 1960s, Yulii~A.~Shreider, one of the leading researchers
of VINITI, considered the problem of automatic classification of documents
and their retrieval by means of a model featuring a triple $(M,L,f)$,
where $M$ is a set of documents, $L$ is a set of attributes and $f:M \to 2^L$
maps each document to a set attributes from $L$ \cite{Shreider:1968}. There, similarity of two documents was defined via non-emptiness of the intersection of their descriptions $f(d_1) \cap f(d_2)$.
In that paper, Shreider mentioned the relevance of lattices to problems of document classification and
retrieval, where he also cited the work of Soergel \cite{Soergel:1967} on this issue.

Thus, these two introduced mappings, $T$ and $f$ highly resemble to conventional prime operators in FCA for the context of documents and their attributes (keywords, terms, descriptors) with ``document-term containment'' relation. In the middle of 80s, Godin et al.~\cite{Godin:1986} proposed a lattice-based retrieval model for database browsing, where objects (documents, e.g. course syllabi) were described by associated keywords. The resulting (in fact, concept) lattice used for navigation by query modification using its generality/specificity relation.

In 90-s several FCA-based IR models and systems appeared, the reviews can be found \cite{Carpineto:2005c,Priss:2006}.
Thus, in \cite{Carpineto:2005c}, Carpineto and Romano classified main IR problems that can be solved by FCA means through review of their own studies by the year 2005.
Uta Priss described a current state of FCA for IR domain~\cite{Priss:2006} by the year 2004.
Recently, a survey on FCA-based systems and methods for IR including prospective affordances was presented at FCA for IR workshop at ECIR 2013~\cite{Valverde-Albacete:2013}, and our colleagues, Codocedo and Napoli, taking an inspiration, are summarising the latest work on the topic in a separate forthcoming survey.

Below, we shortly overview our own study on applying FCA-based IR methods for describing the state-of-the-art in FCA for IR field. The rest topics are spread among the most representative examples of FCA-based IR tasks and systems including the summary of the author's experience.
	
	\begin{itemize}
	\item Text Mining scientific papers: a survey on FCA-based IR applications \cite{Poelmans:2012a}

  \item FCA-based meta-search engines (FOOCa, SearchSleuth, Credo etc.) \cite{Koester:2006,Carpineto:2005}
	
	\item FCA-based IR visualisation \cite{Carpineto:2005} and navigation (ImageSleuth, Camelis \cite{Ferre:2007})
	
	\item FCA in criminology: text mining of police reports \cite{Poelmans:2012a}
		
	\item FCA-based approach for advertising keywords in web search \cite{Ignatov:2012a}
			
	\item FCA-based Recommender Systems \cite{Nenova:2013}
	
	\item Triadic FCA for IR-tasks in Folksonomies \cite{Hotho:2006}
		
	\item FCA-based approach for near-duplicate documents detection \cite{Ignatov:2013a,Ignatov:2009}
			
	\item Exploring taxonomies of web site users \cite{Kuznetsov:2007a}
	
	\item Concept-based models in Crowdsourced platforms: a recommender system of like-minded persons, antagonists and ideas \cite{Ignatov:2013w}
		
	\end{itemize}

	\subsection{Text Mining scientific papers: a survey on FCA-based IR applications}
	
In~\cite{Poelmans:2012}, we visually represented the literature on FCA and IR using concept lattices, in which the objects are the scientific papers and the attributes are the relevant terms available in the title, keywords and abstract of the papers. We developed an IR tool with a central FCA component that we use to index the papers with a thesaurus containing terms related to FCA research and to generate the lattices. It helped us to zoom in and give an extensive overview of 103 papers published between 2003 and 2009 on using FCA in information retrieval.

\begin{center}
\begin{cxt}%
\cxtName{}%
\atr{browsing}%
\atr{mining}%
\atr{software}%
\atr{web services}%
\atr{FCA}%
\atr{Information Retrieval}%
\obj{xxx.x.}{$Paper 1$}
\obj{..x.xx}{$Paper 2$}
\obj{.x.xx.}{$Paper 3$}
\obj{x.x.x.}{$Paper 4$}
\obj{...xxx}{$Paper 5$}
\end{cxt}
\end{center}

We developed a knowledge browsing environment CORDIET to support our literature analysis process. One of the central components of our text analysis environment is the thesaurus containing the collection of terms describing the different research topics. The initial thesaurus was constructed based on expert prior knowledge and was incrementally improved by analyzing the concept gaps and anomalies in the resulting lattices. The layered thesaurus contains multiple abstraction levels. The first and finest level of granularity contains the search terms of which most are grouped together based on their semantic meaning to form the term clusters at the second level of granularity.
The papers downloaded from the Web were converted to plain text and the abstract, title and keywords were extracted. The open source tool Lucene\footnote{\url{https://lucene.apache.org/core/}} was used to index the extracted parts of the papers using the thesaurus. The result was a cross table describing the relationships between the papers and the term clusters or research topics from the thesaurus. This cross table was used as a basis to generate the lattices. 

The most relevant scientific sources sources that were used in the search for primary studies contain the work published in those journals, conferences and workshops which are of recognized quality within the research com-munity. These sources are: IEEE Computer Society, ACM Digital Library, Sciencedirect, Springerlink, EBSCOhost, Google Scholar, Conference repositories: ICFCA, ICCS and CLA conferences. Other important sources such as DBLP or CiteSeer were not explicitly included since they were indexed by some of the mentioned sources (e.g. Google Scholar). In the selected sources we used various search terms including ``Formal Concept Analysis'', ``FCA'', ``concept lattices'', ``Information Retrieval''. To identify the major categories for the literature survey we also took into account the number of citations of the FCA papers at CiteseerX.

The efficient retrieval of relevant information is promoted by the FCA representation that makes the inherent logical structure of the information transparent. FCA can be used for multiple purposes in IR \cite{Carpineto:2005,Priss:2006}. First, FCA provides an elegant language for IR modeling and is an interesting instrument for browsing and automatic retrieval through document collections. Second, FCA can also support query refinement, ranking and enrichment by external resources. Because a document-term lattice structures the available information as clusters of related documents which are partially ordered, lattices can be used to make suggestions for query enlargement in cases where too few documents are retrieved and for query refinement in cases where too many documents are retrieved. Third, lattices can be used for querying and navigation supporting relevance feedback.  An initial query corresponds to a start node in a document-term lattice. Users can then navigate to related nodes. Further, queries are used to ``prune'' a document-term lattice to help users focus their search (Carpineto et al. 1996b). For many purposes, some extra facilities are needed such as processing large document collections quickly, allowing  more flexible matching operations, allowing ranked retrieval and give contextual answers to user queries. The past years many FCA researchers have also devoted attention to these issues. 

\begin{figure}
	\centering
		\includegraphics[width=1.0\textwidth]{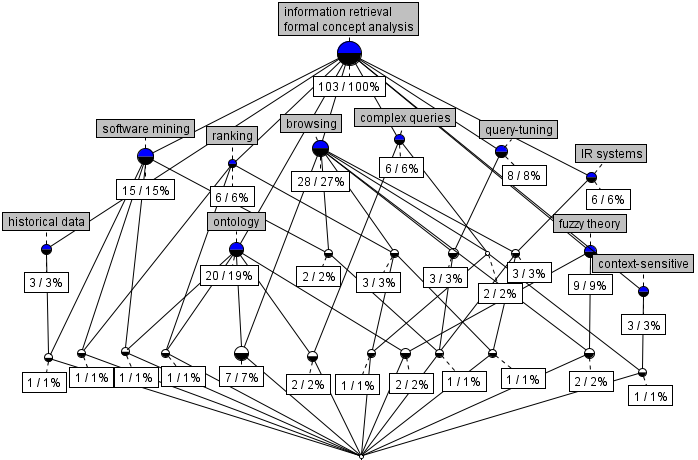}
	\label{fig:IRlattice}
	\caption{Lattice containing 103 papers on using FCA in IR}
\end{figure}

86 \% of the papers on FCA and information retrieval are covered by the research topics in Fig.~\ref{fig:IRlattice}. Further, in our study, we intuitively introduced the process of transforming data repositories into browsable FCA representations and performing query expansion and refinement operations. Then we considered 28 \% of papers on using FCA for representation of and navigation in image, service, web, etc. document collections. Defining and processing complex queries covered 6\% of the papers and was described as well. The review of papers on contextual answers (6\% of papers) and ranking of query results (6\% of papers) concluded the case-study. 

\paragraph{Knowledge representation and browsing with FCA}

In 28\% of the 103 selected papers, FCA is used for browsing and navigation through document collections. In more than half of these papers (18\% of total number of papers), a combination of navigation and querying based on the FCA lattices is pro-posed. Annotation of documents and finding optimal document descriptors play an important role in effective information retrieval (9\% of papers). All FCA-based approaches for information retrieval and browsing through large data repositories are based on the same underlying model. We first have the set $D$ containing objects such as web pages, web services, images or other digitally available items. The set $A$ of attributes can consist of terms, tags, descriptions, etc. These attributes can be related to certain objects through a relation $I\subseteq D \times A$   which indicates the terms, tags, etc. can be used to describe the data elements in $D$. This triple $(D, A, I)$ is a formal context from which the concept lattice can be created.

\paragraph{Query result improvement with FCA}

\begin{figure}[h]
\centering
{\unitlength .5mm
\begin{picture}(60,75)%
\put(0,0){%
\begin{diagram}{60}{75}
\Node{1}{10}{10}  
\Node{2}{10}{50}  
\Node{3}{50}{50}  
\Node{4}{50}{10}
\Node{5}{30}{30}
\Node{6}{70}{35}
\Node{7}{30}{70}
\Node{8}{-10}{35}
\Edge{1}{5}
\Edge{5}{2}
\Edge{5}{3}
\Edge{4}{5}
\Edge{7}{2}
\Edge{7}{3}
\Edge{8}{2}
\Edge{3}{6}
\rightObjbox{4}{10}{2}{$((B_c \cup \{t\})^\prime, (B_c \cup \{t\})^{\prime\prime})$}  
\rightObjbox{5}{10}{2}{$(\{d_8, d_9\},\{t_1, t_2, t_3, t_4, t_5\})$}  
\leftAttbox{2}{2}{2}{$((A_c \cup \{d\})^{\prime\prime}, (A_c \cup \{d\})^\prime)$}  
\end{diagram}}
\put(30,30){\color{darkkhaki}{\circle*{3}}}
\end{picture}}
  \caption{Query modification in a concept lattice: a fish-eye view}
  \label{fig:qtun}
\end{figure}
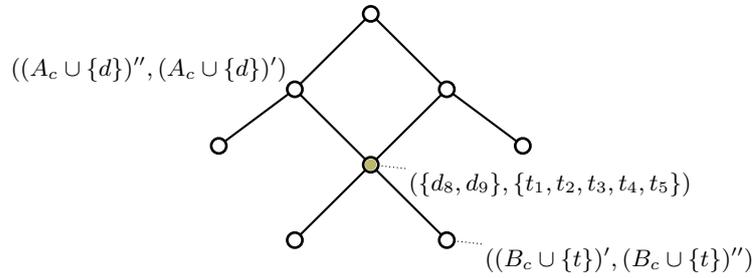

Search engines are increasingly being used by amongst others web users who have an information need. The intent of a concept corresponds to a query and the extent contains the search results. A query $q$ features a set of terms $T$ and the system returns the answer by evaluating  $T^\prime$. Upon evaluating a query $q$ the system places itself on the concept $(T',T'')$ which becomes the current concept $c$. For example, in Fig.~\ref{fig:qtun}, the intent of the current concept  $B_c= \{t_1, t_2, t_3, t_4, t_5\}$ and the extent of the current concept  $A_c = \{d_8, d_9\}$, where $t$ stands for term and $d$ stands for word. Since a query provided by a user only approximates a user's need, many techniques have been developed to expand and refine query terms and search results. Query tuning is the process of searching for the query that best approximates the information need of the user. Query refinements can help the user express his original need more clearly. Query refinement can be done by going to a lower neighbor of the current concept in the lattice by adding a new term to the query items. In \textbf{minimal conjunctive query refinement} the user can navigate for example to a subconcept $((B_c \cup \{t\})^\prime, (B_c \cup \{t\})^{\prime\prime})$ by adding term $t$.

Query enlargement, i.e. retrieving additional relevant web pages, can be performed by navigating to an upper neighbor of the current concept in the lattice by removing a term from the query items. The user can navigate for example to a superconcept $((B_c \cup \{d\})^{\prime\prime}, (B_c \cup \{d\})^\prime)$  by adding document $d$. The combination of subsequent refine and expand operations can be seen as navigation through the query space. Typically, navigation and querying are two completely separate processes, and the combination of both results in a more flexible and user-friendly method. These topics are investigated in 8 \% of the IR papers. See the comprehensive survey on query refinement in~\cite{Carpineto:2012}.

\paragraph{Concept lattice based ranking.}

6\% of IR papers in our study are devoted concept lattice ranking. 

Below we explain the concept lattice based ranking (CLR) proposed in \cite{Carpineto:2000} and compared with hierarchical clustering based (HCR) and best-first matching (BFR) rankings. The experiments with two public benchmark collections showed that CLR outperformed the two competitive methods when the ranked documents did no match the query and was comparable to BMF and better than HCR in the rest cases.

Let $(D,T,I)$ be a document context, where $D$ is the set of documents, $T$ is the set of their terms, and $I \subseteq D \times T$. 
Consider the ordered set of all concepts $(\mathfrak{L}(D,T,I), \succ\prec)$ with the nearest neighbour relation $\succ\prec$, i.e., for $c_1, c_2 \in \mathfrak{L}(D,T,I), c_1 \succ\prec c_2$ iff $c_1 \succ c_2$ or $c_1 \prec c_2$. Then distance between concepts $c_1$ and $c_2$ is defined as the least natural number $n$ as follows: 

$\exists c_{i_1}, \ldots, c_{i_n} \in \mathfrak{L}(D,T,I)$ such that $c_1=c_{i_0} \succ\prec  c_{i_1} \ldots \succ\prec c_{i_n} =c_2.$

For each query $q$ and $d_1,d_2 \in D$, $d_1$ is ranked higher than $d_2$ if the distance between $(d_1^{\prime\prime},d_1^\prime)$ and $(q^{\prime\prime},q^{\prime})$ is shorter than than the distance between $(d_2^{\prime\prime},d_2^\prime)$ and $(q^{\prime\prime},q^{\prime})$. It means that we need less query transformations to obtain $q$ from $d_1$ than to do that from $d_2$.

As the author of CLR admitted in ~\cite{Carpineto:2005}, CLR is sensitive to addition new documents into the search collection. Further analysis of usage of concept neigbourhood based similarity for IR needs is given in~\cite{Eklund:2012}.

\begin{example}

In Fig.~\ref{fig:qrank} we provide an example of concept lattice based ranking for the previous context of papers and their terms. The underlying query is the conjunction of two terms: ``browsing, FCA''. The query concept is as follows: 
$$(\{p_1,p_4\}, \{browsing,FCA,software\}).$$
The resulting ranking yields $p_4<p_1<p_2<p_3=p_5$. A curious reader may admit that concepts with the same ranks lie in concentric circles around the query concept at the same distance. Obviously, for concepts from the same such circle we need their subsequent ranking, e.g. by best-match ranking via dot product of the document and query profiles based on term frequency.  

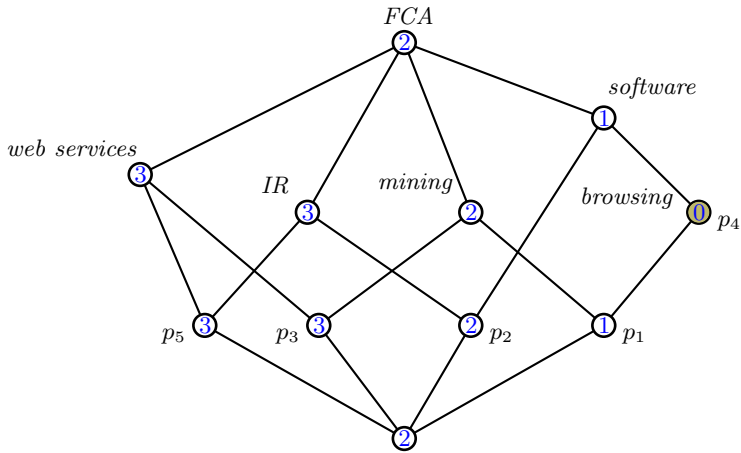
\begin{figure}[h]
\centering
{\unitlength .25mm
\begin{picture}(330,230)%
\put(0,0){%
\begin{diagram}{330}{230}
\Node{1}{170.0}{220.0}
\Node{2}{170.0}{10.0}
\Node{3}{31.0}{150.0}
\Node{4}{65.0}{70.0}
\Node{5}{125.0}{70.0}
\Node{6}{119.0}{130.0}
\Node{7}{205.0}{70.0}
\Node{8}{275.0}{180.0}
\Node{9}{325.0}{130.0}
\Node{10}{275.0}{70.0}
\Node{11}{205.0}{130.0}
\Edge{4}{6}
\Edge{10}{11}
\Edge{9}{8}
\Edge{2}{7}
\Edge{5}{11}
\Edge{3}{1}
\Edge{10}{9}
\Edge{2}{5}
\Edge{8}{1}
\Edge{2}{10}
\Edge{5}{3}
\Edge{4}{3}
\Edge{7}{6}
\Edge{2}{4}
\Edge{7}{8}
\Edge{11}{1}
\Edge{6}{1}
\NoDots
\CircleSize{12}
\leftObjbox{4}{10}{2}{$p_5$}  
\leftObjbox{5}{10}{2}{$p_3$}  
\rightObjbox{7}{10}{2}{$p_2$} 
\rightObjbox{9}{10}{2}{$p_4$} 
\rightObjbox{10}{10}{2}{$p_1$} 
\centerAttbox{1}{2}{10}{FCA}  
\leftAttbox{3}{2}{10}{web services}  
\leftAttbox{6}{10}{10}{IR}  
\rightAttbox{8}{2}{10}{software}  
\leftAttbox{9}{14}{2}{browsing}  
\leftAttbox{11}{10}{10}{mining}
\end{diagram}}
\put(325.0,130.0){\color{darkkhaki}{\circle*{11}}}
\put(167.0,216.0){\color{blue}2}
\put(167.0,6.0){\color{blue}2}
\put(28.0,146.0){\color{blue}3}
\put(62.0,66.0){\color{blue}3}
\put(122.0,66.0){\color{blue}3}
\put(116.0,126.0){\color{blue}3}
\put(202.0,66.0){\color{blue}2}
\put(272.0,176.0){\color{blue}1}
\put(322.0,126.0){\color{blue}0}
\put(272.0,66.0){\color{blue}1}
\put(202.0,126.0){\color{blue}2}
\end{picture}}
  \caption{Concept lattice based ranking for the query ``browsing, FCA''; the distance values are given inside the corresponding circles.}
  \label{fig:qrank}
\end{figure}
$\square$
\end{example}

An interested reader can find the rest sections in our survey:
\bi
\item Web and email retrieval (partially covered in Section~\ref{ssec:searcheng});
\item {Image, software and knowledge base retrieval (partially covered in Section~\ref{ssec:imsearch});}
\item {Defining and processing complex queries with FCA;}
\item {Domain knowledge in search results: contextual answers \& ranking.}
\ei

\subsection{FCA-based IR visualisation and meta-search engines}\label{ssec:searcheng}

From the beginning 2000s many independent IR developers proposed so called meta-search engines also known as search results clustering engines.
To name a few, two project are still alive Carrots$^2$\footnote{\url{http://search.carrot2.org/}} and Nigma.ru\footnote{\url{http://www.nigma.ru/}}. See summarising survey on Web clustered search by Carpineto et al. in~\cite{Carpineto:2009w}.

FCA has been used as the basis for many web-based knowledge browsing systems developed during the past years. Especially its comprehensible visualisation capabilities seem to be of interest to RuSSIR audience. 
The results returned by web search engines for a given query are typically formatted as a list of URLs accompanied by a document title and a a snippet, i.e. a short summary of the document. Several FCA-based systems were developed for analyzing and exploring these search results. CREDO \cite{Carpineto:2004}, FooCA \cite{Koester:2006} and SearchSleuth~\cite{Ducrou:2007ss,Dau:2008} build a context for each individual query which contains the result of the query as objects and the terms found in the title and summary of each result as attributes.

\begin{figure}
	\centering
		\includegraphics[width=1.0\textwidth]{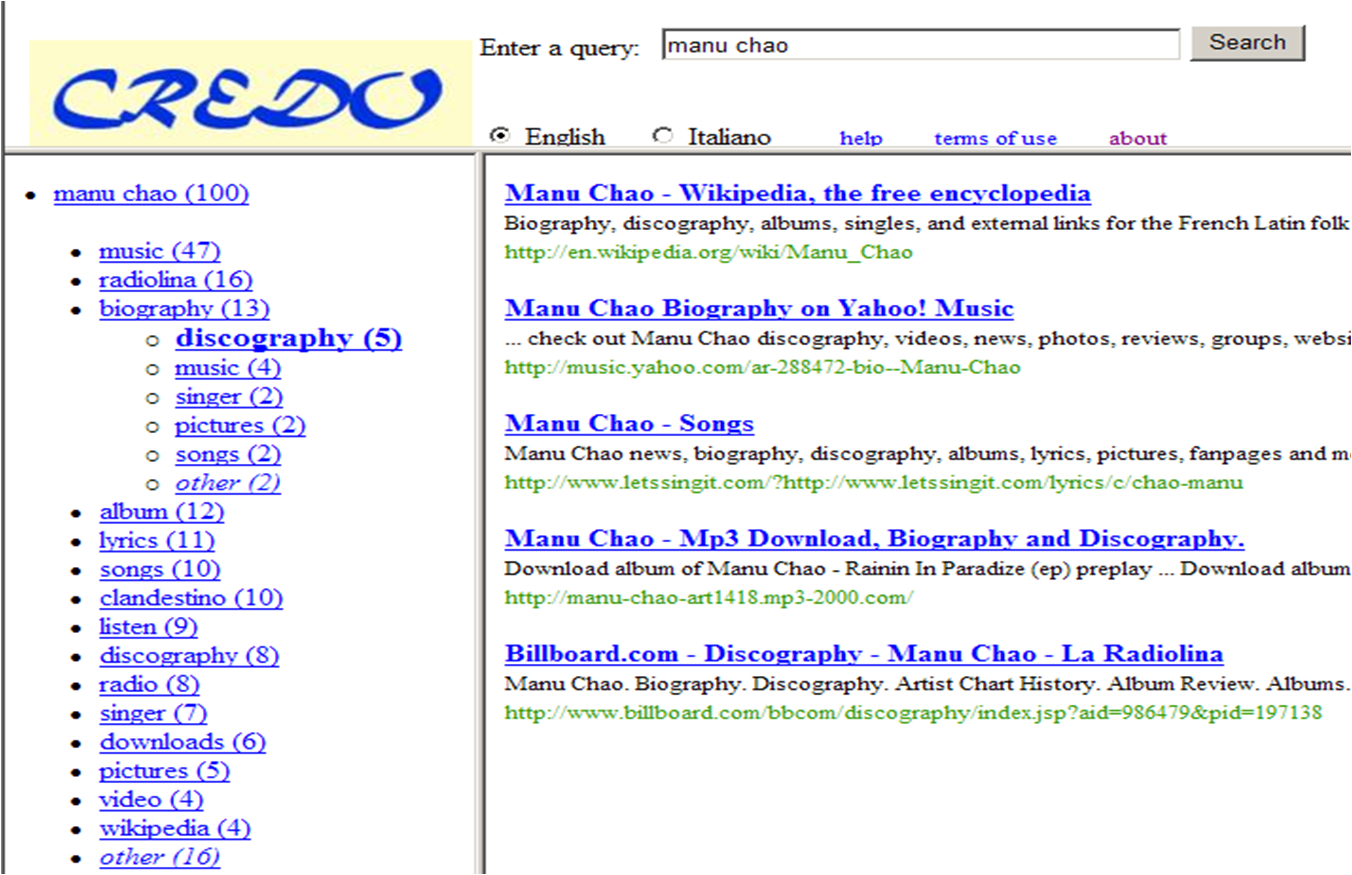}
	\caption{An example of CREDO's web search interface}
	\label{fig:credo}
\end{figure}

The CREDO system\footnote{\url{http://credo.fub.it/}} then builds an iceberg lattice which is represented as a tree and can be interactively explored by the user. 

\begin{figure}
	\centering
		\includegraphics[width=1.0\textwidth]{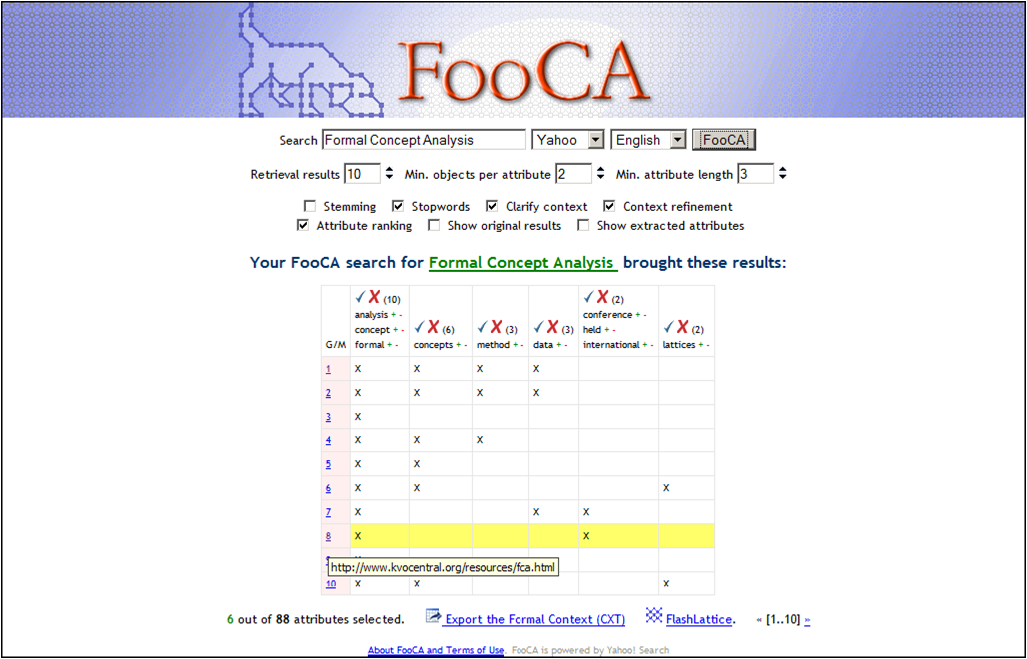}
	\caption{An example of FooCA's web search interface. It processes the results of search queries to Yahoo or Google and organises them into the interactive cross-table.}
	\label{fig:fooca}
\end{figure}

FooCA\footnote{\url{http://www.bjoern-koester.de/}} shows the entire formal context to the user and offers a great degree of flexibility in exploring this table using the ranking of attributes, selecting the number of objects and attributes, applying stemming and stop word removal etc. 

\begin{figure}
	\centering
		\includegraphics[width=1.0\textwidth]{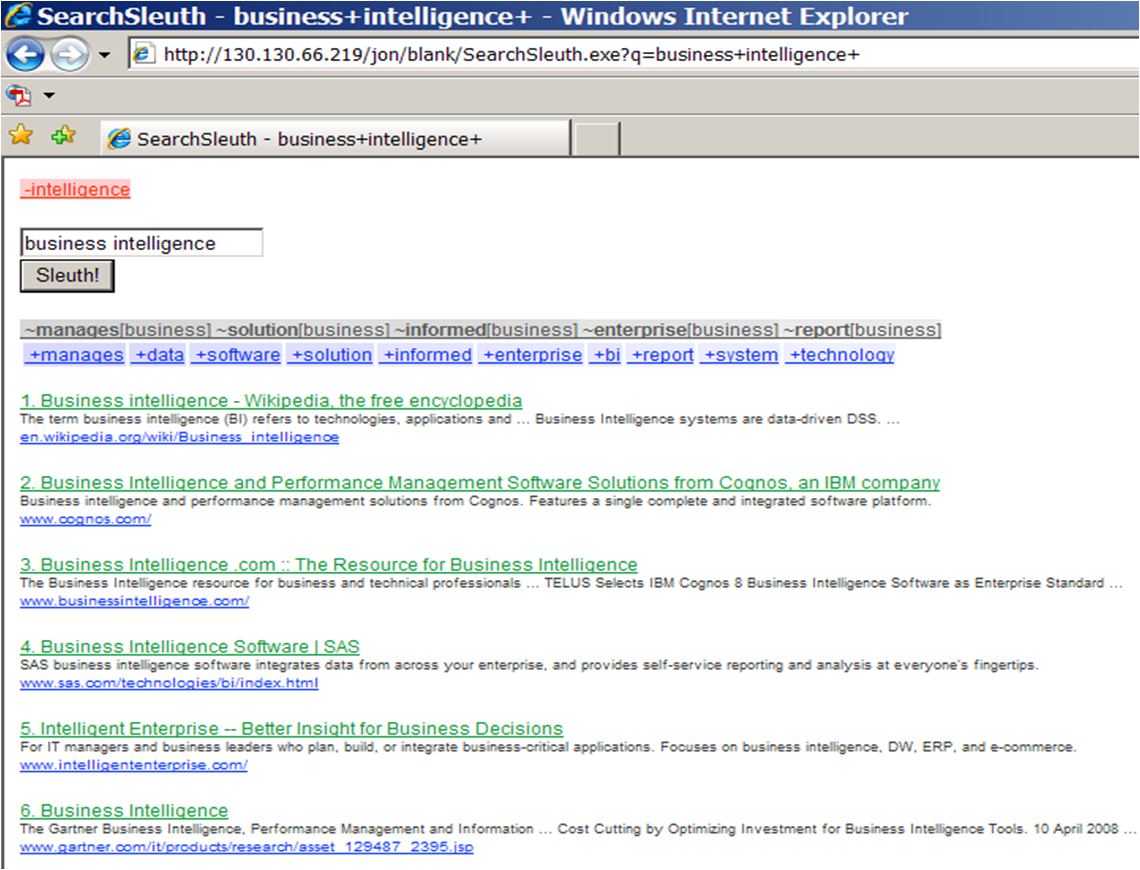}
	\caption{An example of SearchSleuth's web interface. It processes the results of search queries to Yahoo. Passing to more general (more specific) categories is done by clicking -term(+term).}
	\label{fig:ssleuth}
\end{figure}

SearchSleuth does not display the entire lattice but focuses on the search concept, i.e. the concept derived from the query terms. The user can easily navigate to its upper and lower neighbors and siblings. Nauer et al.\cite{Nauer:2009} also propose to use FCA for iteratively and interactively analyzing web search results. The user can indicate which concepts are relevant and which ones are not for the retrieval task. Based on this information the concept lattice is dynamically modified. Their research resulted in the CreChainDo system\footnote{\url{http://orpailleur.loria.fr/index.php/CreChainDo}}. Kim et al. \cite{Kim:2004} presented the FCA-based document navigation system KAnavigator for small web communities in specialized domains. Relevant documents can be annotated with keywords by the users. Kim et al. \cite{Kim:2006} extended the search functionality by combining lattice-based browsing with conceptual scales to reduce the complexity of the visualisation. Cigarran et al. \cite{Cigarran:2004} present the JBrainDead IR System which combines free-text search with FCA to organise the results of a query. 
Cole et al. \cite{Cole:2003} discuss a document discovery tool named Conceptual Email Manager (CEM) which is based on FCA. The program allows users to navigate through emails using a visual lattice. The paper also discusses how conceptual ontologies can support traditional document retrieval systems and aid knowledge discovery in document collections. The development of this software is based on earlier research on retrieval of information from semi-structured texts (\cite{Cole:2001,Eklund:2000}). Building further on this work is the Mail-Sleuth software (Eklund et al. \cite{Eklund:2004}) which can be used to mine large email archives. Eklund et al. \cite{Eklund:2005} use FCA for displaying, searching and navigating through help content in a help system. 
Stojanovic \cite{Stojanovic:2005} presents an FCA-based method for query refinement that provides a user with the queries that are ``nearby'' the given query. Their approach for query space navigation was validated in the context of searching medical abstracts.  Stojanovic \cite{Stojanovic:2005} presents the SMART system  for navigation through an online product catalog. The products in the database are described by elements of an ontology and visualized with a lattice, in which users can navigate from a very general product-attribute cluster containing a lot of products to very specific clusters that seem to contain a few, but for the user highly relevant products. Spyratos et al. \cite{Spyratos:2006} describe an approach for query tuning that integrates navigation and querying into a single process. The FCA lattice serves for navigation and the attributes for query formulation. Le Grand et al.~\cite{Grand:2006} present an IR method based on FCA in conjunction with semantics to provide contextual answers to web queries. An overall lattice is built from tourism web pages. Then, users formulate their query and the best-matching concepts are returned, users may then navigate within the lattice by generalizing or refining their query. Eklund et al.~\cite{Eklund:2008} present AnnotationSleuth to extend a standard search and browsing interface to feature a conceptual neighborhood centered on a formal concept derived from curatorial tags in a museum management system.
Cigarran et al.~\cite{Cigarran:2005} focus on the automatic selection of noun phrases as documents descriptors to build an FCA based IR system. Automatic attribute selection is important when using FCA in a free text document retrieval framework. Optimal attributes as document descriptors should produce smaller, clearer and more browsable concept lattices with better clustering features. Recio-Garcia et al.~\cite{Recio-Garcia:2006} use FCA to perform semantic annotation of web pages with domain ontologies. Similarity matching techniques from Case Based Reasoning can be applied to retrieve these annotated pages as cases. Liu et al.~\cite{Liu:2007} use FCA to optimise a personal news search engine to help users obtain the news content they need rapidly. The proposed technique combines the construction of user background using FCA, the optimisation of query keywords based on the user's background and a new layout strategy of search results based on a ``Concept Tree''. Lungley et al.~\cite{Lungley:2009} use implicit user feedback for adapting the underlying domain model of an intranet search system. FCA is used as an interactive interface to identify query refinement terms which help achieve better document descriptions and more browsable lattices.

\subsection{FCA-based image retrieval and navigation}\label{ssec:imsearch}
	
	FCA-based IR visualisation \cite{Carpineto:2005} and navigation (ImageSleuth, Camelis \cite{Ferre:2007})

Ahmad et al.~\cite{Ahamd:2003} build concept lattices from descriptions associated to images for searching and retrieving relevant images from a database. In the ImageSleuth project \cite{Ducrou:2006}, FCA was also used for clustering of and navigation through annotated collections of images. The lattice diagram is not directly shown to the user. Only the extent of the present concept containing thumbnails, the intent containing image descriptions and a list of upper and lower neighbors is shown. In Ducrou~\cite{Ducrou:2007}, the author built an information space from the Amazon.com online store and used FCA to discover conceptually similar DVDs and explore their conceptual neighborhood. The system was called DVDSleuth. Amato et al.~\cite{Amato:2008} start from an initial image given by the user and use a concept lattice for retrieving similar images. The attributes in this lattice are facets, i.e. an image similarity criterion based on e.g. texture, color or shape. The values in the context indicate for each facet how similar an image in the database is with respect to the user provided initial image. By querying, the user can jump to any cluster of the lattice by specifying the criteria that the sought cluster must satisfy. By navigation from any cluster, the user can move to a neighbor cluster, thus exploiting the ordering amongst clusters. 	

In \cite{Ferre:2007} Ferre et al. proposed to use so called Logical Information Systems (LIS) for navigation through photo collections\footnote{Camelis, \url{http://www.irisa.fr/LIS/ferre/camelis/}}. In fact LIS, similarly to Pattern Structures, exploit partially ordered object descriptions but expressed as logical formulas.

\bi
\item location: Nizhniy Novgorod $\sqsubseteq$ Russia 
\item date: date = 18 Aug 2014 $\sqsubseteq$ date in Aug 2014 .. Jul 2015
\item event: event is ``summer school RuSSIR'' $\sqsubseteq$ event contains ``summer school''
\ei

 Further it was extended for work with document collections\cite{Ferre:2009}. Since Camelis uses lattice-based navigation and queriying by formulas, it overcomes current drawbacks of tree-like navigation imposed by current restrictions of file-systems.

Recently, the previous studies of Eklund et al.~\cite{Eklund:2012} in organising navigation through annotated collections of images in virtual museums resulted in an iPad application that allows users to explore an art collection via semantically linked pathways that are generated using Formal Concept Analysis\footnote{``A place for art'', \url{https://itunes.apple.com/au/app/a-place-for-art/id638054832?mt=8}}. In fact navigation in this application is organised by showing context and relationships among objects in a museum collection.

\subsection{FCA in criminology: text mining of police reports}

 In \cite{Poelmans:2012a}, we proposed an iterative and human-centred knowledge discovery methodology based on FCA. The proposed approach recognises the important role of the domain expert in mining real-world enterprise applications and makes use of specific domain knowledge, including human intelligence and domain-specific constraints. Our approach was empirically validated at the Amsterdam-Amstelland police to identify suspects and victims of human trafficking in 266,157 suspicious activity reports. Based on guidelines of the Attorney Generals of the Netherlands, we first defined multiple early warning indicators that were used to index the police reports. 

\begin{example} This is an example of a police report where some indicator-words are highlighted that are used for its contextual representation.

\begin{minipage}[h]{0.50\linewidth}
\noindent\textbf{Report 1:}

\texttt{On the night of 23 of March 2008 we stopped a car with a \emph{Bulgarian} license plate for routine motor vehicle inspection. It was a \emph{Mercedes GLK} with license plate BL XXX. The car was driving around in circles in a \emph{prostitution area}.  On the \emph{backseat of the car} we noticed two well dressed young girls. We asked for their identification papers but they did not speak English nor Dutch. The driver of the car was \emph{in possession of their papers} and told us that they were \emph{on vacation} in the Netherlands for two weeks etc. 
}\end{minipage}
\hfill  
\begin{minipage}[h]{0.50\linewidth}  
    \begin{center}
\begin{cxt}%
\cxtName{}%
\atr{Expensive cars}%
\atr{Prostitutes}%
\atr{Id-papers}%
\atr{Vacation}%
\atr{Former Eastern Europe}%
\atr{Information Retrieval}%
\obj{xxxxxx}{$Report 1$}
\obj{xxxx.x}{$Report 2$}
\obj{xxx..x}{$Report 3$}
\obj{.x..x.}{$Report 4$}
\obj{x....x}{$Report 5$}
\end{cxt}
\end{center}
 \end{minipage}
 $\square$
\end{example}

Our method based on FCA consists of four main types of analysis which are carried out as
follows:

\be
\item  Concept exploration of the forced prostitution problem of Amsterdam: 
In Poelmans et al.~\cite{Poelmans:2011d}, this FCA-based
approach for automatically detecting domestic violence in unstructured text police
reports is described in detail. 
\item Identifying potential suspects: concept lattices allow for the detection of potentially
interesting links between independent observations made by different police
officers. 
\item Visual suspect profiling: some FCA-based methods such as temporal concept
analysis (TCA) were developed to visually represent and analyse data with a
temporal dimension \cite{Wolff:2005}. Temporal concept lattices were used in Elzinga et al.~\cite{Elzinga:2010} to create visual profiles of
potentially interesting terrorism subjects. Elzinga et al.~\cite{Elzinga:2012} used TCA in combination with nested line diagrams to analyse
pedophile chat conversations. 
\item Social structure exploration: concept lattices may help expose interesting persons
related to each other, criminal networks, the role of certain suspects in these
networks, etc. With police officers we discussed and compared various FCA-based
visualisation methods of criminal networks. 
\ee

In our investigations we also used the model that was developed by Bullens and Van Horn~\cite{Bullens:2010} for the identification of loverboys who typically force girls of Dutch nationality into prostitution.
Loverboys use their love affair with a woman to force her to work in prostitution. Forcing girls and women into prostitution through a loverboy approach is seen as a special kind of human trafficking in the Netherlands (article 250a of the code of criminal law). This model
is a resource used by the Amsterdam-Amstelland police during the trainings of police
officers about this topic. A typical loverboy approach consists of three main phases which give rise to corresponding indicators:

\be
\item Preparatory activities to recruit girls.
\item Forcing her into prostitution.
\item Keeping the girl in prostitution by emotional dependence or social isolation.
\ee

The pimp will also try to protect his organisation.

In our data-set, there were three reports available about girl H. The
reports about this girl led us to the discovery of loverboy suspect B. 
The first report (26 November 2008) contains the notification of the police by a youth
aid organisation in Alkmaar about girl H. They report a suspicious tattoo on her wrist
containing the name ‘B’. This ‘B’ refers to her boyfriend who carries the same first name, is 30 years old, and is of Surinamian origin.
The second report was written by a police officer who works in the red light district and knows many women working in brothels or behind the windows. During a patrol he saw H working as a prostitute, had a conversation with her, and observed suspicious that his included in the report. The next report contains four suspicious facts recorded by the officer. First, an unbelievable story why she works as a prostitute: a bet between girlfriends if someone would dare to work as a prostitute. Second, the tattoos of which one tattoo is mentioned in the first report (‘B’) and a new one on her belly. Third, the injuries, she has scratches on her arm (possibly from a fight) and burns on her leg.

\begin{figure}
	\centering
		\includegraphics[width=0.85\textwidth]{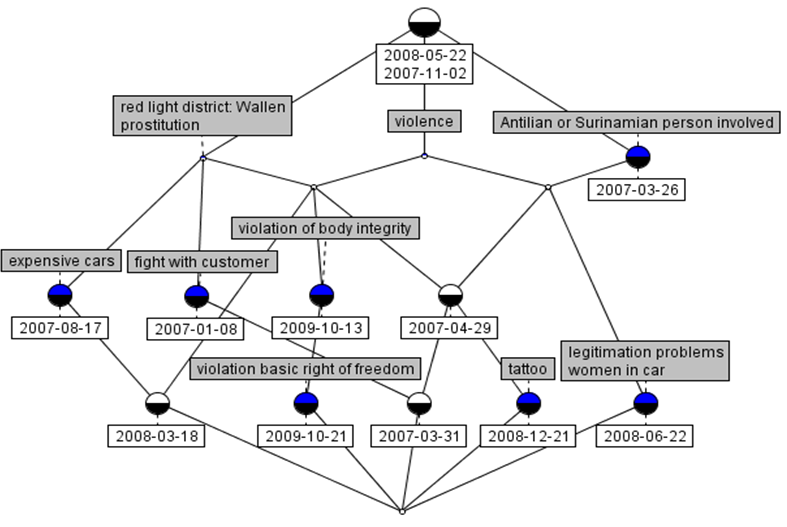}
	\caption{Line diagram for the report context of loverboy suspect B}
	\label{fig:loverboy}
\end{figure}

According to the victim, she has dropped a hot iron on her leg and had an accident with a gourmet set. Fourth is the observation of making long working days. The third document (21 December 2008) showed an observation of the victim walking with the
possible suspect. In this document the police officer reports that he saw the victim and a man walking close to each other. The police officer knows the man and knows that he is active in the world of prostitution. When the man saw the officer, he immediately took some distance of the victim. As soon as they have passed the officer, they walk close together and into a well-known street where prostitutes work behind the windows. The first name of the person is B, the same name which is tattooed on the victim’s wrist, and the description of the person is about the same as described by the youth aid organisation. This information signals that the man is the possible loverboy of the victim. The three reports together give serious presumptions of B being a loverboy with H being the victim. The
next step is investigating B. We need serious indications that B is really involved in forced prostitution. Twelve observational reports were found for B and the resulting lattice is shown in Figure~\ref{fig:loverboy}.

Investigating these reports shows that he frequently visits the red light district and has strong relationships with other pimps. One of these pimps is the suspect of another loverboy case. From the six observations where B was seen in the red light district, four are violence related, including the observation of H’s suspicious burn wounds. The other violence-related observations are situations of fights with customers who are unwilling to leave or pay. Such violence-related observations are related to pimps who want to protect their prostitutes from customers and competing gangs. Since in the Netherlands, prostitution is legal, each prostitute has the right to ask the police to protect her. The violence observations of the suspect strengthened the suspicion of B being the pimp of
H. Moreover, we found another girl R who was also a potential victim of him. These
indications were enough to create a summary report and send a request for using further
investigation techniques to the public prosecutor.

Using concept lattices, we revealed numerous unknown human trafficking and loverboy suspects. Indepth investigation by the police resulted in a confirmation of their involvement in illegal activities resulting in actual arrestments been made. This approach was embedded into operational policing practice and is now successfully
used on a daily basis to cope with the vastly growing amount of unstructured
information.

There are other FCA-based studies in criminology, for example, terrorist activity modeling and analysis~\cite{Koester:2009} and developments of lattice-based access policies for information systems~\cite{Obiedkov:2009,Dau:2009}.

\subsection{FCA-based approach for advertising keywords in web search}
	
Online advertising by keywords matching is bread and butter of modern web search companies like Google and Yandex. For our experimentation we used data of US Overture~\cite{2004:Zhukov:SC} (now, a part of Yahoo), which were first transformed in the standard context form. We consider the following context:  $\K_{FT}=(F,T,I_{FT})$, where $F$ is the set of advertising firms (companies),  $T$ is the set of advertising terms, or phrases, $fI_{FT}t$ means that firm $f \in F$ bought advertising term $t \in T$.
 In the context $|F| =  2000$, $|T| = 3000$, $|I_{FT}| =  92345$.

The data are typically sparse, thus the number of attributes per object is bounded as  follows: $13 \leq |g'| \leq 947$. For objects per attribute we have $18 \leq |m'| \leq 159$.
From this context we computed formal concepts of the form (advertisers, bids) that represent market sectors. Formal concepts of this form can be further used for recommendation to the companies on the market, which did not buy bids contained in the intent of the concept. 

This can also be represented as association rules of the form ``If an advertiser bought bid $a$, then this advertiser may buy term $b$''
See~\cite{2000:Badrul:an_rec} for the use of association rules in recommendation systems.

To make recommendations we used the following steps:

\be
\item D-miner algorithm for detecting large market sectors as concepts and our biclustering algorithm;
\item Coron system for constructing association rules;
\item Construction of association metarules using morphological analysis;
\item Construction of association metarules using ontologies (thematic catalogs).
\ee

\paragraph{Detecting  large market sectors with D-miner and OA-biclustering.}
   The D-miner algorithm~\cite{2005:Besson:CBBSM} constructs the set of concepts satisfying given constraints on sizes of extents and intents (i.e. intersection of icebergs and dual icebergs).  D-miner takes as input a context and two parameters: minimal admissible extent and intent sizes and outputs  a ``band'' of the concept lattice: all concepts satisfying constraints given by parameter values ($|intent| \geq m$ and $|extent| \geq n$, where $m, n \in \mathbb{N}$, see table \ref{table:D-miner_res}).

\begin{table}[h]
\begin{center}
\vskip 0.15in
\begin{footnotesize}
{\renewcommand{\tabcolsep}{3pt}%
\caption{ D-miner results.}\label{table:D-miner_res}
\vskip .1in
  \begin{tabular}{|c|c|c|}
  \hline
    Minimal extent   &	Minimal intent & Number of  \\
    size & size & concepts\\
  \hline
0 &	0	&  8 950 740\\
10	& 10	 &  3 030 335\\
15	& 10   &  759 963\\
15	& 15   &  150 983\\
15	& 20   & 	14 226\\
20	& 15   & 	661\\
20    & 16   & 53\\
20	& 20   & 	0\\
  \hline
\end{tabular}}
\end{footnotesize}
\end{center}
\end{table}

\begin{example}

We provide examples of two intents of formal concepts for the case $|L|=53$, where $|L|$ is a number of formal concepts obtained by D-miner.

\textbf{Hotel market.}

{  \footnotesize
  $\{$ angeles hotel los, atlanta hotel, baltimore hotel, dallas hotel, denver hotel, hotel chicago, diego hotel san, francisco hotel san, hotel houston, hotel miami, hotel new orleans, hotel new york, hotel orlando, hotel philadelphia, hotel seattle, hotel vancouver$\}$}

\indent \textbf{Weight loss drug market.}

$\{$ adipex buy, adipex online, adipex order, adipex prescription, buy didrex, buy ionamin, ionamin purchase, buy phentermine, didrex online, ionamin online, ionamin order, online order phentermine, online phentermine, order phentermine, phentermine prescription, phentermine purchase $\}$
$\square$
\end{example}

Applying the biclustering algorithm to our data we obtained 87 OA-biclusters ($\rho=0.85$), which is much less than the number of concepts found by D-miner. Expert interpretation of these biclusters implies that each market described by formal concepts found by D-miner (where each market can be represented by several formal concepts) corresponds to a bicluster among these 87. The number of formal concepts generated by D-miner becomes feasible for human interpretation if there are no more than 20 firms and about 15 terms. For these thresholds D-miner could find only large markets and ignored important average-size markets. For our data these ignored markets were, e.g. car and flower markets, which were found using biclustering approach.

\begin{example} Flower market OA-bicluster.

{\footnotesize
$(\{$ 24, 130, 170, 260, 344, 415, 530, 614, 616, 867, 926, 1017, 1153, 1160, 1220, 1361, 1410, 1538, 1756, 1893 $\}, \{$'anniversary flower', 'arrangement flower', 'birthday flower', 'bouquet flower', 'buy flower', 'buy flower online', 'delivery flower', 'flower fresh', 'flower gift', 'flower line', 'flower online', 'flower online order', 'flower online send', 'flower online shop', 'flower rose', 'flower send', 'flower shop', 'flower sympathy', 'red rose'$\})$, with $\rho\approx0.84$
}$\square$
\end{example}

\paragraph{Recommendations based on association rules.}
Using the Coron system (see~\cite{2005:Szathmary:CORON}) we construct the informative basis of association rules~\cite{2007:Szathmary:ZART}.

\begin{example}
Here are some examples of association rules:
\bi
\item $\{e vitamin\} \rightarrow \{c vitamin\}$, supp=31 [1.55\%] and conf=0.86;
\item $\{gift \ graduation\} \rightarrow \{anniversary \ gift\}$, supp=41 [2.05\%] and conf=0.82.
\ei
$\square$
\end{example}

The value  $supp=31$ of the first rule means that 31 companies bought phrases ``e vitamin'' and ``c vitamin''. The value $conf=0.861$ means that 86,1\% companies that bought the phrase ``e vitamin'' also bought the phrase ``c vitamin''.

To make recommendations for each particular company one may use an approach proposed in~\cite{2000:Badrul:an_rec}. For company $f$ we find all association rules, the antecedent of which contain all the phrases bought by the company, then we construct the set  $T_u$ of unique advertising phrases not bought by the company  $f$  before. Then we order these phrases by decreasing of confidence of the rules where the phrases occur in the consequences. If buying a phrase is predicted by multiple rules, we take the  largest confidence.


\paragraph{Morphology-based Metarules}

Each attribute of our context is either a word or a phrase. Obviously, synonymous phrases are related to same market sectors. The advertisers companies have usually thematic catalogs composed by experts, however due to the huge number of advertising terms manual composition of catalogs is a difficult task. Here we propose a morphological approach for detecting similar terms.

Let $t$ be an advertising phrase consisting of several words (here we disregard the word sequence): $t=\{w_1, w_2, \ldots, w_n\}$.  A stem is the root or roots of a word, together with any derivational affixes, to which inflectional affixes are added \cite{1991:Crystal:dict_ling}. The stem of word $w_i$ is denoted by $s_i=stem(w_i)$ and the set of stems of words of the phrase $t$ is denoted by $stem(t)=\bigcup\limits_i stem(w_i)$, where $w_i \in t$. Consider the formal context $\K_{TS}=(T, S, I_{TS})$, where $T$ is the set of all phrases and   $S$ is the set of all stems of phrases from  $T$, i.e. $S=\bigcup\limits_i stem(t_i)$. Then $tIs$ denotes that the set of stems of phrase $t$ contains $s$.

In this context we construct rules of the form $t \rightarrow s_i^{I_{TS}}$ for all $t \in T$, where $(.)^{I_{ts}}$ denotes the prime operator in the context $K_{TS}$. Then the a morphology-based metarules of the context  $\K_{TS}$ (we call it a metarule, because it is not based on experimental data, but on implicit knowledge resided in natural language constructions) corresponds to  $t \xrightarrow[]{FT} s_i^{I_{TS}}$, an association rule of the context $\K_{FT}=(F, T, I_{FT})$. If the values of support and confidence of this rule in context $\K_{FT}$ do not exceed certain thresholds, then the association rules constructed from the context $\K_{FT}$ are considered not very interesting.

\begin{example} An example of an input context for morphological association rules.

\begin{table}[h]
\begin{center}
\vskip 0.15in
\begin{footnotesize}
{\renewcommand{\tabcolsep}{3pt}%
\caption{A toy example of context $\K_{FT}$ for ``long distance calling'' market.}\label{table:ex_K_FT}
\vskip .1in
  \begin{tabular}{|c||c|c|c|c|c|}
  \hline
  firm $\setminus$ phrase & call & calling & calling & carrier & cheap \\
  & distance  & distance & distance & distance & distance\\
    & long & long &  long plan & long & long\\
  \hline
	  \hline
$f_1$  & x & & x & & x \\
$f_2$	&	 & x & x & x & \\
$f_3$	&	 & & & x & x	\\
$f_4$	&	 & x & x & & x	\\
$f_5$	&	x &x & & x & x	\\
  \hline
\end{tabular}}
\end{footnotesize}
\end{center}
\end{table}

\begin{table}[h]
\begin{center}
\vskip 0.15in
\begin{footnotesize}
{\renewcommand{\tabcolsep}{3pt}%
\caption{A toy example of context $\K_{TS}$ for ``long distance calling'' market.}\label{table:ex_K_TS}
\vskip .1in
  \begin{tabular}{|c||c|c|c|c|c|c|}
  \hline
  phrase $\setminus$ stem  & call  & carrier & cheap & distanc & long & plan \\
  \hline
	  \hline
call distance long  & x & &  & x & x & \\
calling distance long	& x & &  & x & x & \\
calling distance long plan & x & &  & x & x & x \\
carrier distance long	&	 & x &  & x & x &	\\
cheap distance long	&	 & & x & x & x &	\\
  \hline
\end{tabular}}
\end{footnotesize}
\end{center}
\end{table}
$\square$
\end{example}

Metarules of the following forms seem also to be reasonable. First, one can look for rules of the form  $t \xrightarrow[]{FT} \bigcup\limits_i s_i^{I_{TS}}$, i.e., rules, the consequent of which contain all terms containing at least one word with the stem common to a word in the antecedent term. Obviously, constructing rules of this type may result in the fusion of phrases related to different market sectors, e.g. ``black jack''  and ``black coat''.
Second, we considered rules of the form  $t \xrightarrow[]{FT} (\bigcup\limits_i s_i)^{I_{TS}}$, i.e., rules with the consequent with the set of  stems being the same as the set of stems of the antecedent.
Third, we also propose to consider metarules of the form $t_1 \xrightarrow[]{FT} t_2$, where $t_2^{I_{TS}} \subseteq t_1^{I_{TS}}$. These are rules with the consequent being sets of  stems that contain the set of stems of the antecedent.

\begin{example} An example of metarules.

\bi
\item   $t \xrightarrow[]{FT} s_i^{I_{TS}}$

 $\{last \ minute \ vacation\} \rightarrow \{last \ minute \ travel\}$\\
 	supp=	19	conf=	0,90

\item   $t \xrightarrow[]{FT} \bigcup\limits_i s_i^{I_{TS}}$

$\{mail \ order \ phentermine\} \rightarrow \{adipex \ online \ order,  \ldots,$

 $ phentermine \ purchase, phentermine \ sale\}$\\
 	supp=	19 \quad conf=	0,95

\item $t \xrightarrow[]{FT} (\bigcup\limits_i s_i)^{I_{TS}}$

        $\{distance \ long \ phone\}\rightarrow \{call \ distance \ long \ phone, \ldots,$

$distance \ long \ phone \ rate, distance \ long \ phone \ service\}$\\
 	supp=	37 \quad conf=	0,88

\item $t_1 \xrightarrow[]{FT} t_2$,   $t_2^{I_{TS}} \subseteq t_1^{I_{TS}}$

 $\{ink \ jet\}\rightarrow\{ink\}$, 	supp=	14	\quad conf=	0,7

\ei
$\square$
\end{example}

\paragraph{Experimental Validation}

For validation of association rules and metarules we used an adapted version of cross-validation. The training set was randomly divided into 10 parts, 9 of which were taken as the training set and the remaining part was used as a test set. 
The confidence of rules averaged over the test set is almost the same as the $min\_conf$ for the training set, i.e.,  $(0.9-0.87)/0.9\approx0.03$.

Note that the use of morphology is completely automated and allows one to find highly plausible metarules without data on purchases. The rules with low support and confidence may be tested against recommendation systems such as Google AdWords, which uses the frequency of queries for synonyms. Thus 90\% of recommendations (words) for ontological rules (see \cite{Ignatov:2012a}) were contained in the list of synonyms output by AdWords.

\subsection{FCA-based recommender systems}

Motivated by prospective applications of Boolean Matrix Factorisation (BMF) in the context of Recommender Systems (RS) we proposed an FCA-based approach which follows user-based k-nearest neighbours strategy \cite{Ignatov:2014}. Another approach similar to MF is biclustering, which has also been successfully applied in recommender system domain \cite{Symeonidis:2008,Ignatov:2012a}. As we have mentioned, FCA can be also used as a biclustering technique and there are several examples of its applications in the recommender systems domain \cite{duBoucherRyan:2006,Ignatov:2008}. A parameter-free approach that exploits a neighbourhood of the object concept for a particular user also proved its effectiveness \cite{Alqadah:2014}.

Belowe we discuss our recent studies in application of BMF for RS. In the recommender systems domain, the context is any auxiliary information concerning users (like gender, age, occupation, living place) and/or items (like genre of a movie, book or music), which shows not only a user's mark given to an item but explicitly or implicitly describes the circumstances of such evaluation (e.g., including time and place) \cite{Adomavicius:2005}.

From representational viewpoint an auxiliary information can be described by a binary relation, which shows that a user or an item posses a certain attribute-value pair.

As a result one may obtain a block matrix:
\[
	I =
		\begin{bmatrix}
			R 			& C_{user} \\
			C_{item}	& O
		\end{bmatrix},
\]
where $R$ is a utility matrix of users' ratings to items, $C_{user}$ represents context information of users, $C_{item}$ contains context iformation of items and $O$ is a zero-filled matrix.

\begin{example} An example of a rating matrix enriched by user-feature and item-feature auxiliary information.

\begin{table}[!htb]\footnotesize
\begin{center}
\caption{Adding auxiliary information}\label{tab:context}
\begin{tabular}{|p{13mm}||p{10mm}|p{10mm}|p{10mm}|p{11mm}|p{10mm}|p{10mm}||p{4mm}|p{4mm}|p{7mm}|p{8mm}|p{7mm}|}
	\hline
		&
		\multicolumn{6}{c||} {Movies} &
		\multicolumn{2}{c|} {Gender }&
		\multicolumn{3}{c|} {Age} \\
	\cline{2-12}
		&
		Brave Heart&
		Termi-nator	&
		Gladi-ator	&
		Slum-dog Million-aire&
		Hot Snow &
		God-father	&
		M	&
		F	&
		0-20	&
		21-45	&
		46+	\\
	\hline
	\hline
		Anna 	&
			5	&		&	5	&	5	&		&	2	&
				&	+	&
			+	&		&		\\
	\hline
		Vladimir&
				&	5	&	5	&	3	&		&	5	&
			+	&		&
				&	+	&		\\
	\hline
		Katja&
			4	&		&	4	&	5	&		&	4	&
				&	+	&
				&	+	&		\\
	\hline
		Mikhail	&
			3	&	5	&	5	&		&		&	5	&
			+	&		&
				&	+	&		\\
	\hline
		Nikolay	&
				&		&	2	&		&	5	&	4	&
			+	&		&
				&		&	+	\\
	\hline
		Olga	&
			5	&	3	&	4	&	5	&		&		&
				&	+	&
			+	&		&		\\
	\hline
		Petr	&
			5	&		&		&	4	&	5	&	4	&
			+	&		&
				&		&	+	\\
	\hline
	\hline
		Drama	&
			+	&		&	+	&	+	&	+	&	+	&
				&		&
				&		&		\\
	\hline
		Action	&
				&	+	&	+	&		&	+	&	+	&
				&		&
				&		&		\\
	\hline
		Comedy	&
			+	&		&		&	+	&		&		&
				&		&
				&		&		\\
	\hline		
\end{tabular}

\end{center}
\end{table}
$\square$
\end{example}

\begin{example}

In case of more complex rating's scale the ratings can be reduced to binary scale (e.g., ``\textit{like/dislike}'') by binary thresholding or by FCA-based scaling.

\begin{table}[!htb]\footnotesize
\caption{Derived Boolean utility matrix enriched by auxiliary information}\label{tab:context_matrix}
\begin{center}
\begin{tabular}{r|cccccc|ccccc}
	\rowcolor{Gray}
			&	$m_1$	&	$m_2$	&	$m_3$	&	$m_4$	&	$m_5$	&	$m_6$	&	$f_1$	&	$f_2$	&	$f_3$	&	$f_4$	&	$f_5$	\\
	\hline
		\cellcolor{Gray}
		$u_1$	&
				1	&	0	&	1	&	1	&	0	&	0	&
																0	&	1	&	1	&	0	&	0	\\
		\cellcolor{Gray}
		$u_2$	&
				1	&	0	&	1	&	1	&	0	&	0	&
																1	&	0	&	0	&	1	&	0	\\
		\cellcolor{Gray}
		$u_3$	&
				1	&	0	&	1	&	1	&	0	&	1	&
																0	&	1	&	0	&	1	&	0	\\
		\cellcolor{Gray}
		$u_4$	&
				1	&	0	&	1	&	1	&	0	&	0	&
																1	&	0	&	0	&	1	&	0	\\
		\cellcolor{Gray}
		$u_5$	&
				0	&	0	&	0	&	0	&	1	&	1	&
																1	&	0	&	0	&	0	&	1	\\
		\cellcolor{Gray}
		$u_6$	&
				1	&	0	&	1	&	1	&	0	&	0	&
																0	&	1	&	1	&	0	&	0	\\
		\cellcolor{Gray}
		$u_7$	&
				1	&	0	&	0	&	1	&	1	&	1	&
																1	&	0	&	0	&	0	&	1	\\
	\hline
		\cellcolor{Gray}
		$g_1$	&
				1	&	0	&	1	&	1	&	1	&	1	&
																0	&	0	&	0	&	0	&	0	\\
		\cellcolor{Gray}
		$g_2$	&
				0	&	1	&	1	&	0	&	1	&	1	&
																0	&	0	&	0	&	0	&	0	\\
		\cellcolor{Gray}
		$g_3$	&
				1	&	0	&	0	&	1	&	0	&	0	&
																0	&	0	&	0	&	0	&	0	\\
\end{tabular}

\end{center}
\end{table}
$\square$
\end{example}

Once a matrix of ratings is factorized we need to learn how to compute recommendations for users and to evaluate whether a particular method handles this task well.

Given the factorized matrices already well-known algorithm based on the similarity of users can be applied, where for finding  $ k $ nearest neighbors we use not the original matrix of ratings $ R \in \mathbb {R} ^ {m \times n} $, but the matrix $ I \in \mathbb {R} ^ {m \times f} $, where $ m $ is a number of users, and $ f $ is a number of factors. After the selection of $k$  users, which are the  most similar to a given user, based on  the factors that are peculiar to them, it is possible, based on collaborative filtering formulas to calculate the prospective ratings for a given user.

After generation of recommendations the performance of the recommender system can be estimated by such measures as Mean Absolute Error (MAE), Precision and Recall.

Collaborative recommender systems try to predict the utility (in our case ratings) of items for a particular user based on the items previously rated by other users.

Memory-based algorithms make rating predictions based on the entire collection of previously rated items by the users. That is, the value of the unknown rating $r_{u,m}$ for a user $u$ and item $m$ is usually computed as an aggregate of the ratings of some other (usually, the $k$ most similar) users for the same item $m$:
$$r_{u,m}=aggr_{\tilde{u}\in \tilde{U}}r_{\tilde{u},m},$$

where $\tilde{U}$ denotes a set of $k$ users that are the most similar to user $u$, who have rated item $m$. For example, the function $aggr$ may have the following form \cite{Adomavicius:2005}:

$$r_{u,m}= \frac{\sum\limits_{\tilde{u}\in \tilde{U}} sim(\tilde{u},u)\cdot r_{\tilde{u},m},\label{two}}{\sum\limits_{\tilde{u}\in \tilde{U}} sim(u,\tilde{u})}.$$

Similarity measure between users $u$ and $\tilde{u}$, $sim(\tilde{u},u)$, is essentially an inverse distance measure and is used as a weight, i.e., the more similar users $c$ and $\tilde{u}$ are, the more weight rating $r_{\tilde{u},m}$  will carry in the prediction of $r_{\tilde{u},m}.$

Similarity between two users is based on their ratings of items that both users have rated. There are several popular approaches:  Pearson correlation, cosine-based, and Hamming-based similarities.

We mainly use the cosine-based and normalised Hamming-based similarities.

To apply this approach in case of FCA-based BMF recommender algorithm we simply consider the user-factor matrices obtained after factorisation of the initial data as an input.

\begin{example}

For the input matrix in Table~\ref{tab:context_matrix} one can find the following covering factors:

	\begin{center}
		\begin{tabular}{ll}
			
	$	( \{u_1, u_3, u_6, u_7, g_1, g_2\}, 	\{m_1, m_4\} 			)$, & 	$	( \{u_2, u_4\}, 				\{m_2, m_3, m_6, f_1, f_4\}	)$,\\
	$	( \{u_5, u_7\}, 				\{m_5, m_6, f_1, f_5\}		)$, & 	$	( \{u_1, u_6\}, 				\{m_1, m_3, m_4, f_2, f_3\}	)$,\\
	$	( \{u_5, u_7, g_1, g_3\}, 		\{m_5, m_6\}			)$, &	$	( \{u_2, u_3, u_4\}, 			\{m_3, m_6, f_4\}		)$,\\
	$	( \{u_2, u_4, g_3\}, 			\{m_2, m_3, m_6\}			)$, &	$	( \{u_1, u_3, u_6, g_1\}, 		\{m_1, m_3, m_4\}			)$,\\
	$	( \{u_1, u_3, u_6\}, 			\{m_1, m_3, m_4, f_2\}		)$.& \\

		\end{tabular}
\end{center}

The corresponding decomposition is below:

\[
\left( \begin{array}{ccccccccc}
		1&0&0&1&0&0&0&1&1\\
		0&1&0&0&0&1&1&0&0\\
		1&0&0&0&0&1&0&1&1\\
		0&1&0&0&0&1&1&0&0\\
		0&0&1&0&1&0&0&0&0\\
		1&0&0&1&0&0&0&1&1\\
		1&0&1&0&1&0&0&0&0\\
		1&0&0&0&1&0&0&1&0\\
		0&0&0&0&1&0&1&0&0\\
		1&0&0&0&0&0&0&0&0
	\end{array}
	\right)
	\circ
	\left(\begin{array}{ccccccccccc}
		1&0&0&1&0&0&0&0&0&0&0\\
		0&1&1&0&0&1&1&0&0&1&0\\
		0&0&0&0&1&1&1&0&0&0&1\\
		1&0&1&1&0&0&0&1&1&0&0\\
		0&0&0&0&1&1&0&0&0&0&0\\
		0&0&1&0&0&1&0&0&0&1&0\\
		0&1&1&0&0&1&0&0&0&0&0\\
		1&0&1&1&0&0&0&0&0&0&0\\
		1&0&1&1&0&0&0&1&0&0&0
	\end{array}
	\right)
\]
$\square$
\end{example}

However, in this case in the obtained user profiles vectors most of the components are getting zeros, and thus we lose similarity information.

To smooth the loss effects we proposed the following weighted projection:
\[
	\tilde{P}_{uf} 	= \frac{I_{u\cdot}\cdot Q_{f\cdot}}{||Q_{f\cdot}||_1}
						= \frac{\sum\limits_{v \in V} I_{uv}\cdot Q_{fv}}{\sum\limits_{v \in V} Q_{fv}},
\]
where $\tilde{P_{uf}}$~ indicates whether factor~$f$ covers user~$u$,
	$I_{u\cdot}$~is a binary vector describing profile of user~$u$,
	$Q_{f\cdot}$~is a binary vector of items belonging to factor~$f$ (the corresponding row of $Q$ in decomposition eq.~\eqref{def:bmf}).
The coordinates of the obtained projection vector lie within~$[0; 1]$.

\begin{example}

For Table~\ref{tab:context} the weighted projection is as follows:
\[
	\tilde{P} =
	\left(\begin{array}{ccccccccc}
		1	&	\frac{1}{5}	&	0			&	1			&	0			&	\frac{1}{3}	&	\frac{1}{3}	&	1			&	1			\\
		0	&	1			&	\frac{1}{2}	&	\frac{1}{5}	&	\frac{1}{2}	&	1			&	1			&	\frac{1}{3}	&	\frac{1}{4}	\\
		1	&	\frac{3}{5}	&	\frac{1}{4}	&	\frac{4}{5}	&	\frac{1}{2}	&	1			&	\frac{2}{3}	&	1			&	1			\\
		0	&	1			&	\frac{1}{2}	&	\frac{1}{5}	&	\frac{1}{2}	&	1			&	1			&	\frac{1}{3}	&	\frac{1}{4}	\\
		0	&	\frac{2}{5}	&	1			&	0			&	1			&	\frac{2}{3}	&	\frac{1}{3}	&	0			&	0			\\
		1	&	\frac{1}{5}	&	0			&	1			&	0			&	\frac{1}{3}	&	\frac{1}{3}	&	1			&	1			\\
		1	&	\frac{2}{5}	&	1			&	\frac{1}{5}	&	1			&	\frac{1}{3}	&	\frac{1}{3}	&	\frac{2}{3}	&	\frac{1}{2}	\\
		1	&	\frac{2}{5}	&	\frac{1}{2}	&	\frac{2}{5}	&	1			&	\frac{2}{3}	&	\frac{2}{3}	&	1			&	\frac{3}{4}	\\
		0	&	\frac{2}{5}	&	\frac{1}{2}	&	\frac{1}{5}	&	1			&	\frac{2}{3}	&	1			&	\frac{1}{3}	&	\frac{1}{4}	\\
		1	&	0			&	0			&	\frac{1}{5}	&	0			&	0			&	0			&	\frac{2}{3}	&	\frac{1}{2}		\end{array}\right)
.\]
$\square$
\end{example}

The proposed approach and compared ones have been implemented in C++\footnote{\url{https://github.com/MaratAkhmatnurov/BMFCARS}} and evaluated on MovieLens-100k data set.
This data set features 100000 ratings in five-star scale, 1682 Movies, Contextual information about movies (19 genres), 943 users (each user has rated at least 20 movies), and demographic info for the users (gender, age, occupation, zip (ignored)).
The users have been divided into seven age groups: under 18, 18-25, 26-35, 36-45, 45-49, 50-55,	56+.

Five star ratings are converted to binary scale by the following rule:
\[
	I_{ij} =
	\begin{cases}
		1 , & R_{ij} > 3 ,\\
		0 , & \text{else} 
	\end{cases}
\]

The scaled dataset is split into two sets according to bimodal cross-validation scheme \cite{Ignatov:2012cv}: the training set and the test set with a ratio 80:20, and 20\% of ratings in the test set are hidden.

\begin{figure}[!htb]\label{fig:vsSVD}
\begin{center}
	\includegraphics[width=\textwidth]{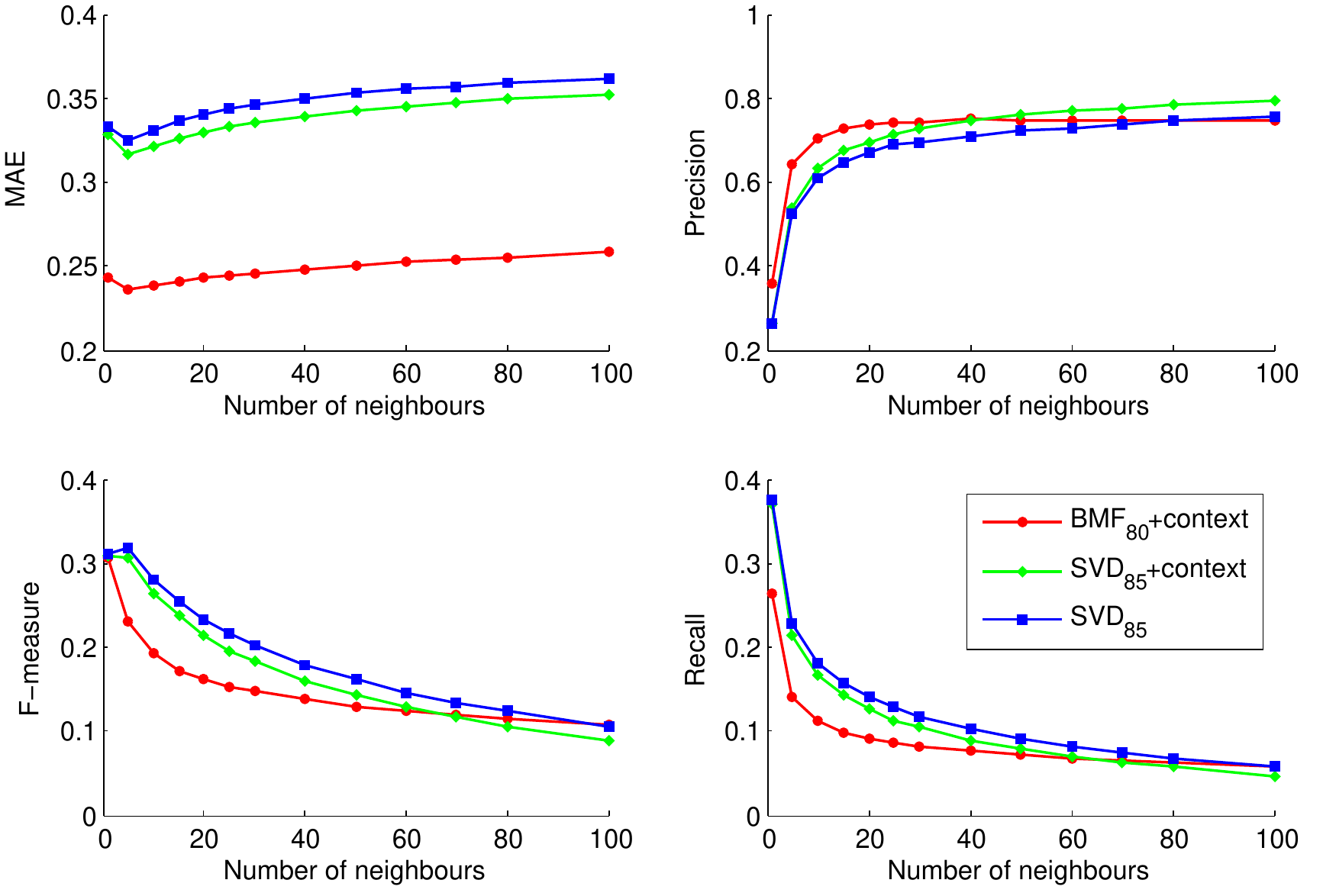}
\end{center}
\caption{Different approaches of matrix factorisation}
\end{figure}	

We found out that MAE of our BMF-based approach is significantly lower than MAE of SVD-based approach for almost the same number of factor at fixed coverage level of BMF and $p$-level of SVD.  The Precision of BMF-based approach is slightly lower when the number of neighbours is about a couple of dozens and comparable for the remaining part of the observed range. The Recall is lower that results in lower F-measure.  It can be explained by different nature of factors in these factorisation models. The proposed weighted projection alleviates the information loss of original Boolean projection resulting in a substantial quality gain. 
We also revealed that the presence of auxiliary information results in a small quality increase (about 1-2\%) in terms of MAE, Recall and Precision.
In our previous study, with the original BMF-based scheme (weighting is not used),  we obtained comparable results in terms of MAE, and both Precision and Recall~\cite{Ignatov:2014,Nenova:2013}.

\subsection{FCA-based approach for near-duplicate documents detection}

From the dawn of web search engines, the problem of finding near-duplicate documents in the web search results is crucial for providing users with relevant documents \cite{Brin:1995,Broder:1997,Ilyinsky:2002}.

Below we shortly describe our studies on near duplicate detection~\cite{Ignatov:2009} within a competition ``Internet mathematics'' organised by Yandex and ROMIP (Russian Information Retrieval Evaluation Seminar) in 2004--2005: our project ``Optimisation of search for near duplicates in the web: images and similarity'' was selected, as the rest 33 projects, out of 252 applications \footnote{\url{https://academy.yandex.ru/events/imat/grant2005/}}.
 
As experimental data the ROMIP collection of web documents from narod.ru domain\footnote{\url{http://romip.ru/en/collections/narod.html}} was provided; it consists of 52 files of general size 4.04 GB. These files contained 530 000 web pages from narod.ru domain. Each document from the collection has size greater or equal to 10 words.
For experiments the collection was partitioned into several parts consisting of three to 24 such files
(from 5\% to 50\% percent of the whole collection). As an evaluation benchmark for recall and precision calculation we use the list of duplicate pairs provided by Yandex; the duplicate pairs were identified for all document pairs via Perl String::Similarity with 85\% similarity threshold.

For composing document images we followed a popular shingling approach \cite{Broder:1997}.
For each text, the program \textbf{shingle} with two parameters (\emph{length} and \emph{offset}) generate contiguous subsequences of size \emph{length} such that the distance between the beginnings of two subsequent substrings is \emph{offset}.  The set of sequences obtained in this way is hashed so that each sequence receives its own hash code. From the set of hash codes that corresponds to the document a fixed size (given by parameter) subset is chosen by means of random permutations described in~\cite{Broder:1997,Broder:1998,Broder:2000}. The probability of the fact that minimal elements in permutations on hash code sets of shingles of documents $A$ and $B$ (these sets are denoted by $F_A$ and $F_B$, respectively) coincide, equals to the similarity measure of these documents sim($A,B$):
$$sim(A, B)=P[min\{\pi(F_A)\}=min\{\pi(F_B)\}]=\frac{|F_A\cap F_B|}{|F_A\cup F_B|}.$$

Further we used FCA to define cluster of near duplicate documents.

Let $\K_{DF}=(D,F,I\subseteq D\times F)$ be a \textbf{context of documents}, where $D$ is a set of documents, $F$ is a set of hash codes (fingerprints), and  $I$ shows that a document $d$ has an attribute $f$ whenever $dIf$.

 For a subset of documents $A \subseteq D$, $A'$ describe their similarity in terms of common fingerprints, and the closed set $A''$ is a \textbf{cluster of similar documents}.
  
To find all near duplicate clusters, we need to enumerate all intents of the context $\K_{FD}=(F,D,I\subseteq F\times D)$ such that their common set of fingerprints exceeds a threshold set by user.

In fact, to this end we need to use nothing but frequent itemsets of documents. A set of documents $A \subseteq D$ is called \textbf{$k$-frequent} if $|A'| > k$, where $k$ is a parameter.

\paragraph{Program Implementation}

Software for experiments with syntactical representation comprise the units that perform the following operations:

\bi
\item[1.] XML Parser (provided by Yandex): it parses XML packed collections of web documents

\item[2.] Removing html-markup of the documents

\item[3.] Generating shingles with given parameters length-of-shingle, offset

\item[4.] Hashing shingles

\item[5.] Composition of document image by selecting subsets (of hash codes) of shingles by means of methods {\it $n$ minimal
elements in a permutation} and {\it minimal elements in $n$ permutations}.

\item[6.] Composition of the inverted table the list of identifiers of documents shingle thus preparing data to the format of programs for computing closed itemsets.

\item[7.] Computation of clusters of {\it $k$-similar documents} with FPmax* algorithm: the output consists of strings, where the first elements are names (ids) of documents and the last element is the number of common shingles for these documents.

\item[8.] Comparing results with the existing list of duplicates (in our experiments with the ROMIP collection of web documents, we were supplied by a precomputed list of duplicate pairs).
\ei

This unit outputs five values: 1) the number of duplicate pairs in the ROMIP collection, 2) the number of duplicate pairs for our realisation, 3) the number of unique duplicate pairs in the ROMIP collection, 4) the number of unique duplicate pairs in our results, 5) the number of common pairs for the ROMIP collection and our results.

In step 7, we used a leader in time efficiency, the algorithm FPmax*~\cite{Grahne:2003}, from the competition organised in series of workshops on Frequent  Itemset Mining Implementations (FIMI)\footnote{\url{http://fimi.ua.ac.be/}}. 

\paragraph{Experimental results}

In our experiments we used Cluto \footnote{\url{http://glaros.dtc.umn.edu/gkhome/views/cluto}}, a software package for clustering high-dimensional datasets including those from information retrieval domain, for comparison purposes. We chose the repeated-bisecting algorithm that uses the cosine similarity function with a 10-way partitioning (ClusterRB), which is mostly scalable according to its author~\cite{Karypis:2003}. The number of clusters was a parameter, documents were given by sets of attributes, fingerprints in our case. The algorithm outputs a set of disjoint clusters. Algorithms from FIMI repository can process very large datasets, however, to compare with Cluto (which is much more time consuming as we show below) we took collection narod.1.xml that contains 6941 documents.

Shingling parameters used in experiments were as follows:
the number of words in shingles was 10 and 20, the offset was always taken to be 1
(which means that the initial set of shingles contained all possible contiguous word sequences of a given length).
The sizes of resulting document images were taken in the interval 100 to 200 shingles.
As frequency thresholds defining {\it frequent closed sets}
(i.e., the numbers of common shingles in document images from one cluster) we experimentally
studied different values in intervals, where the maximal value is equal to the number of shingles
in the document image. For example, the interval [85, 100] for document images with 100 shingles,
the interval [135, 150] for document images of size 150, etc.
Obviously, choosing the maximal value of an interval, we obtain clusters where document images coincide completely.

For parameters taking values in these intervals we studied the relation between resulting clusters of duplicates and ROMIP collection of duplicates, which consists of  pairs of web documents that are considered to be near duplicates. Similarity of each pair of documents in this list is based on Edit Distance measure, two documents were taken to be duplicates by authors of this testbed if the value of the Edit Distance measure exceeds threshold 0.85. As we show below, this definition of a duplicate is prone to errors, however making a testbed by manual marking duplication in a large web document collection is hardly feasible. Unfortunately, standard lists of near-duplicates were missing at that period even for standard corpora such as TREC or Reuters collection~\cite{Potthast:2007}. For validating their methods, researchers create ad-hoc lists of duplicates using slightly transformed documents from standard collections. Now the situation is drastically better, see, for example, workshop series on Plagiarism Analysis,
Authorship Identification, and Near-Duplicate Detection (PAN)\footnote{\url{http://www.uni-weimar.de/medien/webis/events/pan-15/pan15-web/plagiarism-detection.html}}.

In our study for each such pair we found an intent that contains both elements of the pair, and vice versa,
for each cluster of {\it very similar documents}
(i.e., for each corresponding closed set of documents with more than $k$ common description units)
we take each pair of documents in the cluster and looked for the corresponding pair in the ROMIP collection.
As result we obtain the number of common number of near duplicate pairs found by our method
 and those in the ROMIP collection, and the number of unique pairs of HSE duplicates
(document pairs occurring in a cluster of ``very similar documents" and not occurring in the ROMIP collection).
The results of our experiments showed that the ROMIP collection of duplicates, considered to be a benchmark,
is far from being perfect.
First, we detected that a large number of false duplicate pairs in this collection due to similar framing of documents.
For example the pages with the following information in table \ref{tbl:histpers} about historical personalities 1 and 2 were declared to be near duplicates.

\begin{table}[!h]
\centering
\caption{Information about historical personalities}\label{tbl:histpers}
\begin{tabular}{|cc|c|}
\hline

\begin{tabular}{l}

{\bf 1. Garibald II, Duke of Bavaria }\\

Short information:\\

Full Name: Garibald\\

Date of birth: unknown\\

Place of birth: unknown\\

Date of death: 610\\

Place of death: unknown\\

Father: Tassilo I Duke of Bavaria\\

\\

Mother: uknown \\
\end{tabular} & \quad &
\begin{tabular}{l}
2. {\bf Giovanni, Duke of Milan}\\

Short information:\\

Full Name: Giovanni Visconti\\

Date of birth: unknown\\

Place of birth: unknown\\

Date of death: 1354\\

Place of death: unknown\\

Father: Visconti Matteo I,\\
 the Great Lord of Milan\\

Mother: uknown\\
\end{tabular}
\\
\hline
\end{tabular}
\\
\end{table}

However these pages, as well as many other analogous false duplicate pairs in ROMIP collection do not belong to concept-based (maximal frequent) clusters generated in our approach.

Second, in our study we also looked for {\it false duplicate clusters} in the ROMIP collection, caused by transitive closure of the binary relation ``$X$ is a duplicate of $Y$" (as in the typical definition of a document cluster in~\cite{Broder:2000}). Since  the similarity relation is generally not transitive, the clusters formed by transitive closure of the relation may contain absolutely dissimilar documents. Note that if clusters are defined via maximal frequent itemsets (subsets of attributes) there cannot be effects like this, because documents in these clusters share necessarily large itemsets (common subsets of attributes).

We analysed about 10000 duplicate document pairs and found four rather big \emph{false duplicate clusters} about 50-60 documents each.
Further studies on this collection see in~\cite{Zelenkov:2007}.

We shortly summarise experimental results below:

\begin{itemize}
\item FPmax* (F-measure=0.61 and elapsed time 0.6 seconds), ClusterRB (F-measure=0.63 and elapsed time 4 hours);
\item For FPMax* the number of single document cluster is 566, for ClusterRB 4227;
\item The total number of clusters for FPmax* is 903 versus 5000 for Cluto 903;
\item The number of NDD clusters for FPmax* is 337 versus 773 Cluto.
\end{itemize}

\begin{table}[!h]
\caption{Comparison of the obtained clusters in terms of pairs of near duplicate documents}
\begin{center}

\begin{tabular}{lc}

\hline
The number of ROMIP duplicates: & 2997\\
The number of NDD found by FPmax*: & 2722\\
The number of NDD found by Cluto: & 2897\\
\hline
The number of unique NND pairs of ROMIP: & 1155\\
The number of unique NDD pairs found by FPmax*: & 1001\\
The number of unique NDD pairs found by  Cluto: & 1055\\
\hline
The number of common NDD pairs for FPmax* and ROMIP: & 1721\\
The number of common NDD pairs for Cluto and ROMIP: & 1842\\
\hline

\end{tabular}

\end{center}
\end{table}

Graphs and tables show that for 5000 clusters the output of ClusterRB has almost the same value of F-measure (0.63) as FPmax* for threshold 150 (F1=0,61). However, computations took 4 hours for ClusterRB and half a second for FPmax*.

We continued our developments of that time and developed GUI and duplicate document generator (for a provided text collection) for testing purposes \cite{Ignatov:2009s}.The archive of these projects is freely available at Bitbucket\footnote{\url{https://bitbucket.org/dimanomachine/nearduplicatesarch}}.

Later on, we proposed a prototype of near-duplicate detection system for web-shop owners. It's a typical situation for this online businesses to buy description of their goods from so-called copyrighters. A copyrighter may cheat from time to time and provide the owner with some almost identical descriptions for different items. In that study we demonstrated how we can use FCA for revealing and fast online clustering of such duplicates in a real online perfume shop.	Our results were also applicable for near duplicate detection in collections of R\&D project's documents~\cite{Ignatov:2008b}.
	
\subsection{Triadic FCA for IR-tasks in Folksonomies} 

Four our data mining studies on triclustering (see Section~\ref{ssec:tric} and \cite{Ignatov:2015,Gnatyshak:2014,Gnatyshak:2013}) folksonomic data became a shootingrange since the first efficient FCA-based algorithm to mine tiradic data was proposed for mining communities in folksonomies~\cite{Jaschke:2006}.

But it is a rich field with interactive resource-sharing systems like Bibsonomy\footnote{\url{http://www.bibsonomy.org/}},  CiteULike\footnote{\url{http://www.citeulike.org/}}, Flickr\footnote{\url{https://www.flickr.com/}} and Delicious\footnote{\url{https://delicious.com/}} that need fully-fledged IR functionality including retrieval, ranking and recommendations. For example, Bibsonomy is a social bookmarking and publication management system. Apart from DBLP\footnote{\url{http://dblp.uni-trier.de/}}~\cite{Ley:2009} that only collects, stores bibliographic data and provide publication and author search, Bibsonomy allows to create user's own lists of bibliographic bookmarks, use tags and social interactions.

As we have mentioned in Section~\ref{ssec:tric}, the underlying Folksonomy structure is a formal tricontext $\K=(U,T,R,Y)$ with $U$ being a set of users, $T$ a set tags, and $R$ a set of resources, where $Y\subseteq U \times T \times R$ relates entities from these three sets. Sometimes, a user-specific subtag/supertag-relation $\succ$ is also included into the definition, i.e. $\succ \subseteq U \times T \times T$.

We shortly discuss main IR tasks that folksonomic data give rise.

First of all, we have to say that traditional PageRank cannot be directly applied to folksonomies. The authors of paper~\cite{Hotho:2006} modified the PageRank algorithm for folksonomic data by considering the input triadic data as an undirected tripartite graph. The weights for each type of edge were assigned according to the occurrences of the third entity, e.g. an edge ${u, t}$ being weighted with $|{r \in
R : (u, t, r) \in Y }|$, the number occurrences of the related tags.

Formally, the weight spreading condition looks as follows:

$$w \gets \alpha w + \beta Aw + \gamma p, where$$

$A$ is the row-stochastic version of the graph adjacency, $p$ is a preference
vector, $\alpha, \beta, \gamma \in [0, 1]$ are initial parameters with $\alpha + \beta + \gamma = 1$. Thus, $\alpha$ 
regulates the speed of convergence, while the proportion between $\beta$ and $\gamma$ controls the
influence of the preference vector.

However, the first results on Delicious data were rather discouraging even with a term-frequency ranker combination, the resulting ranking was similar (though not identical) to the the initial edge weights.
It resulted in authors' own ranking algorithm FolkRank, which takes into account the difference in the resulting rankings
with and without preference vector~\cite{Hotho:2006}.

In that paper the authors formulated peculiar tasks:

\bi
\item Documents that are of potential interest to a user can be suggested to him. 
\item Other related tags can be suggested to a user. Thus, FolkRank additionally considers the tagging behavior of other users and can be used for tag recommendations.
\item Other users that work on related topics can be made explicit and this facilitates
knowledge transfer and formation of user communities.
\ei

\begin{figure}[!tbh]
	\centering
		\includegraphics[width=1.0\textwidth]{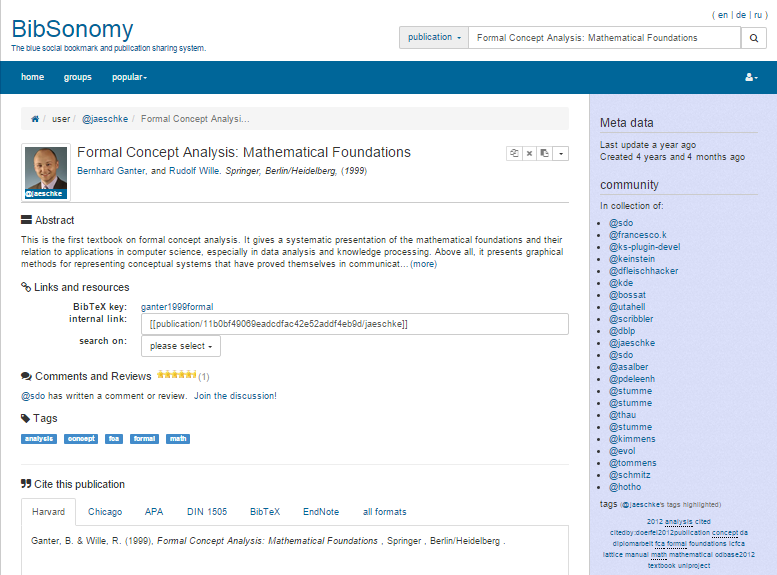}
	\caption{An example of Bibsonomy interface}
	\label{fig:bibsonomy}
\end{figure}

Later on, they implemented (not only) all this features in the Bibsonomy systems~\cite{Benz:2010}.

Moreover, during those studies they admitted that search query logs naturally forms folksonomic data, $(users,queries,resources)$, where the resources are those that were clicked by a user after performing a query~\cite{Benz:2010q}. Predictably, they gave a name logsonomy to this new data structure. When  Bibsonomy was at the early stages, it faced spam  abuse problem and in 2008 ECML PKDD discovery challenge \footnote{\url{http://www.kde.cs.uni-kassel.de/ws/rsdc08/}} addressed this problem. The year after the challenging problem \footnote{\url{http://www.kde.cs.uni-kassel.de/ws/dc09/}} were recommendations for Bibsonomy and it resulted in new fruitful algorithms~\cite{Doerfel:2013}.

\subsection{Exploring taxonomies of web site users}

In 2006 we participated in analysis of web sites audience in collaboration with SpyLog company (now OpenStat\footnote{\url{https://www.openstat.com/}})\cite{Ignatov:2007}.

Owners of a web-site are often interested in analysing groups of
users of their site. Information on these groups can help to
optimise the structure and contents of the site. For example,
interaction with members of each group may be organized in a
special manner. In the performed study we used an approach based on formal
concepts for constructing taxonomies of groups of web users.

For our experiments we have chosen four target websites: the site
of the State University Higher School of Economics,
an e-shop of household equipment, the site of a large bank, and
the site of a car e-shop (the names of the last three sites cannot
be disclosed due to legal agreements).

Users of these sites are described by attributes that correspond
to other sites, either external (from three groups of sites:
finance, media, education) or internal (web-pages of the site).
More precisely, initial ``external" data consists of user records
each containing the user id, the time when the user first entered
this site, the time of his/her last visit, and the total number of
sessions during the period under consideration. An ``internal" user
record, on the other hand, is simply a list of pages within the
target website visited by a particular user.

By ``external" and ``internal" taxonomies we mean (parts of) concept
lattices for contexts with  either ``external" or ``internal"
attributes. For example, the external context has the form
$\K_e= (U,S_e, I_e),$ where $U$  is the set of all users of the target site,  $S_e$  is the
set of all sites from a sample (not including the target one), the
incidence relation $I_e$ is given by all pairs $(u,s)\colon  u\in U, s\in
S_e$, such that user $u$ visited site $s$. Analogously, the internal
context is of the form $\K_i=(U,S_i, I_i)$, where $S_i$ is the set
of all own pages of the target site.

A concept of this context is a pair $(A, B)$ such that $A$ is a
group of users that visited together all other sites from $B$.

As we have mentioned, one of the target websites was the site of our university\footnote{\url{www.hse.ru}}. 

We received ``external" data with the following  fields for each
user-site pair: ({\bf user id, time of the first visit, time of the last visit,
total number of sessions during the period}).
``Internal" data  have almost the same format with an additional field
{\bf url page}, which corresponds to a particular visited page of the target site.

The provided information was gathered from about 10000 sites of Russian
segment of Internet (domain .ru). Describing users in terms of sites they visited, we
had to tackle the problem of dimensionality, since the resulting concept
lattices can be very large (exponential in the worst case in terms
of objects or attributes). To reduce
the size of input data we used the following techniques.

For each user we selected only those
sites that were visited by more than a certain number of times
during the observation period. This gave us information about
permanent interests of particular users. Each target web site was
considered in terms of sites of three groups: newspaper sites, financial sites, and
educational sites. 

However, even for large reduction of input size, concept lattices
can be very large. For example, a context of size $4125\times225$ gave
rise to a lattice with 57 329 concepts.

To choose  interesting groups of users we employed stability
index of a concept defined in \cite{Kuznetsov:1990,Kuznetsov:2007} and considered in
\cite{Roth:2008} (in slightly different form) as a tool for
constructing taxonomies. On one hand, stability index shows the independence
of an intent on particular objects of extent (which may appear or not appear in the context depending
on random factors). On the other hand, stability index of a concept
shows how much extent of a concept is different from similar smaller extents
(if this difference is very small, then its doubtful that extent refers to a
``stable category''). For detailed motivation of stability indices see~\cite{Kuznetsov:1990,Kuznetsov:2007,Roth:2008}.
\bigskip

\bd

Let $\K = (G, M, I)$ be a formal context and $(A,
B)$ be a formal concept of $K$. The stability index $\sigma$ of
$(A, B)$ is defined as follows:

$$\sigma(A,B) = \frac{|\{C\subseteq A | C^\prime=B \}|}{2^{|A|}}.$$
\bigskip

\ed

Obviously, $0 \leq \sigma (A, B) \leq 1$.

The stability index of a
concept indicates how much the concept intent depends on
particular objects of the extent. A stable intent (with stability
index close to 1) is probably ``real" even if the description of
some objects is ``noisy". In application to our data, the stability
index shows how likely we are to still observe a common group of
interests if we ignore several users. Apart from being noise-resistance,
a stable group does not collapse (e.g., merge with a different
group, split into several independent subgroups) when a few
members of the group stop attending the target sites.

In our experiments we used ConceptExplorer for computing and visualising lattices
and their parts.

We compared results of taking most stable concepts (with stability
index exceeding a threshold) with taking an ``iceberg" 
lattice. The results look correlated, but
nevertheless, substantially different. The set of stable extents
contained very important, but not large groups of users.

\begin{figure}
	\centering
		\includegraphics[width=0.9\textwidth]{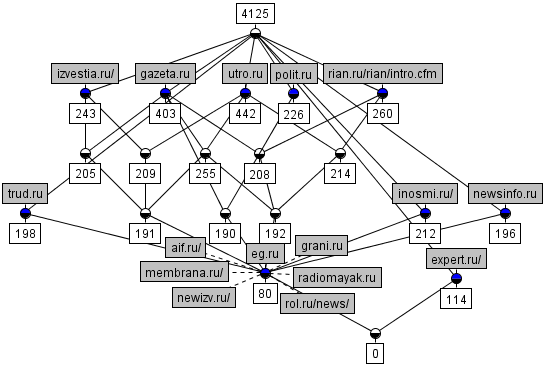}
	\caption{Ordered set of 25 concepts with largest stability} \label{fig:hse_news_20_sep_u_20}
\end{figure}

In Figs.\ref{fig:25iceberg},\ref{fig:hse_news_20_sep_u_20}  we present parts of a concept lattice for the HSE web site described by ``external" attributes which were taken to be Russian
e-newspapers visited by users of  www.hse.ru during one month (September 2006) more than 20 times.
Fig.~\ref{fig:25iceberg}  presents an iceberg
with 25 concepts having largest extent. Many of the concepts correspond to
newspapers that are in the middle of political spectrum, read ``by everybody" and thus,
not very interesting in characterising social groups.

Fig.~\ref{fig:hse_news_20_sep_u_20} presents an ordered set of
25 concepts having largest stability index. As compared to the iceberg,
this part of the concept lattice contains several sociologically important groups
such as readers of ``ExpressGazeta'' (``yellow press"), Cosmopolitan, Expert (high professional
analytical surveys) etc.

\subsection{FCA-based models for crowdsourcing}

The success of modern collaborative technologies is marked by the appearance of many novel platforms for holding distributed brainstorming or carrying out so called ``public examination''. There are a lot of such crowdsourcing companies in the USA (Spigit \footnote{\url{http://spigit.com/}}, BrightIdea \footnote{\url{http://www.brightidea.com/}}, InnoCentive \footnote{\url{http://www.innocentive.com/}} etc.) and Europe (Imaginatik \footnote{\url{http://www.imaginatik.com/}}). There is also the Kaggle platform \footnote{\url{http://www.kaggle.com}} which is the most beneficial for data practitioners and companies that want to select the best solutions for their data mining problems. In 2011 Russian companies launched business in that area as well. The two most representative examples of such Russian companies are Witology \footnote{\url{http://witology.com/}} and Wikivote \footnote{\url{http://www.wikivote.ru/}}. Several all-Russian projects have already been finished successfully (for example, Sberbank-21\footnote{\url{http://sberbank21.ru/}}, National Entrepreneurial Initiative~\footnote{\url{http://witology.com/en/clients_n_projects/3693/}} etc.). The core of such crowdsourcing systems is a socio-semantic network \cite{Roth:2010,Yavorsky:2011}, which data requires new approaches to analyze. Before we tried to accommodate FCA as a methodological base for the analysis of data generated by such collaborative systems~\cite{Ignatov:2013w}.

As a rule, while participating in a project, users of such crowdsourcing platforms \cite{Howe:2006} discuss and solve one common problem, propose their ideas and evaluate ideas of each other as experts. Finally, as a result of the discussion and ranking of users and their ideas we get the best ideas and users (their generators). For deeper understanding of users's behavior, developing adequate ranking criteria and performing complex dynamic and statistic analyses, special means are needed. Traditional methods of clustering, community detection and text mining need to be adapted or even fully redesigned. Earlier we described models of data used in crowdsourcing projects in terms of FCA. Furthermore, we presented the collaborative platform data analysis system CrowDM (Crowd Data Mining), its architecture and methods underlying the key steps of data analysis~\cite{Ignatov:2013w}.
  
 The principles of these platforms' work are different from the work of online-shops or specialized music/films recommender websites. Crowdsourcing projects consist of several stages and results of each stage substantially depend on the previous stage results. That's why the existing models of the recommender systems should be adapted properly. In the accompanion paper~\cite{Ignatov:2014r} or in its shorter predecessors~\cite{Ignatov:2014c,Ignatov:2014w}, we present new methods for making recommendations based on FCA and OA-biclustering (see Section~\ref{sssec:oabic}): The original methods of idea recommendation (for voting stage), like-minded persons recommendation (for collaboration) and antagonists recommendation (for counteridea generation stage). The last recommendation type is very important for stimulating user's activity on Witology platform during the stage of counteridea generation.

\section{FCA in Ontology Modeling and Attribute Exploration}

	Applications of FCA in ontology modeling and its relations with Semantic Web deserve a special treatment.
	However, we shortly mention several interesting approaches and showcase an interactive technique which can be used for ontology  and knowledge bases refinement and building.

	\begin{itemize}
	
	\item Attribute exploration as an expert knowledge acquisition method \cite{Ganter:1999a}
	
	\item FCA in ontology building and refining \cite{Stumme:2001,Cimiano:2005}
	
	\end{itemize}

\subsection{Attribute Exploration}

Attribute exploration is an interactive knowledge acquisition procedure based on implications and counter examples~\cite{Ganter:1999a} that was initially applied for knowledge acquisition in mathematics itself and still a suitable tool up to date~\cite{Revenko:2012}.

The basic algorithm is as follows:
\bi
		\item Start with any (possibly empty) set of objects.
		\item Generate an implication valid in the current subcontext.
		\item If the implication is not valid in the entire context, provide an object that violates it (a counterexample).
		\item Go to the next implication and so on.
\ei
		
		A sophisticated algorithm implementation can follow the Duquenne-Guigues base to ask minimal number of questions.

\begin{example}\label{ex:attexp}	 Attribute exploration for the context of transportation means.
\begin{figure}
	\centering
		\includegraphics[width=0.85\textwidth]{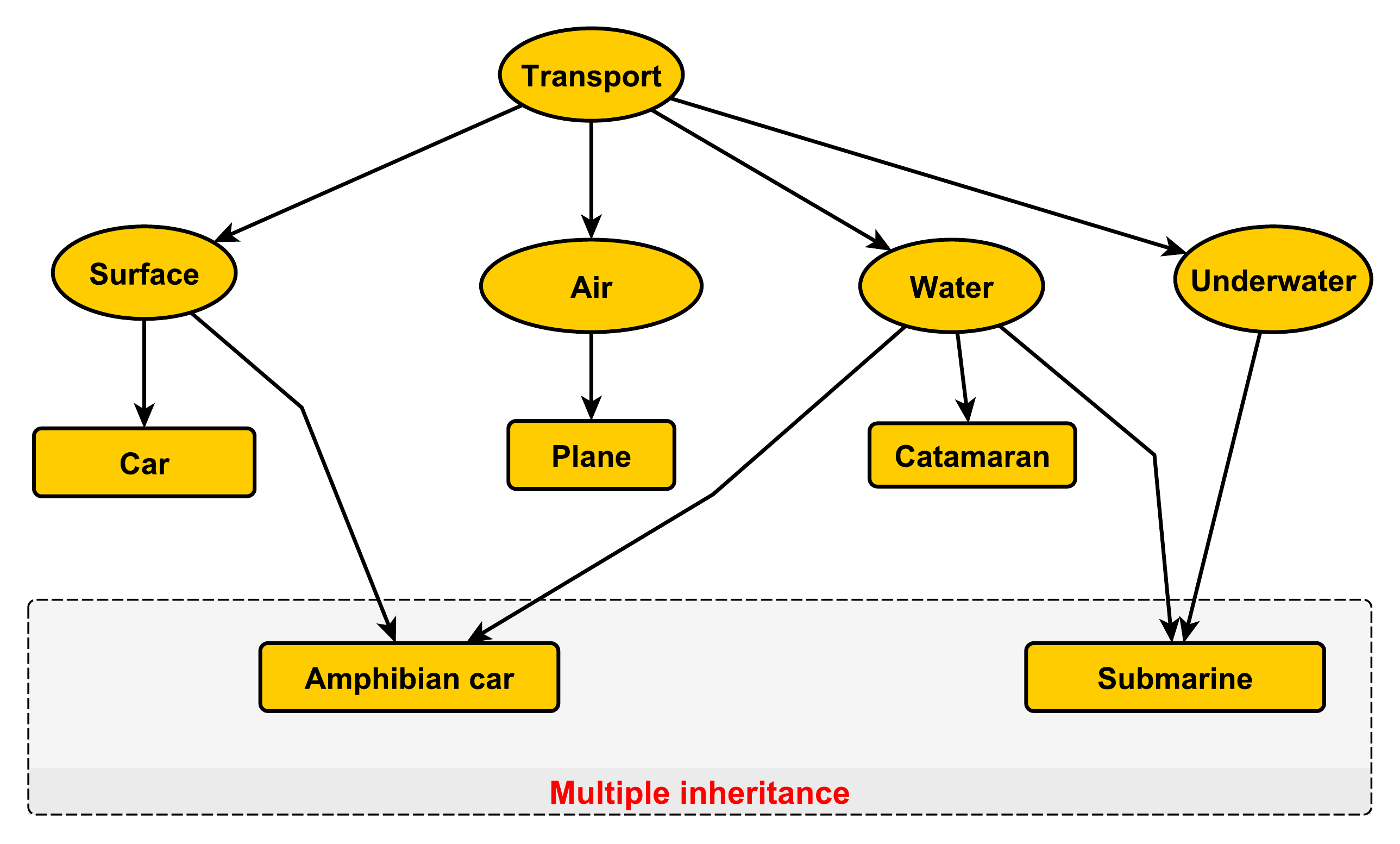}
	\caption{The taxonomy of transportation means as an example of not a tree-like (multiple) inheritance}
	\label{fig:Transportation}
\end{figure}

Whether all possible types of transportation means are enumerated with respect to their attributes (surface, air, water, underwater)?

To this end we start attribute exploration with composing the corresponding formal context.

\begin{center}
\begin{cxt}%
\cxtName{}%
\atr{surface}%
\atr{air}%
\atr{water}%
\atr{underwater}%
\obj{.x..}{plane}
\obj{x.x.}{amphibian car}
\obj{..x.}{catamaran}
\obj{..xx}{car}
\obj{..xx}{submarine}
\end{cxt}
\end{center}

The main steps of attribute exploration, as a dialog between system $A$ and expert $E$ for transport context, is as follows:
\bi
\item{Step 1.} $A$ Question: Is it true that, when an object has attribute ``Can move underwater'', it also has attribute ``Can move by water''?
\item{Step 1.} $E$ Answer: Yes, it is. The expert knows that it is true for submarines and there are no other types of underwater transport.
\item{Step 2.} $A$ Question: Is it true that, when an object has attributes ``Can move by air'' and ``Can move by water'' have attributes ``Can move by surface'' and ``Can move underwater''?
\item{Step 2.} $E$ Answer: No, it is not. There is a counterexample, $\{hydroplane\}^\prime = \{air, water\}$.
\item{Step 3.} $A$ Question: Is it true that, when an object has attributes ``Can move by air'', ``Can move by water'' и ``Can move underwater'' have attributes ``Can move by surface''?
\item{Step 3.} $E$ Answer: Yes, it is.
$\{air, water, underwater\}^\prime = \emptyset$.
\item{Steps 4, 5, 6} Trivial questions.
\ei
$\square$
\end{example}

The resulting concept lattice can be considered as a non-tree like taxonomy of transportation means since it allows multiple inheritance in the concept hierarchy. If the expert suppose that the not only objects but attributes are missed then object exploration can be done in similar manner, e.g. by the same procedure on the transposed context.

\begin{exercise}\label{exercise:attrexp} 1. Compare the concept lattices from the previous example before starting and after completion of the attribute exploration. What is/are new concept(s) that we have obtained? How can it/they be interpreted?
2. Perform attribute exploration with ConceptExplorer for a slightly modified context from~\cite{Sertkaya:2010}

\begin{center}
\begin{cxt}%
\cxtName{}%
\atr{Asian}%
\atr{EU}%
\atr{European}%
\atr{G7}%
\atr{Mediterranean}%
\obj{.xxxx}{France}
\obj{x.x.x}{Turkey}
\obj{.xxx.}{Germany}
\end{cxt}
\end{center}

$\square$

\end{exercise}

\subsection{FCA in ontology building and refining}

Often, the notion of Ontology in Computer Science is introduced as related sets of concepts and the typical relation can be ``is-a'', ``has-a'', ``part-of'', or super/subconcept relation. Concept lattices could be seen as ontology-like structures since they feature hierarchically related concepts by super/subconcept order (cf. subsumption order in Descriptive logic).  However, because of their simplicity tree-like ontologies seem to be more popular, thus in the early paper of Cimiano et al.\cite{Cimiano:2005}, the way to transform concept lattices built from text collections to tree-like ontologies was proposed.  

\begin{exercise} Build concept lattice from the context of terms extracted from texts (left). Find the transformation that resulted in the tree-like ontology of terms on the right side.

\begin{minipage}[h]{0.25\linewidth}
 \begin{center}
\begin{cxt}%
\cxtName{}%
\atr{bookable}%
\atr{rentable}%
\atr{rideable}%
\atr{driveable}%
\atr{joinable}%
\obj{x....}{hotel}
\obj{xx...}{apartment}
\obj{xxx..}{car}
\obj{xxxx.}{bike}
\obj{x...x}{excursion}
\obj{x...x}{trip}
\end{cxt}
\end{center}
\end{minipage}
\hfill  
\begin{minipage}[h]{0.60\linewidth}  
\Tree [.bookable [.joinable [.\node[draw]{excursion}; ] [.\node[draw]{trip}; ] ]
				[.\node[draw]{hotel}; ]
				[.rentable [.driveable [.rideable [.\node[draw]{bike}; ] ] [.\node[draw]{car}; ] ]  [.\node[draw]{apartment}; ] ] ]  
\end{minipage}

$\square$

\end{exercise}

Another example where FCA can help is ontology merging: The authors of~\cite{Stumme:2001} successfully tested their FCA-based merging approach on two text collection from touristic domain.

There is also strong connection between Description Logic, Ontologies and Formal Concept Analysis~\cite{Sertkaya:2010}.

Thus OntoComP\footnote{\url{http://code.google.com/p/ontocomp/}}~\cite{Sertkaya:2009} is a Prot{\'{e}}g{\'{e}}\footnote{\url{http://www.co-ode.org/downloads/protege-x/}} 4 plugin for OWL ontologies completion. It enables the user to check whether an OWL ontology contains ``all relevant information'' about the application domain, and extend the ontology appropriately otherwise. It asks the users questions like ``are instances of classes $C_1$ and $C_2$ also instances of the class $C_3$?''. If the user replies positively, then a new axiom of the application domain (that does not follow from the ontology) has been discovered, and this axiom should be added to the ontology. If the user provides a counterexample to this question, i.e., an object that is an instance of $C_1$, $C_2$ and not $C_3$. When all such questions (about the initially given classes) have been answered, the ontology is supposed to be complete.

Obviously, this approach that was originally introduced in~\cite{Baader:2007} for completing Description Logic knowledge bases uses attribute exploration.

It seems that attribute exploration is a fruitful technique for ontology building and refinement. Two more examples,  Rudolph~\cite{Rudolph:2006} proposed its extension for relation exploration in ontological modeling for knowledge specification and recently in combination with machine learning techniques attribute exploration was used for ontology refinement~\cite{Potoniec:2014}. You probably have seen from exercise\ref{exercise:attrexp}, that attribute exploration may be uneasy because of laborious fact checking.  However, to help potential users, in ~\cite{jaschke:2013} the authors paired attribute exploration with web information retrieval, in particular by posing appropriate queries
to search engines\footnote{\url{https://github.com/rjoberon/web-attribute-exploration}}.

\section{Conclusion}

In the end of the invited talk at the ``FCA meets IR'' workshop 2013, Prof. Carpineto has summarised strengths and limitations of
FCA for IR.  It seems to be evident that IR will be increasingly relying on contextual
knowledge and structured data and FCA can improve both query pre-processing
and query post-processing of modern IR systems. Among the mentioned technologies that could benefit from FCA are query expansion, web search diversification, ontology-based information retrieval, querying and navigating RDF (there is a progress to this date~\cite{Codocedo:2014}), and many others. However, the community needs to endeavour (by theoretical advances
and system engineering) to deploy a comprehensive FCA-based tool for information retrieval and integrate it with existing search and indexing taking into account both the intrinsic complexity issues and the problem of good features generation.

Even in an extensive tutorial it is not possible to cover all models and applications of Formal Concept Analysis. For example, concept lattices and its applications in social sciences including Social Network Analysis deserve a special treatment. The grounding steps have been done by Vincent Duquenne~\cite{Duquenne:1996}, Linton Freeman~\cite{Freeman:1996} and their collaborators (see also \cite{Gnatyshak:2012} for our SNA-related study).
Another large and interesting domain is Software Engineering~\cite{Tilley:2005,Arevalo:2009}.
For these two and many other topics, we also refer the readers to the recent surveys~\cite{Poelmans:2013a,Poelmans:2013b}.

Overall, we hope that this introductory material with many examples and exercises will help the reader not only to understand the theory basics, but having this rich variety of tools and showcases to use FCA in practice.

\subsubsection{Acknowledgments.} The author would like to thank all colleagues who have made this tutorial possible: Jaume Baixeries, Pavel Braslavsky, Peter Becker, Radim Belohlavek, Aliaksandr Birukou, Jean-Francois Boulicaut, Claudio Carpineto, Florent Domenach, Fritjhof Dau, Vincent Duquenne, Bernhard Ganter, Katja Hofmann, Robert Jaeshke, Evgenia Revne (Il'ina), Nikolay Karpov, Mehdy Kaytoue, Sergei Kuznetsov, Rokia Missaoui, Elena Nenova,  Engelbert Mephu Nguifo, Alexei Neznanov,  Lhouari Nourin, Bjoern Koester, Natalia Konstantinova, Amedeo Napoli, Sergei Obiedkov, Jonas Poelmans, Nikita Romashkin, Paolo Rosso, Sebastian Rudolph, Alexander Tuzhilin, Pavel Serdyukov, Baris Serkaya, Dominik Slezak, Marcin Szchuka, and, last but not least, the brave listeners. 
The author would also like to commemorate Ilya Segalovich who inspired the author's enthusiasm in Information Retrieval studies, by giving personal explanations of near duplicate detection techniques in 2005, in particular.

Special thank should go to my grandmother, Vera, who has been hosting me in a peaceful countryside place, Prechistoe, during the last two weeks of the final preparations. 

The author was partially supported by the Russian Foundation for Basic Research grants no. 13-07-00504 and 14-01-93960 and prepared the tutorial within the project ``Data mining based on applied ontologies and lattices of closed descriptions'' supported by the Basic Research Program of the National Research University Higher School of Economics.

\bibliographystyle{splncs}
\bibliography{tutbib}

\end{document}